\documentclass[12pt,english]{article}
\usepackage[english]{babel}
\usepackage{amsmath,amssymb,amscd,xcolor}
\usepackage{amsfonts}
\usepackage{mathrsfs}
\usepackage{mathtools}
\usepackage{float}
\usepackage{graphicx}
\usepackage{url}
\usepackage{hyperref}
\usepackage{cite}
\usepackage{physics}
\usepackage{braket}
\usepackage{verbatim}
\usepackage[font=footnotesize,labelfont=bf,width=.9\textwidth]{caption}
\usepackage{multirow}
\usepackage{boldline}
\usepackage{enumitem}
\usepackage{wrapfig}
\usepackage[normalem]{ulem}
\allowdisplaybreaks

\hypersetup{
    colorlinks,%
    citecolor=blue,%
    filecolor=blue,%
    linkcolor=blue,%
    urlcolor=blue
}

\makeatletter
\renewcommand\section{\@startsection {section}{1}{\z@}%
                                   {-5.5ex \@plus -1ex \@minus -.2ex}%nn
                                   {2.3ex \@plus.2ex}%
                                   {\normalfont\large\bfseries}}
\renewcommand\subsection{\@startsection{subsection}{2}{\z@}%
                                     {-3.25ex\@plus -1ex \@minus -.2ex}%
                                     {1.5ex \@plus .2ex}%
                                     {\normalfont\bfseries}}

\numberwithin{equation}{section}

\makeatother

\newcommand{\bea}{\begin{eqnarray}}
\newcommand{\eea}{\end{eqnarray}}
\newcommand{\beq}{\begin{equation}}
\newcommand{\eeq}{\end{equation}}
\newcommand{\be}{\begin{equation}}
\newcommand{\ee}{\end{equation}}
\newcommand{\nn}{\nonumber}

\newcommand{\mf}[1]{\mathfrak{#1}}
\newcommand{\ms}[1]{\mathsf{#1}}
\newcommand{\hmf}[1]{\hat{\mathfrak{#1}}}
\newcommand{\tmf}[1]{\tilde{\mathfrak{#1}}}
\newcommand{\mc}[1]{\mathcal{#1}}
\newcommand{\blangle}{\big\langle}
\newcommand{\brangle}{\big\rangle}

\newcommand{\pa}{\partial}
\newcommand{\mTr}{\mathrm{Tr}}
\newcommand{\mtr}{\mathrm{tr}}
\newcommand{\Li}[1]{\text{Li}_{#1}}
\newcommand{\gev}[1]{\big\langle #1\big\rangle_{\rm grav}}
\newcommand{\barmc}[1]{\bar{\mathcal{#1}}}
\newcommand{\bs}[1]{\boldsymbol{#1}}
\newcommand{\Zs}{Z_\text{scalar}}
\newcommand{\Zg}{\mathcal{Z}_\text{grav}}

\renewcommand{\title}[1]{\vbox{\center\LARGE{#1}}\vspace{5mm}}
\renewcommand{\author}[1]{\vbox{\center#1}\vspace{5mm}}
\newcommand{\address}[1]{\vbox{\center\footnotesize\em#1}}
\newcommand{\email}[1]{\vbox{\center\footnotesize\tt#1}\vspace{5mm}}

\begin{document}
\begin{titlepage}
 \begin{flushright}

\end{flushright}

\begin{center}

\hfill \\
\hfill \\
\vskip 1cm
\title{Keeping matter in the loop\\ 
in dS$_3$ quantum gravity}

\author{Alejandra Castro$^{a}$, Ioana Coman$^{b,c}$, Jackson R. Fliss$^{a,c}$, and Claire Zukowski$^{d}$}
\address{
{$^{a}$Department of Applied Mathematics and Theoretical Physics, University of Cambridge,\\ Cambridge CB3 0WA, United Kingdom}\\
$^{b}${Kavli Institute for the Physics and Mathematics of the Universe (WPI), University of Tokyo,\\ Kashiwa, Chiba 277-8583, Japan}\\
$^{c}$Institute for Theoretical Physics, University of Amsterdam,\\ Science Park 904, 1090 GL Amsterdam, The Netherlands\\
$^{d}${Department of Physics and Astronomy, University of Minnesota Duluth,\\ Duluth, MN 55812, USA}
}
  
\email{ac2553@cam.ac.uk, ioana.coman@ipmu.jp, jf768@cam.ac.uk, czukowsk@d.umn.edu}

\end{center}

\vfill
\abstract{
We propose a mechanism that couples matter fields to three-dimensional de Sitter quantum gravity. 
Our construction is based on the Chern-Simons formulation of three-dimensional Euclidean gravity, and it centers on a collection of Wilson loops winding around Euclidean de Sitter space. We coin this object a Wilson spool. 
To construct the spool, we build novel representations of $\mathfrak{su}(2)$. 
To evaluate the spool, we adapt and exploit several known exact results in Chern-Simons theory. 
Our proposal correctly reproduces the one-loop determinant of a free massive scalar field on $S^3$ as $G_N\to 0$.
Moreover, allowing for quantum metric fluctuations, it can be systematically evaluated to any order in perturbation theory.  
}
\vfill

\end{titlepage}

\newpage
{
  \hypersetup{linkcolor=black}
  \tableofcontents
}
\newpage 
\section{Introduction}
\label{sect:intro}

de Sitter (dS) spacetime is famous for its prominence in theoretical cosmology. It is also infamous for its elusiveness in quantum gravity. One obstruction arises from the fact that it is notoriously difficult to realize in string theory. Another point of frustration is that despite its similarities with Anti-de Sitter (AdS) spacetime, basic and fundamental aspects of a holographic description for de Sitter remain unsettled. The arguments supporting and reflecting these obstructions can be found in, e.g., \cite{Witten:2001kn,Strominger:2001pn,Banks:2005bm,Anninos:2012qw}.   

There is however a context for which quantum gravity in dS has an advantage relative to its AdS cousin. In three space-time dimensions one has the luxury of casting general relativity (with or without a cosmological constant) in terms of a Chern-Simons theory \cite{Achucarro:1987vz,Witten:1988hc}. For a theory with a positive cosmological constant, i.e., dS$_3$ gravity,  the gauge group of the Chern-Simons theory is $SO(4)$ in Euclidean signature. This is an advantage since several aspects of Chern-Simons theory are known exactly when the gauge group is compact. In the absence of other degrees of freedom---for a pure theory of gravity---this partnership with Chern-Simons has been leveraged successfully to discuss quantum aspects of dS$_3$ gravity at the perturbative and non-perturbative level \cite{Carlip:1992wg,Guadagnini:1995wv,Banados:1998tb, Park:1998yw, Govindarajan:2002ry,Castro:2011xb,Anninos:2020hfj,Anninos:2021ihe,Hikida:2021ese}.

Here we will push this advantage one step further. We will demonstrate that the Chern-Simons formulation of dS$_3$ gravity can address the following question: \emph{How does quantum gravity alter the physics of matter fields in de Sitter?}

It is useful to contrast this question between the metric and Chern-Simons formulations. The metric couples naturally to matter through minimal coupling to a metric-compatible connection.  In the Chern-Simons formulation it is not clear how to write down such a coupling while also maintaining the topological nature that makes it attractive as a theory of gravity.  In the context of effective field theory this situation is natural: Chern-Simons theories arise as effective theories precisely by integrating out degrees of freedom above some mass gap.  These massive degrees of freedom are not invisible to the low-energy theory: their remnants are topological line operators, Wilson lines, which are regarded, loosely, as the worldlines of charged matter.  With this tenet in mind we will build an effective coupling that captures the physics of massive matter directly in the Chern-Simons theory.

In brief, we will show how to leverage exact results of Wilson loops with the purpose of building a field that couples to Chern-Simons connections. This involves showing that a path integral of a massive field can be viewed as a Wilson loop that ``winds'' arbitrarily many times around the Euclidean spacetime. We call the resulting winding operator a Wilson spool, which has a precise definition in classical and quantum gravity, and can be computed to any order in the gravitational coupling, Newton's constant, $G_N$.  Our proposal is synergetic with the work of \cite{Ooguri:1999bv}, however ours is constructed to have a broader validity for Chern-Simons gravity\footnote{Namely, the na\"ive application of  \cite{Ooguri:1999bv}  will lead to an incorrect scalar one-loop determinant in the classical  limit ($G_N\rightarrow 0$). In section \ref{sec:testofspool} we illustrate how our proposal reproduces the one-loop determinant of massive scalar fields correctly.}. There are several ingredients and steps that go into our construction, so let us elaborate on the specifics of our approach.

At a schematic level, let us first consider writing the gravitational partition function as a sum of metric path-integrals weighted by the Einstein-Hilbert action, $I_{\rm EH}$, and performed about saddle-point geometries $M$:
\beq
\mc Z_{\rm grav}=\sum_{M}\mc Z_{\rm grav}[M]=\sum_{M}\int \left[\mc Dg_{\mu\nu}\right]_{M}e^{-I_{\rm EH}[g_{\mu\nu}]}~,
\eeq
where to leading order in a $G_N$ expansion, the saddle-point is given by a metric, $g_{\mu\nu}^{(M)}$, that satisfies the Einstein equations.  We can then consider coupling matter to this gravity path-integral about each saddle\footnote{We are being a little bit misleading here: $Z_{\rm matter}$ cannot be strictly decoupled from the sum as including matter will shift the saddle, $M$.  However this shift is at order $G_N$.  We will make more sense of this shortly in the Chern-Simons context where the $O(G_N^0)$ background fields will be connected to the {\it topology} upon which we quantize the Chern-Simons theory.  All ``back-reactive'' effects however will be handled in localization.} via
\beq\label{eq:Zgmsaddlesum}
\mc Z_{\rm grav+matter}=\sum_{M}\mc Z_{\rm grav+matter}[M]=\sum_{M}\int[\mc Dg_{\mu\nu}]_Me^{-I_{\rm EH}[g_{\mu\nu}]}Z_{\rm matter}[g_{\mu\nu}]~,
\eeq
where 
\beq
Z_{\rm matter}[g_{\mu\nu}]=\int [\mc D\phi]e^{iS_{\rm matter}[\phi,g_{\mu\nu}]}~
\eeq
is the path-integral of a quantum field $\phi$  minimally coupled to a background geometry, $g_{\mu\nu}$.  Whether or not a sum over geometries converges, or what it converges to,  depends sensitively on the UV completion of the theory of matter and quantum gravity.  However, on a more pragmatic level, we will restrict \eqref{eq:Zgmsaddlesum} to a given saddle $M$, and focus on the quantum gravity corrections to fields quantized on the background, $g_{\mu\nu}^{(M)}$. That is, we will focus on
\beq
\gev{Z_{\rm matter}[M]}:=\int[\mc Dg_{\mu\nu}]_Me^{-I_{\rm EH}[g_{\mu\nu}]}Z_{\rm matter}[g_{\mu\nu}]~.
\eeq
In this regard, let us define an object, $\mathbb W$, such that
\beq
Z_{\rm matter}[g_{\mu\nu}]=\exp\left(\frac{1}{4}\mathbb W[g_{\mu\nu}]\right)~.
\eeq
Formally expanding this out in the gravity path-integral gives
\be
\begin{aligned}
\gev{Z_{\rm matter}[M]}&=\int[\mc Dg_{\mu\nu}]_Me^{-I_{\rm EH}[g_{\mu\nu}]}\left(1+\frac{1}{4}\mathbb W[g_{\mu\nu}]+\ldots\right)\\&=\mc Z_{\rm grav}+\frac{1}{4}\gev{\mathbb W}+\ldots ~.
\end{aligned}
\ee
$\mathbb W$ is what we coin the Wilson spool.%\footnote{\cz{See also~\cite{Ooguri:1999bv} for an example of a spool-like object that is related to one-loop determinants. Note that a naive definition of the spool as a sum over winding would fail to describe gravity in de Sitter spacetime. Our construction is more general to encompass this case.}} 
What we will show is that $\gev{\mathbb W}$ has a very precise definition in Chern-Simons theory, and it can be evaluated to any order in $G_N$. From this point, 
we take on the task of evaluating the gravity path-integral with $\mathbb W$ inserted to calculate
\beq\label{eq:Wgev1}
\gev{\log Z_{\rm scalar}[M]}=\frac{1}{4}\gev{\mathbb W}~.
\eeq

Ultimately our ability to define and evaluate \eqref{eq:Wgev1} relies on celebrated exact methods for the $SU(2)$ Chern-Simons path-integral on a topologically $S^3$ manifold. However, in order to follow through with this task, we need to address several aspects about the relationship between gravity and Chern-Simons that make these methods somewhat unnatural. In particular, there are three aspects we will incorporate:
\begin{description}[leftmargin=0.4cm]
    \item[Non-trivial background connections.] One key aspect of the gravitational interpretation of Chern-Simons theory is that the connections are non-trivial: this is the minimal requirement to have a locally invertible metric. We take this into account.  
    \item[Complex Chern-Simons levels.] $G_N$ controls the imaginary part of the Chern-Simons level $k$. Following the arguments in \cite{Witten:1989ip}, we incorporate complex couplings together with non-trivial background connections. 
    \item[Non-standard $\mathfrak{su}(2)$ representations.] As illustrated in \cite{Castro:2020smu}, fields in dS$_3$ do not fall into the traditional unitary representations of $\mathfrak{su}(2)$. We will extend the representations constructed in \cite{Castro:2020smu} to cover heavy and light scalar fields in dS$_3$. 
\end{description}
With these facts in mind, we then alter two well-known exact methods---Abelianisation and supersymmetric localization---to accommodate these three aspects.  We provide detailed derivations that show how to evaluate the path integral, and Wilson loops, on $S^3$ of $SU(2)_k$ Chern-Simons theory. There are several moving parts to this story and much of the volume of this paper is devoted to giving these elements solid foundation.

Having assembled all of these pieces, we are then able to utilize the above ingredients to define $\mathbb W$ in terms of Wilson loops living in non-standard representations and to evaluate it in the gravitational path-integral.  This evaluation is more than schematic: we will show that $\gev{\mathbb W}$ reduces to a simple integral and has a well-defined $G_N$ expansion about the $S^3$ saddle.  The tree-level, i.e. $O(G_N^0)$, contribution is precisely the one-loop determinant of a scalar field with mass $\mathsf m$ minimally-coupled to a background $S^3$, i.e.,
\beq
\lim_{G_N\rightarrow 0}\frac{\gev{\mathbb W}}{\mc Z_{\rm grav}[S^3]}=4\log Z_{\rm scalar}[S^3]=4\log\det(-\nabla^2_{S^3}+\mathsf{m}^2\ell^2)^{-1/2}~.
\eeq
We also show that at any order of $G_N$ perturbation theory, the integral defining $\gev{\mathbb W}$ is finite and can be practically evaluated.  This provides a useful framework for calculating quantum-gravitational corrections to $\log Z_{\rm scalar}[S^3]$.

\subsection{Overview}

In a nutshell, this work is divided into two parts. The first part, described in Sec.\,\ref{sec:CS-grav-main}, is devoted to the gravitational path integral on dS$_3$ in the absence of matter fields.  The second part tackles our central question, how to quantify the coupling of matter to quantum gravity, which we present in Sec.\,\ref{sec:looping-matter-in}. In the following we describe in more detail the contents of each section.

We will begin with a review of some basic features of three-dimensional Euclidean de Sitter space in Sec.\,\ref{sect:dSintro}, and then review the classical relation between three-dimensional gravity and Chern-Simons theory in Sec.\,\ref{sect:CSgrav}. This will establish our notation and our basic ingredients.  We will then move on to tackle quantum aspects of dS$_3$ gravity using the Chern-Simons formulation focusing on the $S^3$ topology. The advantage is that Chern-Simons theory can provide an all-loop answer for the path integral around a fixed topology. However, to exploit this advantage we need to revisit and rederive results in $SU(2)_k$ Chern-Simons theory: we have to incorporate a non-trivial background connection and a complex level in the path integral. This is done in Sec.\,\ref{sect:CStheory}, where we show carefully and in detail how to adapt the methods of Abelianisation and supersymmetric localization. From here, we obtain an exact result for the Chern-Simons path integral on $S^3$. In Sec.\,\ref{sec:all-loop-grav} we place these exact results in a gravitational context and compare with prior literature. 

In Sec.\,\ref{sec:looping-matter-in}, we loop matter in. That is, we propose a method to incorporate matter fields in the gravitational path integral. Our strategy is again to use the Chern-Simons formulation, and the outcome is what we coin a Wilson spool. This is a delicate procedure, but the payoff is that we will obtain an all-loop result that can be easily evaluated to any order perturbatively in $G_N$. To this end, in Sec.\,\ref{sect:dSrepintro}, we first construct  non-standard representations of $\mf{su}(2)_L\oplus\mf{su}(2)_R$: these mimic the single-particle representations of $\mf{so}(1,3)$, while being well-adapted to Chern-Simons theory. In Sec.\,\ref{sec:Wilson-SU2}, we evaluate a Wilson loop for the non-standard representations while also incorporating the new ingredients of Sec.\,\ref{sect:CStheory}. This again is done using Abelianisation and supersymmetric localization. And finally, in Sec.\,\ref{sect:spool}, by making use of Wilson loops of non-standard representations, we present the Wilson spool. This object captures the path integral of a massive scalar field coupled to dS$_3$ gravity. We first provide various tests of our proposal and subsequently provide a detailed construction of our proposal. We then follow by quantifying quantum properties of the Wilson spool.

We end with an extended discussion in Sec.\,\ref{sect:disc}, where we present new directions and open questions based on our proposal. We have also included several appendices with complementary material: App.\,\ref{app:conventions} covers conventions of $\mf{su}(2)$ and $\mf{so}(1,3)$, App.\,\ref{sec:locApp} discusses non-Abelian localization (and its shortcoming for our purposes), App.\,\ref{app:HK} provides a review of the heat kernel on $S^3$, and App.\,\ref{app:casimir} adds details on how to interpret the Laplacian on $S^3$ in terms of a local $\mf{su}(2)_L\oplus\mf{su}(2)_R$ action. 

% Lastly, in the discussion, Sec.\,\ref{sect:disc} we comment on the historical and scientific accuracy of the new Oppenheimer movie.

\section{dS\texorpdfstring{$_3$}{3} gravity and Chern-Simons theory}\label{sec:CS-grav-main}
In this section we will review the relation between Chern-Simons theory and three-dimensional general relativity with a positive cosmological constant, i.e., dS$_3$ gravity \cite{Witten:1988hc}. Our presentation follows the work of \cite{Castro:2011xb,Anninos:2020hfj}, which discusses the tree-level and loop relation between the two theories.

In the first half of this section we review geometrical properties of dS$_3$, and then classical (tree-level) aspects of the theories at hand. The second half is devoted to quantum aspects of dS$_3$. Our aim is to capture perturbative corrections to all orders in $G_N$ via the Chern-Simons formulation. An important and novel portion of our analysis is to alter known methods to quantize Chern-Simons theory such that we meet the basic features that give Chern-Simons theory a gravitational interpretation. These alterations are to incorporate a non-trivial background connection and a complex level in the path integral. This is done in Sec.\,\ref{sect:CStheory}, where we show how to adapt the derivation of the exact  Chern-Simons path integral on $S^3$ via two different methods commonly used in the literature: Abelianisation and supersymmetric localization. In Sec.\,\ref{sec:all-loop-grav} we discuss how these modifications are in perfect agreement with perturbative results in the metric formulation of the gravitational theory.

\subsection{A primer on dS\texorpdfstring{$_3$}{3} spacetime}\label{sect:dSintro}

Three-dimensional Lorentzian de Sitter space can be realised as the hypersurface in $\mathbb R^{1,3}$ (what we will call embedding space) given by
\beq\label{eq:embeddSdef}
\eta_{AB}X^AX^B=\ell^2~,\qquad\eta=\text{diag}(-1,1,1,1)~,
\eeq
where $A,B\in\{0,1,2,3\}$ and $\ell$ is the dS$_3$ radius.  Embedding space makes it manifest that the isometry group of dS$_3$ is $SO(1,3)$ which is generated by Killing vectors preserving this hypersurface
\beq
L_{AB}=X_A\frac{\pa}{\pa X^B}-X_B\frac{\pa}{\pa X^A}~.
\eeq
More details on the group algebra, $\mf{so}(1,3)$, as well as our conventions are given in App.\,\ref{app:conventions}.

\begin{figure}
\centering
\includegraphics[width=.4\textwidth]{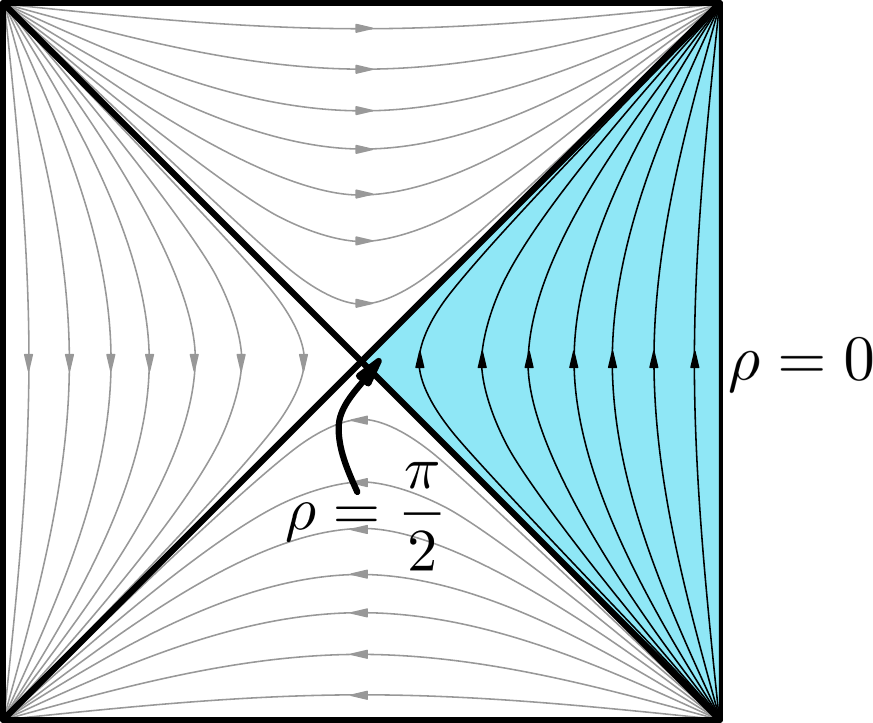}
\caption{\label{fig:penrose}{The Penrose diagram of dS$_3$.  The blue region is the static patch covered by the coordinates \eqref{eq:SPcoords}.  The observer defining the patch lies at $\rho=0$.  Their causal horizon lies at $\rho=\frac{\pi}{2}$.  Also depicted are the flow lines of $D$ which are time-like in this patch.}}
\end{figure}

Different parametrizations of \eqref{eq:embeddSdef} give different coordinate patches of global de Sitter.  A particular coordinate patch of interest to us in this paper is the coordinate patch available to an observer moving along a timelike geodesic, called the {\it static patch}.  Due to the accelerated expansion of the spacetime, individual observers lose causal contact with increasing portions of space which become hidden behind a causal horizon.  Thus the static patch covers a finite causal diamond, depicted as the blue region of the Penrose diagram found in Fig.\,\ref{fig:penrose}. The parametrization for this static patch is given by
\be
\begin{aligned}\label{eq:SPcoords}
X^0=&\ell\cos(\rho)\sinh(t/\ell)~,\\
X^1=&\ell\cos(\rho)\cosh(t/\ell)~,\\
X^2=&\ell\sin(\rho)\cos(\varphi)~,\\
X^3=&\ell\sin(\rho)\sin(\varphi)~,
\end{aligned}
\ee
for which the metric takes the following form
\beq\label{eq:LorSPmetric}
ds^2=\eta_{AB}\dd X^A\dd X^B=-\cos^2\rho\,\dd t^2+\ell^2\dd\rho^2+\ell^2\sin^2\rho\,\dd\varphi^2~.
\eeq
The coordinates range over $t\in(-\infty,\infty)$, $\rho\in[0,\pi/2)$, $\varphi\in[0,2\pi)$ which covers the right-wedge (``north-pole'') of the static patch.  The point $\rho=0$ corresponds to the worldline of the observer defining the static patch, while $\rho=\pi/2$ corresponds to this observer's causal horizon.  The metric \eqref{eq:LorSPmetric} has an obvious time-like Killing vector, $\zeta=\pa_t$.  This Killing vector is in fact the same as the ``dilatation'' Killing vector, $D=L_{03}$, of $\mf{so}(1,3)$.   As depicted in Fig.\,\ref{fig:penrose}, this Killing vector is not globally time-like, however.

Euclidean de Sitter can be defined through the Wick rotation $X^0=-iX^0_E$, which at the level of the static patch coordinates can be achieved through $t=-i\ell\tau$.  The defining equation \eqref{eq:embeddSdef} then defines a three-sphere and indeed the Lorentzian static patch metric rotates to
\beq\label{eq:EucSPmetric}
\frac{ds^2}{\ell^2}=\cos^2\rho\,\dd\tau^2+\dd\rho^2+\sin^2\rho\,\dd\varphi^2~,
\eeq
which is the metric for $S^3$ in torus coordinates.  Regularity at the horizon, $\rho=\pi/2$, requires the identification $\tau\sim\tau+2\pi$, consistent with it being a coordinate for $S^3$.  The isometry group of Euclidean de Sitter is easily seen to be $SO(4)\simeq SU(2)\times SU(2)/\mathbb Z_2$.  The $SU(2)$'s are the left and right group actions acting on $S^3$ which itself is diffeomorphic to $SU(2)$ (the explicit form of this left and right action is given in App.\,\ref{app:casimir}).  As such we will label these two groups by subscripts $L$ and $R$.  The Wick rotation of $\mf{so}(1,3)$ to $\mf{so}(4)$ and the subsequent splitting into $\mf{su}(2)_L\oplus\mf{su}(2)_R$ is given explicitly in App.\,\ref{app:conventions}.

\subsection{Chern-Simons theory and dS\texorpdfstring{$_3$}{3} gravity: tree-level}\label{sect:CSgrav}
 
Now let us briefly review the Chern-Simons formulation of three-dimensional gravity \cite{Witten:1988hc}; see \cite{Carlip:1992wg,Castro:2011xb} for more details and complementary aspects.  Much like the previous subsection, this portion is intended to lay out the necessary ingredients and to establish our notation.  

As we noted in Sec.\,\ref{sect:dSintro}, the splitting $\mf{so}(4)\simeq\mf{su}(2)_L\oplus\mf{su}(2)_R$ of the isometry algebra of Euclidean dS$_3$ indicates that we will be interested in quantizing a pair of $SU(2)$ Chern-Simons theories
\beq\label{eq:SL+SR}
S=k_LS_{\rm CS}[A_L]+k_RS_{\rm CS}[A_R]~,
\eeq
where
\beq\label{eq:CSaction}
S_{\rm CS}[A]=\frac{1}{4\pi}\Tr\int_M\left(A\wedge \dd A+\frac{2}{3}A^3\right)~,
\eeq
and the trace is taken in the fundamental representation.  The levels, $k_{L/R}$, will be non-integer and ultimately related to $G_N^{-1}$.  Following \cite{Witten:1989ip}, the correct framework for approaching this theory is through its complexification $\mf {sl}(2,\mathbb C)$ with $\mf{su}(2)$ taken as a real form.  As emphasized in that paper a decomposition of levels consistent with reality of the action and with Euclidean gravity with positive cosmological constant is given by\footnote{Strictly speaking, $\mf{sl}(2,\mathbb C)$ Chern-Simons theory parameterized in this way describes {\it Lorentzian} gravity with positive cosmological constant and with $\mf{sl}(2,\mathbb R)$ as its real form.  We obtain the Euclidean theory from the Wick rotation: i.e. (supposing the negative sign of the metric is associated with $e^3$) $e^3\rightarrow ie^3$, $L_3\rightarrow iL_3$.}
\beq\label{eq:kLRdef}
k_L=\delta+is~,\qquad k_R=\delta-is~,
\eeq
where $\delta\in\mathbb Z$ and $s\in\mathbb R$.  As further discussed in \cite{Witten:1989ip}, quantum effects lead to a finite renormalization of the levels
\beq\label{eq:kLRrenorm}
k_L\rightarrow r_L=k_L+2~,\qquad k_R\rightarrow r_R=k_R+2~.
\eeq
Importantly these are renormalized in the same way and can be regarded as a renormalization of $\delta$ to $\hat\delta=\delta+2$.  For the rest of this section we will work with the renormalized levels.

To see that indeed this can be related to a theory of gravity, we can decompose the connections as
\beq\label{eq:ALAR}
A_L=i\left(\omega^a+\frac{1}{\ell}e^a\right)L_a~,\qquad A_R=i\left(\omega^a-\frac{1}{\ell}e^a\right)\bar L_a~,
\eeq
where $\{L_a\}$ and $\{\bar L_a\}$ generate  $\mf{su}(2)_L$ and $\mf{su}(2)_R$, respectively.\footnote{With respect to this basis, we have
\beq
\Tr(L_aL_b)=\Tr(\bar L_a\bar L_b)=\frac{1}{2}\delta_{ab}~.
\eeq}  It is natural to interpret $e^a$ as the dreibein and $\omega^a=\frac{1}{2}\varepsilon^{abc}\omega_{bc}$ is the (dual) spin-connection.  Indeed the action \eqref{eq:SL+SR} is equivalent to
\begin{align}\label{eq:StoPalatini}
iS=-I_{\rm EH}-i\hat\delta I_{\rm GCS}~,
\end{align}
where $I_{\rm EH}$ is the Einstein-Hilbert action written in first-order (or Palatini),
\begin{align}
I_{\rm EH}=-&\frac{s}{4\pi\ell}\int\varepsilon_{abc}e^a\wedge\left(R^{bc}-\frac{1}{3\ell^2}e^b\wedge e^c\right)~.
\end{align}
Here $R^{ab}={\varepsilon^{ab}}_c\left( \dd\omega^c-\frac{1}{2}{\varepsilon^c}_{de}\omega^d\wedge \omega^e\right)$ is the Riemann two-form, and we have a positive cosmological constant, $\Lambda=\ell^{-2}$.  This identifies the imaginary part of the levels with Newton's constant
\beq\label{eq:sGN}
s=\frac{\ell}{4G_N}~,
\eeq
which establishes that the semi-classical regime of this theory is the large-$s$ limit.  The second part of this action, once restricted to torsion-free spin connections, is the gravitational Chern-Simons action:
\begin{align}\label{eq:GCS-action}
%I_{GCS}=&\frac{1}{2\pi}\int\mTr\left(\omega\wedge d\omega+\frac{2}{3}\omega^3\right)+\frac{1}{4\pi\ell^2}\int\delta_{ab}e^a\wedge T^a~,
I_{\rm GCS}=&\frac{1}{2\pi}\int\mTr\left(\omega\wedge d\omega+\frac{2}{3}\omega^3\right)+\frac{1}{2\pi\ell^2}\int\mTr\left(e\wedge T\right)~,
\end{align}
where $T^a=\dd e^a-{\varepsilon^a}_{bc}\omega^b\wedge e^c$ is the torsion two-form.   

It is also simple to establish a relation at the level of the equations of motion. The classical equations of motion of the Chern-Simons theories \eqref{eq:SL+SR} are
\beq\label{eq:ALAREoM}
\dd A_L+A_L\wedge A_L=0~,\qquad \dd A_R+ A_R\wedge A_R=0~.
\eeq
The sum and difference of these equations translate, in terms of $e^a$ and $\omega^a$, to the vacuum Einstein equation (with positive cosmological constant) and the vanishing of the torsion two-form:
\beq
R^{ab}=\frac{1}{\ell^2}e^a\wedge e^b~,\qquad T^a=0~.
\eeq
These derivations establish a correspondence between classical solutions in the metric formulation of dS$_3$ gravity and classical solutions in the Chern-Simons theory.

\paragraph{Background configuration.} It will be important to make explicit how to cast Euclidean dS$_3$ space, i.e., the three-sphere, in the language of Chern-Simons theory. We start by constructing the appropriate flat connections on $S^3$, which we will coin $(a_L,a_R)$. 
Given the metric \eqref{eq:EucSPmetric}, a convenient choice of dreibein is 
\beq\label{eq:bgframe}
e^1=\ell \dd\rho~,\qquad e^2=\ell \sin\rho \,\dd\varphi~,\qquad e^3=\ell \cos\rho\, \dd\tau~,
\eeq
with associated torsion-free spin connection
\beq\label{eq:wgframe}
\omega^1=0~,\qquad \omega^2=-\sin\rho\, \dd\tau~,\qquad \omega^3=-\cos\rho\, \dd\varphi~.
\eeq
From these expressions, we find 
\be
\begin{aligned}\label{eq:bgconns}
a_L=&iL_1\dd\rho+i\left(\sin\rho\, L_2-\cos\rho\, L_3\right)(\dd\varphi-\dd\tau)=g_\rho^{-1}g_-^{-1}\dd\left(g_-g_\rho\right)~,\\ a_R=&-i\bar L_1\dd\rho-i\left(\sin\rho\, \bar L_2+\cos\rho\, \bar L_3\right)(\dd\varphi+\dd\tau)=-\dd\left(g_\rho g_+\right) g_+^{-1}g_\rho^{-1}~,
\end{aligned}
\ee
where we used \eqref{eq:ALAR}.
 The second equality of each line above emphasizes that $a_L$ and $a_R$ are pure gauge with 
 \be
 g_\rho=e^{iL_1\rho}~, \qquad g_\pm=e^{-iL_3(\tau\pm \varphi)}~. 
 \ee
The connections \eqref{eq:bgconns} are locally flat, however they possess a point-like singularity. %(simply because a single coordinate chart cannot cover $S^3$).
These are singularities for $\dd\tau$ and $\dd \varphi$ at $\rho=\pi/2$ and $\rho=0$, respectively. These will be treated, as distributions, by
\beq\label{eq:ddvarphi-tau}
\dd(\dd\varphi)= \delta(\rho)\dd\rho\wedge \dd\varphi~,\qquad \dd(\dd\tau)=-\delta(\rho-\pi/2)\dd\rho\wedge \dd\tau~.
\eeq 

It is simple to extract the holonomies of  $a_L$ and $a_R$, which are important to record for later use. For any cycle $\gamma$ wrapping the singular points of the connections,  the connections possess holonomies
\beq\label{eq:holalar}
\mathcal P\exp\oint_{\gamma}a_L=g_\rho^{-1}e^{i2\pi L_3\ms h_L}g_\rho~,\qquad\mathcal P\exp\oint_{\gamma}a_R=g_\rho e^{i2\pi \bar L_3\ms h_R}g_{\rho}^{-1}~.
\eeq
Requiring that the above group elements' action on $S^3\simeq SU(2)$ itself is single-valued implies that $\ms h_L,\ms h_R\in\mathbb Z$ with either both even or both odd.\footnote{Namely, this geometric action, detailed in App.\,\ref{app:casimir}, is in the fundamental representation. In that case $e^{i2\pi L_3\ms h_{L/R}}$ is obviously the identity if $\ms h_{L/R}$ is even.  If $\ms h_L$ and $\ms h_R$ are both odd this yields the group element $(-1,-1)\in SU(2)_L\times SU(2)_R$ which is also the identity inside the $\mathbb Z_2$ quotient.}  In particular, for cycles wrapping the causal horizon at $\rho=\frac\pi 2$, we have\footnote{The holonomy about this point will play a special role in Section~\ref{sec:constructionspool}, when we invoke regularity at the horizon as a condition on scalar one-loop determinants.}
\beq\label{eq:holalar-v}
\ms h_L=1~,\qquad\ms h_R=-1~.
\eeq
 
 Finally, we report on the value of the on-shell action for this background. A short calculation, which uses \eqref{eq:ddvarphi-tau}, shows that they have non-trivial action
\beq\label{eq:SLSRtree}
r_LS_{\rm CS}[a_L]=-\pi r_L=-\pi\hat\delta-i\pi s~,\qquad
r_RS_{\rm CS}[a_R]=\pi r_R=\pi\hat\delta-i\pi s~,
\eeq
and thus
\be
\begin{aligned}
\left.iS\right|_{\rm tree-level}&=  ir_LS_{\rm CS}[a_L] + ir_RS_{\rm CS}[a_R]\\
&=\frac{\pi\ell}{2G_N}~,
\end{aligned}
\ee
where we used \eqref{eq:sGN}. This is the correct on-shell action for dS$_3$ \cite{Carlip:1992wg,Castro:2011xb}. Note that the gravitational Chern-Simons term of $S^3$ vanishes identically.

\subsection{\texorpdfstring{$SU(2)$}{SU(2)} Chern-Simons theory: the partition function}\label{sect:CStheory}

We now turn to quantum aspects of dS$_3$ gravity. Our aim is to perform the gravitational path-integral about a fixed background $S^3$ saddle.\footnote{We will briefly discuss about other topologies in Sec.\,\ref{sect:disc}.} This will be done in the Chern-Simons formulation of the theory which we introduced in Sec.\,\ref{sect:CSgrav}.

It is well-known that many observables in Chern-Simons theory can be evaluated exactly, i.e., to all orders in perturbation theory and also including  non-perturbative effects. However, for our gravitational purpose, some caution is needed since these results are not always applicable due to the subtle relation between Chern-Simons and gravity. Here we will address these subtleties at the level of evaluating the path integral on $S^3$. In a nutshell, we will re-derive $Z_k[S^3]$ for  $SU(2)$ Chern-Simons theory with level $k$, while allowing the level to be complex and also allowing non-trivial background connections. These are two key features that are persistent in the relation among the two theories, as we reviewed in the previous subsection.

Let us therefore begin by reviewing some basic facts and definitions. The Chern-Simons partition function over a three-manifold, $M$, is the path-integral
\beq \label{S3:CSPI}
Z_k[M]=\int \frac{\mc DA}{\mc V}\,e^{ikS_{\rm CS}[A]}~
\eeq
over the action \eqref{eq:CSaction}.  Here $A=A^{a}L_a$ is to be regarded as a connection one-form of a principal $SU(2)$ bundle over $M$, where $\{L_a\}$ generates the $\mf{su}(2)$ Lie algebra.\footnote{Note that given the form of \eqref{eq:CSaction}, we are working with the convention that $A$ is anti-Hermitian in the fundamental representation, i.e. $A^a_\mu\in i\mathbb R$. We will use this convention consistently throughout.}  In the measure we indicate, schematically, a division by the gauge group as $1/\mc V$.

There are three remarks that will be important in what follows. First, the action \eqref{eq:CSaction} is clearly topological and the quantum theory itself is almost topological: its sole geometric input is a choice of {\it framing} which arises from regularizing the phase of $Z_k.$  While there is no ``rule" for establishing the framing, partition functions differing by choices of framing are related by well-established phases \cite{Witten:1988hf}.  In this paper we will be careful to work with a fixed convention for the phase of $Z_k$.\footnote{Which is ultimately related to two-units away from so-called ``canonical framing".}

Second, our evaluation of \eqref{S3:CSPI} will cover complex values of the level $k$. In particular, our derivations will hold for a decomposition as in done in \eqref{eq:kLRdef}-\eqref{eq:kLRrenorm}.

Third, we will incorporate a flat background connection to the path integral \eqref{S3:CSPI}. To that end we will write
\beq
A=a+B~.
\eeq
Here $a$ is a flat background connection on $M$---for most of our purposes $M=S^3$.  It is important to emphasize at this point that, unlike what is typical for Chern-Simons theory quantized on $S^3$, we will {\it not} take the trivial background $a=0$: such a background leads to a degenerate metric which is an unnatural saddle for a theory of gravity. Instead we want connections corresponding to a round $S^3$ metric, i.e., they will be \eqref{eq:bgconns}, with holonomies \eqref{eq:holalar}-\eqref{eq:holalar-v}, for each copy of the $SU(2)$ theory.  The field $B$ captures the quantum fluctuations that we will integrate over in the path-integral shortly afterward.

\paragraph{Adapting exact results.}
We now turn to the tools we will use for evaluating $Z_k[S^3]$. There are several ways to obtain $Z_k[S^3]$, and we do not attempt to describe them all. We selected methods for which the choice of background connections and background topology (and later, expectation values for Wilson loops) are tractable in the path integral of $SU(2)_k$ Chern-Simons theory. The two methods we will discuss in detail are:
\begin{description}[font=\normalfont, leftmargin=0.3cm]
  \item[\emph{Abelianisation.}] The process of Abelianisation was developed in \cite{Blau:1993tv, Blau:2006gh, Blau:2013oha}. In a nutshell, it demonstrates how the non-Abelian Chern-Simons path integral can be reduced to a two-dimensional Abelian theory, under suitable conditions present on the manifold $M$. 
\item[\emph{Supersymmetric localization.}] As a complementary method, we will show how one obtains $Z_k[S^3]$ via supersymmetric localization techniques \cite{Kapustin:2009kz} (see also \cite{Marino:2011nm}). The biggest penalty here is the introduction of fermions in the path integral. Still, the outcome is robust and completely agrees with Abelianisation.  
\end{description}
Both methods will be capable of successfully accommodating the features necessary for dS$_3$ gravity, and we stress that they report the same result (up to a trivial normalization). This subsection will summarize the main steps of both methods, highlighting in particular the features that need to be altered to accommodate gravity.
  
It is worth mentioning a third method. One can also contrast the two methods above with the process of \emph{non-Abelian localization} \cite{Beasley:2005vf,Beasley:2009mb}. Unfortunately, we find an obstruction to utilizing this approach in dS$_3$ gravity rigorously.  The details of this obstruction are detailed in App.\,\ref{sec:locApp}.

\subsubsection{Abelianisation}\label{sect:Abelianisation}

Abelianisation is a powerful method for evaluating the Chern-Simons path-integral for compact, connected and simply-connected Lie groups with Lie algebra $\mathfrak{g}$ on particular types of three-manifolds  \cite{Blau:1993tv, Blau:2006gh, Blau:2013oha}.  In particular, Abelianisation is useful when it is possible to choose the background $M$ to be a circle fibration over a two-dimensional base, $\Sigma_g$, i.e. $M=M_{(g,p)}$  can seen as a principal $U(1)$-bundle: $U(1)\to M_{(g,p)} \overset{\pi}{\rightarrow} \Sigma_g$ with monopole degree $p$.  This is obviously relevant for us by considering $M=S^3$ as a Hopf fibration. 

The approach in \cite{Blau:1993tv, Blau:2006gh, Blau:2013oha} reduces computations from non-Abelian Chern-Simons theory in three dimensions to computations in a two-dimensional Abelian $q$-deformed Yang-Mills theory on $\Sigma_g$ in the following way.  Using the geometry of $M$ we decompose\footnote{This decomposition is similar to that adopted in \cite{Beasley:2005vf}.} the Chern-Simons connection $A\in \Omega^1(M,\mathfrak{g})$ into ``vertical" and ``horizontal" parts,
\begin{equation} \label{AbelianisationsplitA}
    A= \bs\sigma \kappa + A_H~,
\end{equation} 
with respect to a globally-defined real-valued one-form $\kappa$ on $M$, and where $\bs\sigma$ is a $\mf g$-valued scalar.\footnote{We will differ in notation slightly from previous literature \cite{Blau:1993tv, Blau:2006gh, Blau:2013oha} where $\bs\sigma$ is called $\phi$.  This is to keep notation throughout the paper uniform and to make comparison of results clearer.}  Abelianisation works by adding BRST-exact terms to the action to fix the gauge so that $\bs\sigma$ is a $U(1)$-invariant section of $M\times\mf g$. This allows us to ``push" $\bs\sigma$ down to the base, $\Sigma_g$, where it can be diagonalised, setting $\bs\sigma\in\mf t$ (where $\mf t\subset\mf g$ is a Cartan subalgebra). %by setting to zero the components along the complement to the Cartan subalgebra $\mathfrak{t}\subset \mathfrak{g}$. 
The result of the gauge-fixing and the Abelianisation is that $\bs\sigma$ is $\mathfrak{t}$-valued and constant along the $U(1)$ fibers of $M$.  The remaining fields can then be easily integrated out.

With an eye towards applying Abelianisation \cite{Blau:2006gh} to a background saddle relevant for gravity, we will expand the Chern-Simons action, \eqref{eq:CSaction}, about a flat background connection, $a$, which is generically non-zero:
\begin{equation}
A=a+B~,\qquad \dd a+a\wedge a=0~.
\end{equation}
The difference between the Chern-Simons action for $A$ and that of the background connection is then 
\begin{equation}\label{modifiedCSaction0}
S_{\rm CS}[A]-S_{\rm CS}[a]=\frac{k}{4\pi}\int \mathrm{Tr}   \left(  B\wedge \dd_a B + \frac{2}{3} B^3 \right) ~,
\end{equation}
where we have imposed flatness for $a$ and dropped a total derivative.  We have defined above a ``background exterior derivative" acting on $p$-forms as
\begin{equation}\label{eq:BGextder}
\dd_a \omega_p = \dd\omega_p + a\wedge \omega_p - (-1)^p \omega_p \wedge a ~ .
\end{equation}
We now will try to adapt Abelianisation to $B$, however we need to address the non-canonical kinetic term in \eqref{modifiedCSaction0}.  We will write the background connection in terms of a group element, $g$, as 
\begin{equation}
a:=g^{-1}\dd g~.
\end{equation}
In writing above, it might be the case that $g$ is not single-valued on $M$.  This fact manifests itself in the possible existence of holonomies of $a$ around the closed curve, $\gamma$, along a $U(1)$ fibre of $M$:
\beq\label{eq:AbBGholo}
\mc P\exp\left(\oint_\gamma a\right)=g_f^{-1}g_i=\exp 2\pi\mf m~,
\eeq
where $\mf m\in\mf g$.  Performing the field redefinition 
\begin{equation}\label{fieldredefinition}
\tilde{B}:= gBg^{-1} ~,
\end{equation}
we can recover a canonical kinetic form for $\tilde B$:
\begin{equation}\label{modifiedCSaction1}
 \int \mathrm{Tr}   \left(  B\wedge \dd_a B + \frac{2}{3} B^3 \right)  = \int \mathrm{Tr}   \left(  \tilde{B}\wedge \dd \tilde{B} + \frac{2}{3} \tilde{B}^3 \right) ~.
\end{equation}
The cost of this, however, is that $\tilde B$ now possesses twisted boundary conditions: going around the cycle $\gamma$ defining \eqref{eq:AbBGholo} gives
\beq
\tilde B_f=g_f B g_f^{-1}=e^{-2\pi\tmf m}\,\tilde B_i\,e^{2\pi\tmf m}~,
\eeq
where $e^{2\pi\tmf m}=g_i^{-1}e^{2\pi\mf m}g_i$. 
We can state these boundary conditions more clearly by decomposing $B$ into a root-space compatible with $\tmf m$.  That is writing $\tilde B=\tilde B^{(i)}T_i+\tilde B^{(\alpha)}T_\alpha$ where $T_i$ is a basis of a Cartan subalgebra containing $\tmf m$ and $T_\alpha$ is a basis of the root space for this Cartan, then
\beq\label{eq:twistedBCsrootspace}
\tilde B_f^{(i)}=\tilde B_i^{(i)}~,\qquad\tilde B_f^{(\alpha)}=e^{-2\pi\alpha\cdot\tmf m}\tilde B_i^{(\alpha)}~,
\eeq
that is the fields aligned with the Cartan defined by $\tmf m$ retain their periodicity along $\gamma$ while fields aligned with roots transform by phases.  In terms of $\mf g=\mf{su}(2)$ we can write $\tmf m=i\ms h\,L_3$ in which case
\beq
\tilde B_f^{(3)}=\tilde B_i^{(3)}~,\qquad \tilde B_f^{(\pm)}=e^{\mp i2\pi\ms h}\tilde B_i^{(\pm)}~.
\eeq
At this point, we will procede to adapt the Abelianisation procedure \cite{Blau:2006gh} to $\tilde{B}$.  To be explicit, we will specialize to the case where $M=S^3$.

\subsubsection*{Abelianisation on $\tilde{B}$}

From here many of the steps mirror those in \cite{Blau:2006gh}.  Namely, the connection $\tilde{B}$ is split into 
\begin{equation}
\tilde{B}=B_\kappa+B_H:=\bs\sigma\kappa+B_H ~,
\end{equation}
and similarly the exterior derivative on $M$ is split into a ``horizontal" piece (that is, along the base, $\Sigma$) and an action along the fibre
\begin{equation}\label{splitexteriorderivative}
\dd= (\pi^\ast \dd_\Sigma) +\kappa\wedge \mathscr L_\xi ~,
\end{equation}
where $\mathscr L_\xi=\{\dd,\iota_\xi\}$ is the Lie derivative along the fundamental vector field generating the $U(1)$ action.  The action \eqref{modifiedCSaction1} can then be massaged to the form   
\begin{multline}
    \label{CSactionBT0}
\frac{k}{4\pi}\int \mathrm{Tr}   \left(  \tilde{B}\wedge \dd \tilde{B} + \frac{2}{3} \tilde{B}^3 \right)  =\frac{k}{4\pi}\int \mathrm{Tr}   \left(  \bs\sigma^2\kappa\wedge \dd \kappa + 2\bs\sigma\kappa \wedge \dd B_H + B_H \wedge \kappa\wedge \mathcal{L}_{\sigma} B_H\right) ~,
\end{multline}
up to total derivative.  Above we have also defined 
\beq\label{eq:phiLie}
\mc L_\sigma=\mathscr L_\xi+[\bs\sigma,\cdot]~.
\eeq

\subsubsection*{Fixing a gauge}

The choice of gauge that allows the Abelianisation procedure to be applied is \footnote{Note that $\iota_\xi \kappa=1$ and $\iota_\xi \dd\kappa=0$, so ${\cal L}_\xi \kappa = 0$.}
\begin{equation}
\mathscr L_\xi (\bs\sigma\kappa) = 0 \quad \Leftrightarrow \quad \mathscr L_\xi \bs\sigma = \iota_\xi \dd\bs\sigma = 0~,
\end{equation}
which states that $\bs\sigma$ is $U(1)$-invariant.  We additionally gauge-fix that $\bs\sigma$ is valued in the Cartan, $\mf t$:
\begin{equation}\label{eq:gauge1}
\bs\sigma^\mathfrak{l}=0 ~,
\end{equation}
where $\mathfrak{g}=\mathfrak{t}\oplus\mathfrak{l}$. Without loss of generality we will choose this Cartan to align with that defined by the holonomy of the background connection, $a$, \eqref{eq:AbBGholo} so that $\bs\sigma$ remains single-valued on $M$.  This gauge is fixed \cite{Blau:2006gh} by adding the following 
BRST-exact action
\begin{equation}
\int_M \mTr\left(E\star \bs\sigma + \bar{c} \star \mathcal{L}_\sigma c \right)= \int_M \mTr\left(E\star \bs\sigma + \kappa\wedge \dd\kappa ~ \bar{c} \mathcal{L}_\sigma c \right)~,
\end{equation}
where $E$ is a Lagrange-multiplier, and $c$ and $\bar c$ are ghosts.  It is understood that the $U(1)$ invariant modes of these fields (i.e. those satisfying $\mathscr L_\xi E^{\mf t}=\mathscr L_\xi c^{\mf t}=\mathscr L_\xi\bar c^{\mf t}=0$) are not path-integrated \cite{Blau:2006gh}.  We can now describe integrating out modes.

\subsubsection*{Effect of integrating over fields}

\begin{itemize}[leftmargin=0.4cm]
\item The part of $B_H$ valued in the Cartan sub-algebra, denoted $B_H^\mathfrak{t}$, contains the $U(1)$-invariant modes $\hat{B}_H^\mathfrak{t}$ obeying $\mathscr L_\xi\hat B_H^{\mf t}=0$.   Because they are $\mathfrak{t}$-valued, and since $\bs\sigma=\bs\sigma^\mathfrak{t}$ as a result of the gauge-fixing, \eqref{eq:gauge1}, the term $\mathcal{L}_{\sigma} \hat{B}_H^\mathfrak{t} $ vanishes from \eqref{CSactionBT0}. Thus the only term of \eqref{CSactionBT0} in which the fields $\hat{B}_H^\mathfrak{t} $ enter is 
\begin{equation}
2\bs\sigma\kappa \wedge \dd \hat{B}_H^\mathfrak{t} ~,
\end{equation}
and integrating over these fields imposes the constraint that $\bs\sigma=\text{constant}$ on $M$.
\item The Gaussian integrals over the fields $B_H^\mathfrak{l},~ B_H^\mathfrak{t}$, (where $B_H^\mathfrak{t}$ are not $U(1)$-invariant), and over the ghost fields $c^\mathfrak{l},~c^\mathfrak{t}$ give ratios of determinants \cite{Blau:2006gh}:
\begin{equation}\label{determinant0}
\frac{\mathrm{Det} \left( i\mathcal{L}_\sigma  \right)_{\Omega^0_H(S^3,\mathfrak{l})} \mathrm{Det'} \left( i \mathscr L_\xi \right)_{\Omega^0_H(S^3,\mathfrak{t})}}{\mathrm{Det}^{1/2} \left( \star\kappa\wedge i \mathcal{L}_{\sigma} \right)_{\Omega^1_H(S^3,\mathfrak{l})} 
\mathrm{Det'}^{1/2} \left( \star\kappa\wedge i \mathscr L_\xi \right)_{\Omega^1_H(S^3,\mathfrak{t})} }~.
\end{equation}
At this point we need to pause to emphasize that these determinants are, in principle, to be taken over fields with twisted boundary conditions defined by \eqref{eq:twistedBCsrootspace} along the $U(1)$ fibre.  These boundary conditions do not affect the determinants over the Cartan-valued fields.  For fields living in the $\alpha$ root-space, given the form of $\mc L_\sigma$, \eqref{eq:phiLie}, the effect of the twisted boundary conditions, \eqref{eq:twistedBCsrootspace}, is to shift the eigenvalues of $i\mc L_\sigma$
\beq
2\pi n+i\alpha\cdot\bs\sigma~~\rightarrow ~~2\pi n+i\alpha\cdot\bs\sigma-i2\pi\alpha\cdot\tmf m~,\qquad n\in\mathbb Z~.
\eeq
The absolute value of the ratio of these determinants can then be evaluated in manner similar to \cite{Blau:2006gh} to give
\begin{equation}\label{absvaldeterminants}
\mathrm{Abs}\left[\frac{\mathrm{Det} \left( i\mathcal{L}_\sigma  \right)_{\Omega^0_H(S^3,\mathfrak{l})} \mathrm{Det'} \left( i \mathscr L_\xi \right)_{\Omega^0_H(S^3,\mathfrak{t})}}{\mathrm{Det}^{1/2} \left( \star\kappa\wedge i \mathcal{L}_{\sigma} \right)_{\Omega^1_H(S^3,\mathfrak{l})} 
\mathrm{Det'}^{1/2} \left( \star\kappa\wedge i \mathscr L_\xi \right)_{\Omega^1_H(S^3,\mathfrak{t})} } \right] = T_{S^1} (\bs\sigma-2\pi\tmf m) ~,
\end{equation}
where 
\begin{equation}
T_{S^1} (\bs x) = {\det}_\mathfrak{l} \left(1-\mathrm{Ad}e^{\bs x} \right) = \prod_{\alpha>0} 4 \sin^2(i\alpha\cdot\bs x/2)~
\end{equation}
is the Ray-Singer torsion of the connection along the fibre. For $\mf g=\mf{su}(2)$ we can set $\bs\sigma=-i2\pi\sigma L_3$ (where $\sigma$ is now a real constant) and $\tmf m=i\ms h L_3$; for this choice we have 
\beq
T_{S^1}(\bs\sigma-2\pi\tmf m)=4\sin^2\left(\pi(\sigma+\ms h)\right)~.
\eeq
We've stated this result somewhat generally, however it's useful to keep in mind that for the backgrounds of interest for this paper, $\ms h$ will always be integer valued (as discussed at the end of Sec.\,\ref{sect:CSgrav}) and so $\eqref{absvaldeterminants}$ reduces to $T_{S^1}=4\sin^2(\pi\sigma)$.  This is of course consistent with \eqref{eq:twistedBCsrootspace} reducing to single-valued boundary conditions when $\ms h\in\mathbb Z$.  The phase of the determinants \eqref{determinant0} can be defined through a regularized eta invariant and is responsible for the renormalization of level $k\to r=k+2$ for $\mathfrak{g}=\mathfrak{su}(2)$ \cite{Blau:2006gh}.
\end{itemize}

The end result of this is the expression\footnote{We shift the level of the classical action trivially when $S_{\rm CS}[a]\in\mathbb Z$.} of the Chern-Simons partition function as a simple integral over a Cartan-valued field $\bs\sigma$
\begin{equation}
 Z_k[S^3]=e^{irS_{\rm CS}[a]}\int_{\mathfrak{t}} \dd\bs\sigma \, T_{S^1}(\bs\sigma-2\pi\tmf m) \exp 
    \left[-i \frac{r}{4\pi}\mathrm{Tr} \bs\sigma^2 \right]~,
\end{equation}
up to inessential overall normalization and constant ($r$-independent) phase.  We fix this normalization/phase by fiat.  More explicitly setting $\bs\sigma=-i2\pi\sigma L_3$ and $\tmf m=i\ms h\,L_3$, we normalize $Z_k$ as
\beq\label{eq:ZkAb}
Z_k[S^3] = e^{irS_{\rm CS}[a]} \int_{\mathbb R} \dd\sigma \, \mathrm{sin}^2 \left( \pi (\sigma+\ms h)\right) e^{i\frac{\pi }{2}r \, \sigma^2} ~.
\eeq
While the integral over $\sigma$ begins life along the real axis, this integral is Gaussian and we can formally define it through appropriate contour deformations depending on the phase of $r$.  It is simple to perform the integral (letting $\ms h\in\mathbb Z$)
\beq
Z_k[S^3]=e^{irS_{\rm CS}[a]}\,e^{i\phi}\,\sqrt{\frac{2}{r}}\sin\frac\pi r~,
\eeq
where the phase of $Z_k$
\beq\label{eq:Zphase}
\phi=\frac{3\pi}{4}-\frac\pi r=\frac\pi 6\mf c+\frac\pi 4~,\qquad \mf c\equiv\frac{3(r-2)}{r}~,
\eeq
can be identified with a framing phase \cite{Witten:1988hf} (two-units away from canonical framing) plus a phase stemming from a $\sigma$ contour rotation.\footnote{This latter phase is entirely a by-product of our conventions and does not occur in usual Chern-Simons formulas.  However we will give it a gravitational interpretation below.}

\subsubsection{\texorpdfstring{$\mathcal{N}=2$}{N=2} supersymmetric localization }\label{sect:susylocalizationS3}

We now describe an alternative route to the exact calculation of the Chern-Simons partition through localization techniques.  We will focus particularly on $\mc N=2$ supersymmetric localization \cite{Kapustin:2009kz}.   One benefit of this approach is that much of the basic machinery has been established with a non-trivial background connection, $a$, in mind allowing a fairly straightforward incorporation of $a\neq 0$. However: the situations with non-trivial background connections have historically arisen on manifolds with interesting topology (e.g. Lens spaces) and many of the explicit results for $S^3$ have been established with $a=0$ (one notable exception: \cite{Drukker:2012sr}).  Below we collect and synthesize these results in a way that is useful for dS$_3$ gravity.

Before jumping in, let us also make the following brief comments.  Supersymmetry in the context of de Sitter is a contentious subject, with much of the difficulty arising from realizing unitary representations of the supersymmetry algebra in Lorentzian signature \cite{Anous:2014lia,Pilch:1984aw}.  In this paper we will take a somewhat agnostic stance on this topic\footnote{For discussions on the utility of supersymmetric localization to two-dimensional de Sitter space see \cite{Anninos:2022ujl}.}: by working directly in Euclidean signature, we are ultimately discussing $SU(2)_k$ Chern-Simons theory on $S^3$ whose $\mc N=2$ supersymmetric extension is well-established.  We use the existence of this symmetry to our advantage to localize the path-integral all while verifying that the extension to $\mc N=2$ does not alter essential features of the original partition function.  Ultimately, however, this localization will simply verify the results of Sec.\,\ref{sect:Abelianisation}.

Let us set the stage and collect the necessary background.  Much of what follows mirrors the friendly review \cite{Marino:2011nm}. 
The vector multiplet of three-dimensional $\mc N=2$ gauge theory is given by fields
\beq
\{A_\mu,\;\bs\sigma,\;\mf D,\;\lambda,\;\bar\lambda\}~,
\eeq
where $A$ is a $\mf g=\mf{su}(2)$ connection, $\bs\sigma$, $\mf D$ are scalars,\footnote{The $\bs\sigma$ appearing here is {\it a priori} a different field than what appeared in Sec.\,\ref{sect:Abelianisation}. We give it the same name because, ultimately, it will play the same role in the final result.} and $\lambda$, $\bar\lambda$ are Dirac spinors.  All fields are $\mf g$-valued and by convention we will take them all to be anti-Hermitian,\footnote{In comparison to the notation of \cite{Marino:2011nm}, a field here is related to a field there by $\Phi_{\rm here}=i\Phi_{\rm there}$.} with supersymmetry variations parameterized by two Grassmann variables $\bar\epsilon$ and $\epsilon$ as specified in \cite{Marino:2011nm}. The supersymmetric Chern-Simons action is
\beq\label{scsaction}
S_{\rm SCS}=\frac{1}{4\pi}\int\mTr\left(A\wedge \dd A+\frac{2}{3}A^3\right)-\frac{1}{4\pi}\int d^3x\sqrt{g}\,\mTr\left(\bar\lambda\lambda-2\mf D\bs\sigma\right)~,
\eeq 
and enters the path-integral multiplied by the level $k$
\beq
Z^{\rm SCS}_k[S^3]=\int \frac{\mc DA}{\mc V_G}\mc D\bar\lambda\mc D\lambda\mc D\mf D\mc D\bs\sigma\,e^{ikS_{\rm SCS}} ~.
\eeq
To make subsequent notation less cumbersome, we will drop the ``$[S^3]$" above with it understood that we are always working on the three-sphere.  Note that on a formal level, as far as the function dependence on $k$ is concerned, the addition of the auxiliary fields in the multiplet does not alter $Z_k^{\rm SCS}$ with respect to the non-supersymmetric path-integral, $Z_k$~\cite{Marino:2011nm}. 

The deformation that allows us to localize the path-integral $Z^{\rm SCS}_k$  is the super-Yang-Mills action
\begin{multline}
  S_{\rm SYM}= -\int\mTr\left(\frac{1}{2}F\wedge\star F+D\bs\sigma\wedge\star D\bs\sigma\right) \\
 - \int d^3x\sqrt{g}\,\mTr\left(\frac{1}{2}\left(\mf D+\bs\sigma\right)^2+\frac{i}{2}\bar\lambda\gamma^\mu D_\mu\lambda-\frac{1}{2}\bar\lambda[\bs\sigma,\lambda]-\frac{1}{4}\bar\lambda\lambda\right) ~,  
\end{multline}
where $D_\mu$ is the gauge-covariant derivative and $\gamma_\mu$ can be taken to be the Pauli-matrices acting on spinor indices.  $S_{\rm SYM}$ is itself a super-derivative
and therefore $Q$-exact. Adding this to the path-integral with coefficient $\mathsf{t}$, i.e.,
\beq\label{eq:ZSCSSYM}
Z^{\rm SCS+SYM}_k(\mathsf{t})=\int \frac{\mc DA}{\mc V_G}\mc D\bar\lambda\mc D\lambda\mc DD\mc D\bs\sigma\,e^{ikS_{\rm SCS}-\mathsf{t}S_{\rm SYM}}~,
\eeq
is then innocuous: $Z^{\rm SCS+SYM}_k(\mathsf{t})=Z^{\rm SCS}_k$ for any $\mathsf{t}$, including in the limit $\mathsf{t}\rightarrow\infty$ where the path-integral localizes on the saddle of $S_{\rm SYM}$.  

\subsubsection*{Localization locus}

In the $\mathsf{t}\rightarrow\infty$ limit, the path-integral localizes on the following equations of motion
\beq\label{eq:locsaddleeqs}
F=0~,\qquad D\bs\sigma=\dd\bs\sigma+[A,\bs\sigma]=0~,\qquad \mf D+\bs\sigma=0 ~.
\eeq
We expand the solutions around a flat connection $a=g^{-1}\dd g$, for some group element $g$.  Again, $g$ may not be single-valued and $a$ may possess a holonomy, \`a la \eqref{eq:AbBGholo},
\beq\label{eq:holonomy}
\mc P\exp\left(\oint _{\gamma} a\right)=\exp(2\pi\mf m)~,
\eeq
for some curve $\gamma$.  The other fields that have saddle solutions to \eqref{eq:locsaddleeqs} are given by
\beq\label{saddlesusy}
\bs\sigma^{(g)}_0=g^{-1}\bs\sigma_0 g~,\qquad \mf D_0=-\bs\sigma^{(g)}_0~,\qquad \lambda_0=0~,\qquad \bar\lambda_0=0~,
\eeq
for $\bs\sigma_0$ a constant element of $\mf g$. We require $\bs\sigma_0^{(g)}$ to be single-valued and so the constant element defining the saddle must obey
\beq
[\mf m,\bs\sigma_0]=0~.
\eeq
With this we can take $\bs\sigma_0$  to be in a Cartan subalgebra containing $\mf m$.  We will scale fluctuations as  
\be
\begin{aligned}
A=&a+\frac{1}{\sqrt \mathsf{t}}B~, \qquad 
\bs\sigma=\bs\sigma_0^{(g)}+\frac{1}{\sqrt \mathsf{t}}\hat\sigma~, \qquad 
\mf D=-\bs\sigma_0^{(g)}+\frac{1}{\sqrt \mathsf{t}}\hat{\mf D}~,\\
& \qquad \qquad \lambda=\frac{1}{\sqrt \mathsf{t}}\hat\lambda~, \qquad \qquad
\bar\lambda=\frac{1}{\sqrt \mathsf{t}}\hat{\bar\lambda}~,
\end{aligned}
\ee
and perturb the action \eqref{scsaction} around the saddle \eqref{saddlesusy} as $\mathsf{t}\rightarrow\infty$.  The leading contribution to $S_{\rm SCS}$ is 
 \beq
 \lim_{\mathsf{t}\rightarrow\infty}S_{\rm SCS}=S_{\rm CS}[a]-\frac{\text{Vol}(S^3)}{2\pi}\mTr\bs\sigma_0^2 ~.
 \eeq
Meanwhile the leading contribution to $\mathsf{t}\,S_{\rm SYM}$ is
\be
\begin{aligned}
\mathsf{t}\,S_{\rm SYM}=&-\int\mTr\left(\frac{1}{2}\dd_aB\wedge\star \dd_aB+(\dd_a\hat\sigma+[B,\bs\sigma_0^{(g)}])\wedge\star(\dd_a\hat\sigma+[B,\bs\sigma_0^{(g)}])\right)\\
&-\int d^3x\sqrt{g}\mTr\left(\frac{1}{2}\left(\hat{\mf D}+\hat\sigma\right)^2+\frac{i}{2}\hat{\bar\lambda}\gamma^\mu D^{(a)}_\mu\hat\lambda-\frac{1}{2}\hat{\bar\lambda}[\bs\sigma_0^{(g)},\hat\lambda]-\frac{1}{4}\hat{\bar\lambda}\hat\lambda\right)+\ldots ~
\end{aligned}
\ee
where $\dd_a$ is the background exterior derivative \eqref{eq:BGextder}, and $D_\mu^{(a)}$ is the spinor covariant derivative with fixed connection, $a$.  This action can be made Gaussian under a suitable gauge-fixing and then path-integrated in standard fashion.  We (very) briefly highlight the main points of that procedure below, but many details can be found \cite{Marino:2011nm} and references therein.

\subsubsection*{Gauge choice}

We will choose the gauge\footnote{This gauge-fixing is only consistent when $a$ is a flat-connection, implying that $\dd_a^2=0$ defines an equivariant cohomology.}
\beq
\mc G_a[B]=\dd^\dagger_a B\equiv-\star \dd_a\star B=0~,
\eeq
whose Fadeev-Popov determinant, $\Delta_a[B]$, can be enacted through adding ghosts $\bar c,c$:
\begin{multline}
    \label{eq:ZSCSghost}
Z^{\rm SCS+SYM}_k=e^{ikS_{\rm CS}[a]}  \int \dd\bs\sigma_0\,e^{-i\frac{k}{2\pi}\text{vol}(M_3)\mTr\bs\sigma_0^2} \\
\times\int\frac{\mc DB}{\mc V}\mc D{\hat{\bar \lambda}}\mc D{\hat \lambda}\mc D{\hat{\mf D}}\mc D\hat{\sigma}\mc D\bar c\mc Dc\,\delta[\dd^\dagger_aB]\,e^{-\mathsf{t}S_{\rm SYM}-S_{\rm ghost}}~,
\end{multline}
with action
\beq
S_{\rm ghost}=\int \mTr\left(\bar c\wedge\star \dd^\dagger_a\dd_{a+\mathsf{t}^{-1/2}B}\,c\right)=\int d^3x\sqrt{g}\,\mTr\left(\bar c\wedge\star\Delta^0_a\,c\right)+O(\mathsf{t}^{-1/2}) ~,
\eeq
where $\Delta^0_a=\dd^\dagger_a\dd_a$ is the $a$-deformed Laplacian acting on $\mf g$-valued zero-forms.\footnote{It is tacit in \eqref{eq:ZSCSghost} that the zero modes of $\bar c,c$ under $\Delta^0_a$ are not to be integrated over.} 
The ghost determinants simply cancel the determinants from $\hmf D$ and $\hat\sigma$ (as well as a Jacobian from $\delta[\dd^\dagger_aB]$) and so we arrive at the promised Gaussian path-integral:
\be
Z^{\rm SCS+SYM}_k=e^{ikS_{\rm CS}[a]}\int \dd\bs\sigma_0\,e^{-i\frac{k}{2\pi}\text{vol}M_3\mTr\bs\sigma_0^2}\,Z_{\rm Gauss}[\bs\sigma_0]~,
\ee
with
\begin{multline}
    Z_{\rm Gauss}[\bs\sigma_0]:=\\\int[\mc DB]_{\text{ker}\dd_a^\dagger}\mc D\hat{\bar\lambda}\mc D\hat\lambda\,e^{\frac{1}{2}\int\mTr (\dd_aB)^2+\int\mTr[B,\bs\sigma_0^{(g)}]^2-\int\mTr\left(\frac{i}{2}\hat{\bar\lambda}\gamma_\mu D_\mu^{(a)}\hat\lambda-\frac{1}{2}\hat{\bar\lambda}[\bs\sigma_0^{(g)},\hat\lambda]-\frac{1}{4}\hat{\bar\lambda}\hat\lambda\right)} ~.
\end{multline}

\subsubsection*{One-loop determinants}

The remaining task is now to compute the one-loop determinants from integrating out $\{B,\hat{\bar\lambda},\hat\lambda\}$. Recalling the procedure from Sec.\,\ref{sect:Abelianisation}, the first step is to ``canonicalize" the kinetic terms by redefining the fluctuating fields $\{B,\hat{\bar\lambda},\hat\lambda\}\rightarrow\{\tilde B,\tilde{\bar\lambda},\tilde{\lambda}\}$ via
\beq
\Phi=g^{-1}\tilde \Phi g~,\qquad \Phi\in\{B,\hat{\bar\lambda},\hat\lambda\}~.
\eeq
As a result the one-loop integration becomes ostensibly simpler 
\beq
Z_{\rm Gauss}[\bs\sigma_0]
=\int[\mc D\tilde B]_{\text{ker}d^\dagger}[\mc D\tilde{\bar\lambda}\mc D\tilde\lambda]e^{\frac{1}{2}\int\mTr\left(\dd\tilde B\right)^2+\int\mTr[\tilde B,\bs\sigma_0]^2-\int\mTr\left(\frac{i}{2}\tilde{\bar\lambda}\gamma^\mu \nabla_\mu\tilde\lambda-\frac{1}{2}\tilde{\bar\lambda}[\bs\sigma_0,\tilde\lambda]-\frac{1}{4}\tilde{\bar\lambda}\tilde\lambda\right)} ~,
\eeq
however, as we saw earlier, this is at the cost of twisting the fields along the curve $\gamma$:
\beq \label{eq:112}
\tilde\Phi_f=\exp(-2\pi\tmf m)\tilde\Phi_i\exp(2\pi\tmf m)~.
\eeq
In terms of a root-space decomposition $\tilde\Phi=\tilde\Phi^{(i)}T_i+\tilde\Phi^{(\alpha)}T_\alpha$, then \eqref{eq:112} reads 
\beq\label{eq:susytwistedBCsrootspace}
\tilde\Phi^{(i)}_f=\tilde\Phi^{(i)}_i~,\qquad \tilde\Phi^{(\alpha)}_i=\exp(-2\pi\alpha\cdot\tmf m)\tilde\Phi^{(\alpha)}_i~,
\eeq
where $T_i$ is a basis of the Cartan subalgebra containing $\bs\sigma_0$ and $\mf m$, and $T_\alpha$ are a basis of the $\alpha$ root space; $\{\tilde\Phi^{(i)},\tilde\Phi^{(\alpha)}\}$ are honest fields and not elements of $\mf g$. The one-loop determinants of the $\mc N=2$ vector multiplet with twisted boundary conditions, \eqref{eq:susytwistedBCsrootspace}, was described in \cite{Drukker:2012sr} and we happily cite that result here
\beq
Z_{\rm Gauss}[\bs\sigma_0]=\prod_{\alpha>0}\frac{\sin\left(i\alpha\cdot( \bs\sigma_0-2\pi\tmf m)\right)}{\pi^2}~.
\eeq
Again, we've written $Z_{\rm Gauss}$ rather generally, but for the purposes of this paper, we can let $\tmf m=i\ms h\,L_3$ with $\ms h\in\mathbb Z$ in which case it reduces to the usual expression for the Ray-Singer torsion in terms of $\bs\sigma_0=-i\sigma L_3$, i.e. $Z_{\rm Gauss}=\sin^2(\pi\sigma)/\pi^2$.  The phase of $Z_{\rm Gauss}$ again is responsible for the renormalization $k\rightarrow r=k+2$ as explained in \cite{Marino:2011nm}.  Gathering these results and fixing the normalization, we find again the familiar integral, \eqref{eq:ZkAb},
\beq\label{eq:Zksusy}
Z_k^{\rm SCS}=e^{irS_{\rm CS}[a]}\int_\mathbb R\dd\sigma\,\sin^2(\pi(\sigma+\ms h))\,e^{i\frac{\pi}{2}r\sigma^2}~,
\eeq
where again the integration contour over $\sigma$ should be deformed depending on the phase of $r$.

\subsection{Chern-Simons theory and dS\texorpdfstring{$_3$}{3} gravity: all-loop path-integral}\label{sec:all-loop-grav}

Having assembled these exact results, we now address the gravity path-integral about the $S^3$ saddle which, given the discussion in Sec.\,\ref{sect:intro} and Sec.\,\ref{sect:CSgrav}, we path-integral quantize as the product of Chern-Simons theories
\beq
\mc Z_{\rm grav}[S^3]=Z_{k_L}[S^3]Z_{k_R}[S^3] ~.
\eeq
Utilizing the exact partition function in the form of \eqref{eq:ZkAb} and \eqref{eq:Zksusy}, 
the gravity path-integral can be written as
\begin{multline}\label{eq:Zgravasint}
    \mathcal Z_{\rm grav}[S^3]=e^{ir_LS_{\rm CS}[a_L]+ir_RS_{\rm CS}[a_R]}\\\times\int \dd\sigma_L\dd\sigma_R\,e^{i\frac\pi2 r_L\sigma_L^2+i\frac\pi2 r_R\sigma_R^2}\sin^2\left(\pi(\sigma_L+\ms h_L)\right)\sin^2\left(\pi(\sigma_R+\ms h_R)\right)~.
\end{multline}
The background holonomies, $\ms h_{L,R}$, being $\pm 1$, decouple from this integral which is Gaussian and can be performed exactly:
\begin{align}\label{eq:Zgravfull}
\mathcal Z_{\rm grav}[S^3]=e^{2\pi s}\left(i\,e^{-i\frac{\pi}{r_L}-i\frac{\pi}{r_R}}\right)\frac{2}{\sqrt{r_Lr_R}}\sin\left(\frac{\pi}{r_L}\right)\sin\left(\frac{\pi}{r_R}\right)~.
\end{align}
Let us briefly dissect the phase in the parenthesis: the overall $i$ stems from integration contour deformations.  Given the identifications \eqref{eq:kLRdef} and \eqref{eq:kLRrenorm}, the $\sigma_L$ integral \eqref{eq:Zgravasint} is already damped, however the $\sigma_R$ integral is anti-damped.  Deforming the $\sigma_R$ integration contour to a damped region accounts for this $i$; this is wholly analogous to ``Polchinski's phase'' \cite{Polchinski:1988ua} arising from deforming the integration contour of the conformal mode in the gravitational path-integral.  The exponents are a combined framing phase.

At this point, this result, \eqref{eq:Zgravfull}, is not surprising. Up to a total phase, our expression for $\mc Z_{\rm grav}$ has been arrived at before through analytic continuation of the celebrated $SU(2)$ Chern-Simons partition function \cite{Castro:2011xb,Anninos:2020hfj,Hikida:2021ese} .  Here we have simply justified those analytic continuations, and incorporated the background contributions of $a_{L/R}$, through Abelianisation and supersymmetric localization.

It is instructive to cast \eqref{eq:Zgravfull} in gravitational terms. We recall from Sec.\,\ref{sect:CSgrav} that 
\be
r_{L/R}=\hat\delta \pm i\,s~,\qquad s=\frac{\ell}{4G_N}~.
\ee
First, let's inspect the case when $\hat\delta=0$, i.e., in the absence of the gravitational Chern-Simons term \eqref{eq:GCS-action}. The path integral reads
\be\label{eq:Zgrav-delta0}
\begin{aligned}
 \log\left(\mathcal Z_{\rm grav}[S^3]_{\hat\delta =0}\right)&=\log\left(\frac{8G_N}{i\ell} \exp({\frac{\pi}{2}\frac{\ell}{G_N}})\sinh^2\left(4\pi\frac{ G_N}{\ell}\right)\right)\\
 &=\frac{\pi}{2}\frac{\ell}{G_N} + 3\log(\frac{4G_N}{\ell})+\log(2 \pi^2i) + \frac{16\pi^2}{3}\frac{G_N^2}{\ell^2}+\cdots~.
\end{aligned}
\ee
The first line should be viewed as an exact expression in $G_N$ for the fixed background manifold $S^3$. The second line is the loop expansion as $G_N\to0$, where rather curiously the two-loop correction vanishes. In \cite{Anninos:2020hfj} the real part of $\log\mathcal Z_{\rm grav}$ was shown to match the graviton one-loop determinant on an $S^3$ background at one-loop order (that is, $O(G_N^0)$ and $O(\log G_N)$).  See \cite{Carlip:1992wg,Castro:2011xb} for analogous results and matching on Lens spaces.

Next, when $\hat\delta\neq0$, the structure of the results is slightly different. Casting \eqref{eq:Zgravfull} in terms of the gravitational variables, we find that the perturbative expansion is
\be
\begin{aligned}\label{eq:logZgrav-exp-delta}
 \log \mathcal Z_{\rm grav}[S^3] 
 =&\log\left(e^{2\pi s}\left(i\,e^{-2\pi i\frac{ \hat\delta}{\hat\delta^2+s^2}}\right)\frac{2}{\sqrt{\hat\delta^2+s^2}}\left|\sin\left(\frac{\pi}{\hat\delta+is}\right)\right|^2\right)\\
 =& 2\pi s + \log(\frac{2\pi^2 i}{s^3}) + \frac{\pi^2}{3}\frac{1}{s^2} -2\pi i \frac{\hat\delta}{s^2}- \frac{3}{2}\frac{\hat\delta^2}{s^2} +\cdots ~.
\end{aligned}
\ee
Here we kept $s$ as the coupling for clarity, instead of replacing $G_N$. Again, the first line is an exact result, and in the second line we are doing an expansion in $G_N$ (or equivalently large $s$). It is important to mention that in this expansion $\hat\delta$ is kept fixed in the limit $G_N\to0$. The additional purely imaginary term in the perturbative expansion (as compared to \eqref{eq:Zgrav-delta0}) comes from the framing phase, and it  only starts to contribute to $\mathcal Z_{\rm grav}$ at $O(G_N^2)$. Furthermore, this framing term vanishes if the coefficient of $I_{\rm GCS}$ renormalizes to $\hat\delta=0$.

\section{Looping matter in}\label{sec:looping-matter-in} 

In this section we address our central question: how to couple matter to gravity in the Chern-Simons formulation, and how to quantify this coupling beyond leading order in the gravity path-integral. We will provide a concise and precise answer to these questions. 

To that end, we will first cover the unitary representation theory of $\mf{so}(1,3)$, the isometry algebra of Lorentzian dS$_3$.  We will pay particular attention to representations corresponding to massive scalar particles.  Afterwards we will turn our attention towards the Euclidean rotation, $\mf{su}(2)_L\oplus\mf{su}(2)_R$, and see how to define {\it non-standard} representations of $\mf{su}(2)$ mimicking the single-particle representations of $\mf{so}(1,3)$. This is presented in Sec.\,\ref{sect:dSrepintro}. 

Having understood the particle content in a representation theoretic way, these non-standard representations are begging to be utilized in Wilson loop observables.  Thusly, in Sec.\,\ref{sec:Wilson-SU2} we will revisit our exact methods from Sec.\,\ref{sect:CStheory} to accommodate expectation values of Wilson loops carrying non-standard representations. The two methods used are Abelianisation and supersymmetric localization.

The end result of these analyses, presented in Sec.\,\ref{sect:spool}, is to then propose a gauge-invariant observable built from non-standard representations, an object we will call the {\it Wilson spool} (a nomenclature that will become duly clear below).  This is an object that incorporates quantum gravity effects to a free massive scalar field minimally coupled to dS$_3$ gravity.   This object can be intuitively motivated from the worldline quantum mechanics of a single-particle moving on $S^3$, however we will construct the spool bottom-up through a formula for one-loop determinants as a product over quasi-normal modes.  Lastly, enjoying the fruit of our labors from retooling Abelianisation and localization, we show how the spool can be evaluated order-by-order in $G_N$ perturbation theory to give controlled and finite quantum gravity corrections to scalar one-loop determinants.

\subsection{Single-particle representation theory}\label{sect:dSrepintro}

Unitary representations of the $SO(1,3)$ de Sitter isometry group %\cz{[comment: change to the algebra instead of the group to be consistent with the change below]} 
describe single-particle states propagating on dS$_3$ spacetime~\cite{Joung:2006gj,Joung:2007je,Basile:2016aen,Sun:2021thf} (see \cite{Penedones:2023uqc} for multi-particle states). %\cz{For the purposes of a Chern-Simons description, we find it useful to consider representations associated to Euclidean dS$_3$, with isometry algebra $\mathfrak{so}(4)\simeq su(2)\times su(2)$. This algebra splits into two copies in which the barred and unbarred gauge connection is valued.} 
For the purposes of a Chern-Simons description of Euclidean dS$_3$ it will be useful to cast quantities in terms of the Euclidean isometry algebra $\mf{so}(4)\simeq\mf{su}(2)_L\oplus\mf{su}(2)_R$ and make use of its split structure. Recently, it was shown how to mimic the essential features of light scalars ($\mathsf{m}^2\ell^2<1)$ with novel representations of 
%\sout{the Euclidean dS$_3$ isometry group $SO(4)$} 
the $\mathfrak{so}(4)$ algebra \cite{Castro:2020smu}. 

%\cz{We give analytic continuation between the $\mathfrak{so}(1,3)$ and $\mathfrak{so}(4)$ algebras in Appendix~\ref{app:conventions}. However, we note that the $\mathfrak{so}(4)$ \emph{representations} suitable for Chern-Simons theory do not analytically continue to the standard $\mathfrak{so}(1,3)$ \emph{representations}, even though they do mimic certain features of these representations. In Lorentzian dS$_3$, they are suitable for describing the quasinormal modes rather than the single-particle states.} 
It is important to note that although $\mf{so}(1,3)$ and $\mf{so}(4)$ share a common complexification (as we review in App.~\ref{app:conventions}), the representations constructed in \cite{Castro:2020smu} do not analytically continue to standard representations of $\mf{so}(1,3)$. Instead they furnish a representation of quasi-normal modes of Lorentzian dS$_3$, as opposed to single-particle states.
%\sout{Analytically continued to Lorentzian signature, these describe the quasi-normal mode spectrum of dS$_3$, rather than the typical propagating fields. However,} 
As noted by several authors \cite{Denef:2009kn,Jafferis:2013qia,Castro:2020smu,Anninos:2020hfj,Law:2022zdq}, quasi-normal modes provide a rather useful basis for computing a number of physical quantities.  Particularly, in the context of Chern-Simons gravity, classical Wilson lines carrying these representations have been shown to describe Green's functions and other gravitational probes in dS$_3$ \cite{Castro:2020smu}.  Already in \cite{Jafferis:2013qia} it was emphasized that the quasi-normal mode spectrum of four-dimensional de Sitter is unitarily realized in a non-standard way.  We will connect to (and extend) these ideas further below.

In this subsection we will briefly review both the unitary representation theory of $SO(1,3)$ as well as the non-standard representations of $SO(4)$, emphasizing important differences in how they are realized.  In doing so, we will also extend the construction of non-standard representations constructed in~\cite{Castro:2020smu} to incorporate heavy scalar fields ($\mathsf{m}^2\ell^2>1$) in a unified way.

\subsubsection{Unitary representations of \texorpdfstring{$\mf{so}(1,3)$}{so(1,3)}}\label{sec:unitaryreps}

Taking a real basis of generators $L_{AB}\in\mf{so}(1,3)$, unitary representations will realize anti-Hermiticity
\beq\label{eq:antiHermL}
L_{AB}^\dagger=-L_{AB}~.
\eeq
In the conformal basis (see App.\,\ref{app:conventions} for our conventions) this amounts to the condition
\beq \label{eq:antiHermconf}
D^\dagger = -D~, \indent M_{ij}^\dagger = -M_{ij}~, \indent P_i^\dagger = -P_i~, \indent K_i^\dagger = -K_i~.
\eeq
This is the usual Hermiticity choice that is appropriate for building fields on a de Sitter background~\cite{Sun:2021thf}.

The standard spinless representations are defined starting from a primary state $\left|\bar{\Delta}\right>$ with scaling dimension $\bar \Delta=2-\Delta$ as
\be D \left|\bar{\Delta}\right> = \bar{\Delta}~, \indent K_i \left|\bar{\Delta}\right> = 0~, \indent M_{ij} \left|\bar{\Delta}\right> = 0~. \ee
They have a Casimir
\begin{align}\label{eq:so13cas}
c^{\mf{so}(1,3)}_2 =& \frac{1}{2} L_{AB} L^{AB} = D(2-D) + P_i K^i + \frac{1}{2} M_{ij} M^{ij}\nonumber\\
=&\bar\Delta(2-\bar\Delta)=\Delta(2-\Delta)~. 
\end{align}
For scalars this Casimir determines the mass of the field: $c_{2}^{\mf{so}(1,3)}=\mathsf{m}^2\ell^2$.  A convenient basis for building the representation is given by
\be \left|x\right> = e^{x^i P_i} \left|\bar{\Delta}\right>~.\ee
In terms of this, the de Sitter invariant norm is $\left<x|y\right>$. Requiring anti-Hermiticity, \eqref{eq:antiHermconf}, for scalar representations constrains $\Delta$ and the norm to be
\begin{align}
   \Delta &=1+\nu~, ~\nu\in(-1,1)~,&\hspace{-4em}\left<x|y\right> \propto& \frac{1}{|x-y|^{2\bar \Delta}}~,\label{eq:standardcomp}\\
    \Delta &= 1 - i \mu~, ~\mu\in \mathbb{R}~, &\hspace{-4em}\left<x|y\right> =& \delta^2(x-y)~.\label{eq:standardprinc}
   \end{align}
The former case above, \eqref{eq:standardcomp}, is the {\it complementary series} representation.  It corresponds to a ``light'' massive scalar via
\beq
\nu^2=1-\mathsf{m}^2\ell^2~,\qquad \mathsf{m}^2\ell^2<1~.
\eeq
The latter case, \eqref{eq:standardprinc}, is the {\it principal series} representation.  It corresponds to a ``heavy'' massive scalar via
\beq
\mu^2=\mathsf{m}^2\ell^2-1~,\qquad \mathsf{m}^2\ell^2>1~.
\eeq
There are additional representations of $\mf{so}(1,3)$ (such as spinning principal series) but we will content ourselves to massive scalars for this paper. For a more detailed discussion of unitary representations, see e.g. \cite{Penedones:2023uqc} and references within.

\subsubsection{Non-standard representations of \texorpdfstring{ $\mf{su}(2)_L\oplus\mf{su}(2)_R$}{su(2)LxsU(2)R}}\label{sec:nonstandardreps}

Keeping in mind the incorporation of matter into our Chern-Simons theory, we now want to import these representations, or at least their essential features, into the Euclidean theory.  One potential and obvious obstruction is that the Euclidean isometry algebra is $\mf{su}(2)_L\oplus\mf{su}(2)_R$ and the standard representation theory of $\mf{su}(2)$ does not admit any continuous (much less complex!) parameter that can play the role of $\nu$ or $\mu$.  This problem was partially solved for light scalars in \cite{Castro:2020smu} by altering the inner product (or equivalently, the notion of Hermitian conjugation).  Such representations were deemed ``non-unitary" in that paper to highlight their differences (in particular, Hermiticity) from standard $\mf{su}(2)$ representation theory.  However ``non-unitary" can connote a broad range of meanings.  The philosophy in this paper is that ensuring all states have positive norm, regardless of the choice of Hermiticity, is the minimum requirement for a representation to be deemed physical.  To emphasize the physical importance of these representations, in this paper we diverge in nomenclature from \cite{Castro:2020smu} and call these {\it non-standard} representations.  We will construct them now.

We will begin with some generalities about highest-weight representations.  Recall that the root-space basis of the $\mf{su}(2)$ algebra is generated by $L_3, L_\pm$ satisfying
\be
[L_3, L_\pm] = \pm L_\pm~, \indent [L_+,L_-]=2L_3~.
\ee
The quadratic Casimir is
\be 
c_2= L^2 = \frac{1}{2}\left(L_+ L_-+L_- L_+\right) + L_3^2~.\label{eq:CasimirSU2}
\ee
A highest weight representation for $\mf{su}(2)$ is constructed starting from a highest weight state\footnote{The label $j$ here and throughout is what was called $l$ in the previous paper~\cite{Castro:2020smu}.} $\left|j,0\right>$ that satisfies
\beq L_3 \left|j,0\right>= j \left|j,0\right>~, \indent L_+ \left|j,0\right> = 0~.\eeq
An arbitrary state is constructed from this by successive application of the lowering operator
\beq 
\left|j,m\right> = N_{j,m}(L_-)^m \left|j,0\right>~.\label{eq:nonhigheststates}
\eeq
The Casimir of a highest-weight representation is given by its action on the highest-weight state
\beq 
c_2\left|j,0\right> = L^2\left|j,0\right> =c_j\left|j,0\right>~,\qquad c_j:=j(j+1)~.\label{eq:Casimir}
\eeq
To inform our construction of non-standard representations, it is useful to note the following relations in the continuation from $\mf{so}(1,3)$ to $\mf{su}(2)_L\oplus\mf{su}(2)_R$ (full details are in App.\,\ref{app:conventions}).  Firstly, the dilatation and spin generators of $\mf{so}(1,3)$ map, respectively, to
\beq\label{eq:DMtoLs}
D=-L_3-\bar L_3~,\qquad \varepsilon^{ij}M_{ij}=iL_3-i\bar L_3~,
\eeq
and so highest-weight labels of $\mf{su}(2)_L\oplus\mf{su}(2)_R$ remain good labels for the Cartan of $\mf{so}(1,3)$. Namely, they continue to the conformal dimension and spin as\footnote{Besides here and Appendix \ref{app:conventions}, the label ``$s$" for spin does not make any further appearances.  There should be no confusion with the imaginary part of the Chern-Simons level in \eqref{eq:kLRdef}.}
\beq\label{eq:DeltaStojs}
\Delta=-j_L-j_R~,\qquad s=j_L-j_R~.
\eeq
For spinless representations $j_L=j_R\equiv j$, and so 
\beq\label{eq:Deltatojs}
\Delta=-2j~.
\eeq
Secondly, the $\mf{so}(1,3)$ Casimir is related to the $\mf{su}(2)_L\oplus\mf{su}(2)_R$ Casimir via
\beq\label{eq:castocas}
c_2^{\mf{so}(1,3)}=-2c_2^{\mf{su}(2)_L}-2c_2^{\mf{su}(2)_R}~.
\eeq
We then will have the following properties in mind which allow us to match onto the physics of scalar fields in de Sitter:
\begin{itemize}[leftmargin=0.4cm]
    \item {\bf Continuous highest-weight $j$}. Given the relation, \eqref{eq:Deltatojs}, the highest-weights of the representations must be allowed to vary continuously and possibly be complex to mimic the conformal dimensions corresponding to massive particles.
    \item {\bf Negative Casimir}. Given \eqref{eq:castocas}, the $\mf{su}(2)$ Casimir is in fact related to \emph{minus} the mass squared of a scalar particle and so to match %gravitational physics for de Sitter, as opposed to anti- de Sitter,
    physics appropriate for de Sitter, these representations necessarily have a negative Casimir. From \eqref{eq:Casimir}, this occurs in the window $-1<\mbox{Re}(j)<0$. To satisfy this, we must allow for either negative or complex $j$.
    \item {\bf Paired representations}. Since $c_j = c_{-1-j}$, there are two distinct representations at a fixed Casimir when $j$ lies in the window $-1<\mbox{Re}(j)<0$. Inside \eqref{eq:Deltatojs}, sending $j\rightarrow-1-j$ is akin to the ``shadow'' map in $\mf{so}(1,3)$ sending $\Delta\rightarrow 2-\Delta$.  The use of both representations was crucial to match gravitational physics in~\cite{Castro:2020smu}.
\end{itemize}

As alluded to above, we will accommodate the above features in $\mf{su}(2)$ by altering the notion of Hermitian conjugation.  Let us state what this non-standard Hermitian conjugation looks like.  We will first define a map, $\mc S$, between highest-weight representations by its action on basis states, \eqref{eq:nonhigheststates}, via
\beq \mathcal S \left|j,m\right> = \left|\bar{j},m\right>~,\label{eq:shadow}\eeq
where $\bar j$ is related to $j$ in a way to be determined shortly.  The non-standard representations in this paper are then defined with respect to the following Hermitian conjugation:
\beq\label{eq:genNShermcon}
L^\dagger_3=\mc S^{-1}L_3\mc S~,\qquad L^\dagger_\pm=-\mc S^{-1}L_\mp \mc S~.
\eeq
We will additionally demand reality of the Casimir, $c_2^\dagger=c_2$, which, acting on a highest-weight state, fixes $\bar j$ to be one of two cases
\begin{align}
\bar j=j~,\qquad\text{or}\qquad\qquad \bar j=-1-j~.
\end{align}
The former case, where $\mc S=\text{identity}$, will lead to the representations from \cite{Castro:2020smu} which are in fact related to the complementary series of $\mf{so}(1,3)$.  In this paper we will call them {\it complementary-type} representations. In the latter case, $\mc S$ takes the form of a shadow map (from here on, when we refer to $\mc S$, it is always with this latter case in mind).  Representations obeying \eqref{eq:genNShermcon} in this case will be related to the principal series of $\mf{so}(1,3)$.  We will call them {\it principal-type}.

\paragraph{Complementary-type.}
These representations do not explicitly involve the shadow map in their Hermitian conjugation,
\beq L_3^\dagger = L_3~, \indent L^\dagger_{\pm} = -L_{\mp}~.\label{eq:Hermiticitycomp}\ee
Regardless, they are still non-standard on account of the extra minus sign as compared to the standard $\mf{su}(2)$ Hermitian conjugation.\footnote{See \cite{Guijosa:2003ze} for a similar choice regarding representations of dS$_2$.}  Analytically continued to $\mf{so}(1,3)$ in the conformal basis, this is
\beq D^\dagger = D~, \indent M^\dagger = -M~, \indent P_i^\dagger = K_i~.\eeq
This Hermiticity choice is perhaps familiar as the appropriate choice to describe fields in {Euclidean AdS} space, which also has an $SO(1,3)$ isometry group in three dimensions. In the context of de Sitter however this is in fact the Hermiticity choice appropriate for quasi-normal modes equipped with the so-called ``R-norm" constructed in \cite{Jafferis:2013qia}.\\

Now let us consider the specifics of the highest weight $\mf{su}(2)$ representation given this choice of Hermiticity. To determine the normalization in Eq.~\eqref{eq:nonhigheststates}, we evaluate the matrix elements
\begin{align} \left<j, m-1\right| L_{+} \left| j, m\right> &= \left< j, m\right| L_{+}^\dagger \left|j, m-1\right>^*\nn \\
&= -\left< j, m\right| L_{-} \left|j, m-1\right>^*~,\label{eq:normcond}
\end{align}
iteratively starting from the highest weight state. On the one hand,
\be L_- \left|j, m-1\right> = \frac{N_{j,m-1}}{N_{j,m}} \left|j,m\right>~,\label{eq:Lmval}\ee
while on the other hand
\be L_+ \left|j, m\right> = m(2j+1-m)\frac{N_{j,m}}{N_{j,m-1}}\left|j,m-1\right>~,\label{eq:Lpval}\ee
which can be shown using the identity $[L_+, L_-^k] = 2 \sum\limits_{n=1}^k L_-^{k-n} L_3 L_-^{n-1}$.  Then \eqref{eq:normcond}  implies
\be |N_{j, m+1}| = \sqrt{m(m-2j-1)}\, |N_{j,m}|~.\label{eq:norms}\ee
For complementary-type representations, we can assume without loss of generality that the normalizations are real, i.e. $|N_{j,m}| = N_{j,m}$. We thus find for these representations
\be
\begin{aligned} 
L_{-} \left|j, m\right> &= \sqrt{(m-2j)(1+m)}\left|j,m+1\right>~,\label{eq:Lmwnorm}\\
L_{+}\left|j, m\right> &= -\sqrt{m(m-2j-1)}\left|j, m-1\right>~.
\end{aligned}
\ee
Note that the minus sign in the $L_+$ normalization is also necessary to satisfy the algebra commutation relations.

We must check that there are no negative norm states. We will do this inductively assuming $\langle j,0|j,0\rangle=1$.  For the first excited state, using the Hermiticity condition~Eq.~\eqref{eq:Hermiticitycomp} we have
\be
|L_{-} \left|j, 0\right>|^2 = -2j~,
\ee
and so this will be positive if $j<0$.  Now we proceed inductively.  Assume that $\langle j,m|j,m\rangle>0$ and consider
\begin{align} 
 |L_{-} \left|j, m\right>|^2 &= (m-2j)(1+m)\left<j,m|j,m\right>~.\label{eq:Lnorms}
\end{align}
This is also positive for $j<0$.  It then follows from induction that all basis states are positive.

In the usual $\mf{su}(2)$ representation theory, compactness together with unitarity enforce that representations are finite dimensional. So the successive action of lowering operators truncates. This manifests as a violation of norm positivity at $m\geq 2j+1$. Since the maximum value of $m_{\rm max} = 2j$ is an integer, $j$ must be either integer or half integer. However, as evident from Eq.~\eqref{eq:Lnorms}, for this non-standard representations norm positivity imposes no limit on $m$ and hence the representation is infinite dimensional. This also means that $j$ is no longer required to an be an integer or half integer. Thus, the highest-weight can take continuous values, which was the first requirement for these non-standard representations.

We impose by hand the second requirement, $-1<j<0$ and write 
\beq\label{eq:jtonu}
j=-\frac{1}{2}(1+\nu)~,\qquad \nu\in(-1,1)~.
\eeq
The complementary-type representations provide spinless representations of $\mf{su}(2)_{L}\oplus\mf{su}(2)_R$ which translate to $\mf{so}(1,3)$ conformal dimension via \eqref{eq:Deltatojs} as
\beq
\Delta = 1+\nu~.
\eeq
This is obviously related the complementary series, \eqref{eq:standardcomp}, and corresponds to a light massive field with $\mathsf{m}^2\ell^2=1-\nu^2$.

Lastly, it will be useful to define and evaluate a character associated to these representations. We will use
\be\label{char1} \chi_j(z) \equiv \mbox{Tr}_j\left(e^{2\pi i z L_3}\right) = \sum_{m=0}^\infty e^{2\pi i z (j-m)} = \frac{e^{2\pi i z (j+1)}}{e^{2\pi i z}-1}~.\ee
In terms of $\nu$, we have
\beq\label{eq:compchar}
\chi_\nu(z)=\frac{e^{i\pi z\nu}}{2i\sin\pi z}~.
\eeq

\paragraph{Principal-type representations.}
For the principal-type, we instead use Hermiticity explicitly involving the shadow map, $\mc S$, \eqref{eq:shadow}:
\beq L_3^\dagger = \mathcal S L_3 \mathcal S~, \indent L_\pm^\dagger = -\mathcal S L_\mp \mathcal S~,\ee
with $\mc S$ sending $j\rightarrow\bar{j}= -1-j~$. This implies
\beq j = \left< j,0\right|L_3^\dagger \left|j,0\right>^* = (-1-j)^*~,\eeq
which can be solved as
\beq\label{eq:jtomu}
 j = -\frac{1}{2}(1-i\mu)~,\qquad \mu \in \mathbb{R}~.
 \ee
Given \eqref{eq:Deltatojs}, this will be related to spinless principal series representations with
\beq
\Delta =1-i\mu~,
\eeq
analogous to Eq.~\eqref{eq:standardprinc}. It also satisfies the second requirement for our representations, since the Casimir 
\beq
c_j=-1-\mu^2~,
\eeq
is negative for all $\mu \in \mathbb{R}$. In contrast to the complementary-type series, we are now describing heavy fields with $\mathsf{m}^2\ell^2=\mu^2+1$. 

We can again solve for the normalizations in \eqref{eq:nonhigheststates} using the analogue of \eqref{eq:normcond} but for this different Hermiticity choice. This is
\be
\left<j, m-1\right| L_{+} \left| j, m\right> = -\left< j, m\right| \mathcal S L_{-} \mathcal S \left|j, m-1\right>^*~.\label{eq:normcondp}
\ee
Using \eqref{eq:Lmval} and~\eqref{eq:Lpval}, which hold identically in this case, we have
\be
\begin{aligned}
 \mathcal S L_{-} \mathcal S \left| j, m\right> &= \frac{N_{\bar j, m-1}}{N_{\bar j, m}}\left|j,m\right>~,\\
 L_{+} \left| j, m\right> &= -m(m-i\mu)\frac{N_{j,m}}{N_{j,m-1}} \left|j, m-1\right>~.
\end{aligned}
\ee
Unlike for the complementary-type, constructing the $N_{j,m}$'s here is a bit more involved: they are naturally complex. Let
\be \alpha_{j,m} = \mbox{Arg}(N_{j,m})~, \indent \alpha_{\bar j, m} = \mbox{Arg}(N_{\bar j, m})~,\ee
and also
\be \phi_{j,m} = \mbox{Arg}(m-i\mu)~,\indent m\geq 1~,\ee
with the inner product \eqref{eq:normcondp} requiring
\be \alpha_{\bar j, m} - \alpha_{j,m} - \alpha_{\bar j, m-1}+ \alpha_{j, m-1} = \phi_{j, m} \,\mbox{mod} \,2\pi~. \ee
We are free to set $\alpha_{j,0} = \alpha_{\bar j, 0} = 0$ and also $|N_{\bar j, m}| = |N_{j,m}|$. Then we can solve for the norms to obtain
\beq N_{j, m-1} = e^{i \alpha_{j, m-1} - i \alpha_{j,m}}\sqrt{m|m-i\mu|}\, N_{j,m}~,\eeq
and the same for the barred copy, $N_{\bar j,m}$.  Thus we find for the action on states,
\be
\begin{aligned}
L_- \left|j,m\right> &= e^{i \alpha_{j,m}-i\alpha_{j,m+1}}\sqrt{(m+1)|m+1-i\mu|}\left|j,m+1\right>~,\\
L_+\left|j,m\right> &= -e^{i \alpha_{j,m}-i\alpha_{j,m+1}+i\phi_m}\sqrt{(m+1)|m+1-i\mu|}\left|j,m-1\right>~.
\end{aligned}
\ee
We can now proceed inductively to check that the norm of all states are positive. We have
\be |L_- \left|j,0\right>|^2 = |1-i\mu|>0~.\ee
Next
\be |L_- \left|j,m\right>|^2 = (m+1)|m+1-i\mu|\left<j,m|j,m\right>~,\ee
which is again positive assuming positivity of $\langle j,m|j,m\rangle$.  This completes the induction. As for the complementary series, there is no lower bound imposed by this condition. The representations again do not truncate with successive action of lowering operators, and there is no condition on $j$ being integer or half integer. Thus, we again satisfy the first requirement that the quantum number can take continuous values for these representations.  As mentioned before, the second requirement (negative Casimir) is automatically satisfied.

For these representations we will use the same definition for the character as in the complementary-type series. Using this, we find
\be\label{eq:Zukchars} \chi_j(z) = \mbox{Tr}_j\left(e^{2\pi i z L_3}\right)= \sum_{m=0}^\infty e^{2\pi i z (j-m)} = \frac{e^{2\pi i z (j+1)}}{e^{2\pi i z}-1}~.\ee
In terms of $j$ this character takes the exact same functional form as \eqref{char1}.  In terms of $\mu$ this is
\be \chi_\mu(z) = \frac{e^{-\pi z \mu}}{2 i \sin{(\pi z)}}~.\label{eq:charp}\ee

\subsection{\texorpdfstring{$SU(2)$}{SU(2)} Chern-Simons theory: Wilson loop}\label{sec:Wilson-SU2}

Now we turn to using the above representations in expectation values of a Wilson loop.  To establish some notation, Wilson loops are defined by a trace in a specified representation, $R$, of a path-ordered exponential of the integral of the connection, $A$, pulled back to a closed one-dimensional submanifold, $\gamma$:
\beq \label{Wilsondef}
W_R[\gamma]=\Tr_R\left(\mc P\exp \oint_\gamma A\right) ~,
\eeq 
whose expectation values we will denote as
\beq \label{Wilsonloopexpectation}
\big\langle W_R[\gamma]\big\rangle=\int \frac{\mc DA}{\mc V}\,e^{ikS_{CS}} \, W_R[\gamma] ~. 
\eeq
In principle one could similarly consider the expectation values of multiple Wilson loops $\blangle W_{R_1}[\gamma_1]W_{R_2}[\gamma_2]\ldots W_{R_n}[\gamma_n]\brangle$ along multiple non-intersecting paths, however for the purpose here, we will restrict ourselves to a single observable.

Up to a choice of framing (described in Sec.\,\ref{sect:CStheory}), Wilson loop expectation values can only depend on topological features of the curves and the details of their representations and in fact coincide with the colored link invariants of $\gamma_1\cup \gamma_2\cup\ldots\cup \gamma_n$.  The power of topological field theory is that there are several methods for evaluating these expectation values exactly.  In Sec.\,\ref{sect:CStheory} we outlined two of those techniques, Abelianisation and supersymmetric localization, in the context of evaluating the gravity path-integral.  Below we re-examine their utility for incorporating Wilson loops with non-standard representations.

\subsubsection{Abelianisation}\label{sect:WLAb}

Returning now to Abelianisation, Sec.\,\ref{sect:Abelianisation}, we will place the Wilson loop \eqref{Wilsondef} in the fibre direction 
of $M$, i.e. along $\gamma$ in Eq.~\eqref{eq:AbBGholo}.
Following \cite{Blau:2013oha}, this operator then depends only on the associated representation $R$ and on the $a+B_\kappa=a+\kappa\bs\sigma$ part of the connection, but not on $B_H$. The inclusion of such Wilson loops %therefore 
does not change the evaluation of the path-integral though Abelianisation and $\bs\sigma$ can be taken to be valued in $\mathfrak{t}$ and constant on $M$. The expectation value of a Wilson loop $\mathrm{Tr}_{R}\left(\mc P \, \exp (\oint A) \right)$, whether normalised or not, is the same as evaluating the partition function \eqref{S3:CSPI} with the character 
\begin{equation}
    \mathrm{Tr}_{R}\mc P \, \exp \left(\oint_\gamma (a+ \kappa\bs\sigma)\right )~,
\end{equation}
inserted into the integrand, which leads to 
\begin{equation}\label{eq:WexpAb}
\langle W_R[\gamma] \rangle =  e^{ir S_{\rm CS}[a]} \int_\mathbb{R} \dd\sigma \, \mathrm{sin}^2 \left( \pi\sigma \right) \mathrm{exp}\left(i \frac{\pi}{2}r \, \sigma^2 \right)  \chi_R\left(\sigma+\ms h\right)~.
\end{equation}
Recall that $\ms h$ captures the holonomy of $a$. An advantage of Abelianisation is that the procedure does not depend on the nature of the representation $R$.  Of course, it could be that for infinite-dimensional representations the above integral diverges; we verify that it does not. More explicitly, evaluating \eqref{eq:WexpAb} with $R$ being either of the non-standard representations with $\chi_j$ from \eqref{char1} or \eqref{eq:Zukchars} (stated in terms of $j$, the characters are equivalent for both representations) gives 
\beq\label{eq:WexpAb2}
\langle W_j[\gamma]\rangle=\frac{1}{2}\,e^{irS_{\rm CS}[a]+i2\pi\,\ms h\,j}e^{i\phi-i\frac{2\pi}{r}c_j}\sqrt{\frac{2}{r}}\sin\left(\frac{\pi(2j+1)}{r}\right)~,
\eeq
where we've again taken $\ms h\in\mathbb Z$.  We have grouped the terms above in an intuitive manner.  We firstly have the classical contribution, $\exp(irS_{\rm CS}[a]+i2\pi\ms h\,j)$, associated to the background connection.  Secondly we have the overall phase, $\exp\left(i\phi-i\frac{2\pi}{r}c_j\right)$, where $\phi$ reappears from Eq.~\eqref{eq:Zphase} as the combination of contour rotation phase and a framing phase.  The addition of $c_j=j(j+1)$ is the expected framing phase for $W_j$ \cite{Witten:1988hf}. The last two terms, including the sine-function, are also present when $R$ is a finite-dimensional representation, and therefore resonate with known expressions for a Wilson loop \cite{Witten:1988hf,Beasley:2009mb,Beasley:2010hm}.

Despite the simplicity of the above result, we wish to emphasize that arriving at this point is non-trivial: one na\"ive analytic continuation of the standard $SU(2)$ result might involve declaring $2j+1\equiv d_j$ as the dimension of a finite dimensional $SU(2)$ representation and concluding that $\langle W_j\rangle$ does not have a sensible analytic continuation to our infinite-dimensional non-standard representations.  This is obviously not the case. Abelianisation gives us a first-principles justification to apply the standard $SU(2)$ results to non-standard representations that is manifestly finite and sensible even for $j\in\mathbb C$.

Lastly we note for future utility that the above arguments regarding Abelianisation continue to hold if we allow a general coefficient\footnote{We will discuss potential issues of gauge-invariance related to this in Sec.\,\ref{sect:spool}.}, $\alpha$, inside the path-ordered exponential:
\beq
W^{(\alpha)}_R[\gamma]=\mTr_R\mc P\exp\left(\frac{\alpha}{2\pi}\oint_\gamma A\right)~,
\eeq
allowing us to define an expectation value for $W^{(\alpha)}$, at least formally, as the integral
\begin{equation}\label{eq:WalphaAb}
\langle W_R^{(\alpha)}[\gamma] \rangle =  e^{ir S_{\rm CS}[a]} \int_\mathbb{R} \dd\sigma \, \mathrm{sin}^2 \left( \pi\sigma \right) \mathrm{exp}\left(i \frac{\pi}{2}r \, \sigma^2 \right)  \chi_R\left(\frac{\alpha}{2\pi}(\sigma+\ms h)\right)~.
\end{equation}
Doing this integral exactly is much harder, however it can be performed in perturbation theory for large $r$.

\subsubsection{\texorpdfstring{$\mc N=2$}{N=2} supersymmetric localization}\label{sect:WLsusy}
We can similarly re-examine supersymmetric localization procedure in  Sec.\,\ref{sect:susylocalizationS3} in the context of Wilson loops.  In order to do so, we will need to modify $W_R$ to ensure it maintains supersymmetry.  We will assume that its path, $\gamma$, is taken along a great circle preserving supersymmetry. Note that the background connection may possess holonomy \`a la Eq.~\eqref{eq:holonomy}.  The supersymmetric Wilson loop is given by
\beq
W_R[\gamma]=\mTr_R\mc P\exp\left(\oint_{\gamma}\left(A+\dd s\,|\dot x|\bs\sigma\right)\right)~.
\eeq
Inserting this into the path-integral, \eqref{eq:ZSCSSYM}, it is clear that in the $\mathsf{t}\rightarrow\infty$ limit this operator localizes around its saddle point value
\beq \label{netW-loop-susy}
W_R\,\hat{=}\,\mTr_R\mc P\exp\left(\oint_{\gamma}\left(a+\dd s\,|\dot x|\bs\sigma_0^{(g)}\right)\right)\,\hat{=}\,\mTr_R\exp\left(i2\pi(\ms h+\sigma)L_3\right) ~,
\eeq
where $\hat{=}$ denotes ``true inside the path-integral."  The expectation value then localizes to integral
\begin{equation}\label{eq:Wexpsusy}
\langle W_R[\gamma] \rangle =  e^{ir S_{\rm CS}[a]} \int_\mathbb{R} \dd\sigma \, \mathrm{sin}^2 \left( \pi\sigma \right) \mathrm{exp}\left(i \frac{\pi}{2}r \, \sigma^2 \right)  \chi_R\left(\sigma+\ms h\right)~.
\end{equation}
This matches \eqref{eq:WexpAb}. Again these arguments do not rely on the nature of the representation, $R$, appearing in \eqref{eq:Wexpsusy}.  As such, \eqref{eq:Wexpsusy}, continues to hold for our non-standard representations and the integral can be performed to give the same, finite, results as in Abelianisation, \eqref{eq:WexpAb2}.  Additionally, upon inclusion of an arbitrary parameter, $\alpha$,
\beq
W^{(\alpha)}_R[\gamma]=\mTr_R\mc P\exp\left(\frac{\alpha}{2\pi}\oint_\gamma\left(A+\dd s\,|\dot x|\bs\sigma\right)\right)~
\eeq
remains supersymmetric and the arguments regarding localization continue to hold.  As such $\langle W_R^{(\alpha)}\rangle$ can still be defined through the integral \eqref{eq:WalphaAb}.

\subsection{The Wilson spool: quantum gravity coupled to matter}\label{sect:spool}

We want to now leverage the ability evaluate Wilson loops exactly to give insight into how to couple matter into (and integrate out of) our quantum gravity path-integral.  Said another way, we have the tools to evaluate the expectation values of Wilson loops, however we want to know what gravitational physics lies in those expectation values.

We will state the main result shortly, however let us preface it with a few points of guiding intuition:
\begin{itemize}[leftmargin=0.4cm]
\item At low-energies, Wilson lines represent the worldlines of massive particles that have been integrated out.  This intuition extends, at least at the classical level, to Chern-Simons theories of gravity where Wilson lines can be represented as worldline quantum mechanics with equations of motion equivalent to geodesic motion \cite{Witten:1989sx,Carlip:1989nz,Ammon:2013hba}.
\item In Euclidean signature, looped worldlines also compute one-loop determinants of massive scalar fields via the heat kernel representation.  When the base space is compact, the one-loop determinant includes a sum over worldlines wrapping the compact space arbitrarily many times.  We illustrate this explicitly for $S^3$ in App.~\ref{app:HK}.
\end{itemize}
Relying on these two pieces of intuition we expect that the one-loop determinant of a scalar field will be (roughly) related to an object packaging multiply wound Wilson loop observables.

\begin{figure}[ht]\label{fig:alphacont1}
\includegraphics[height=.3\textwidth]{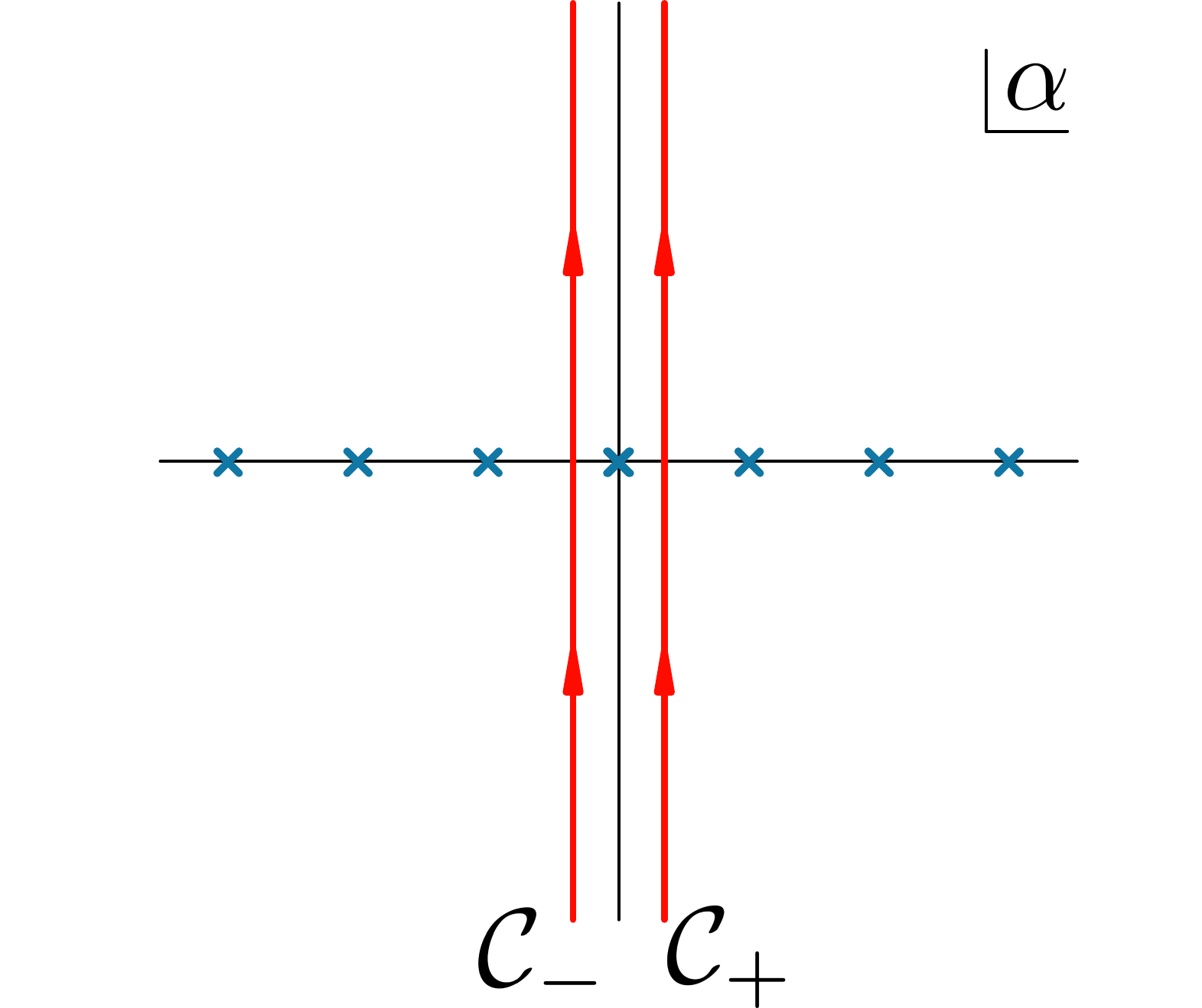}\includegraphics[height=.3\textwidth]{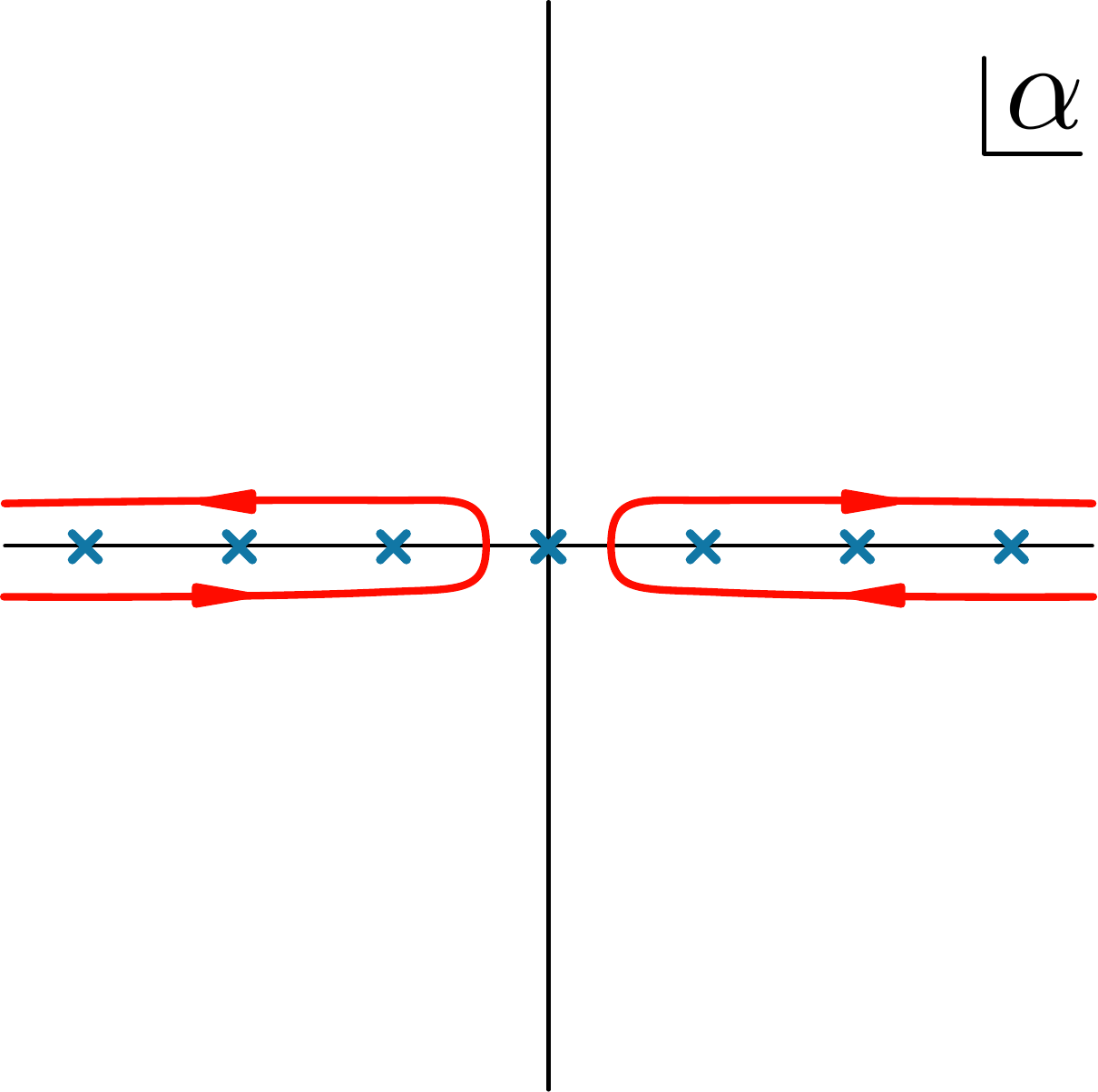}\includegraphics[height=.3\textwidth]{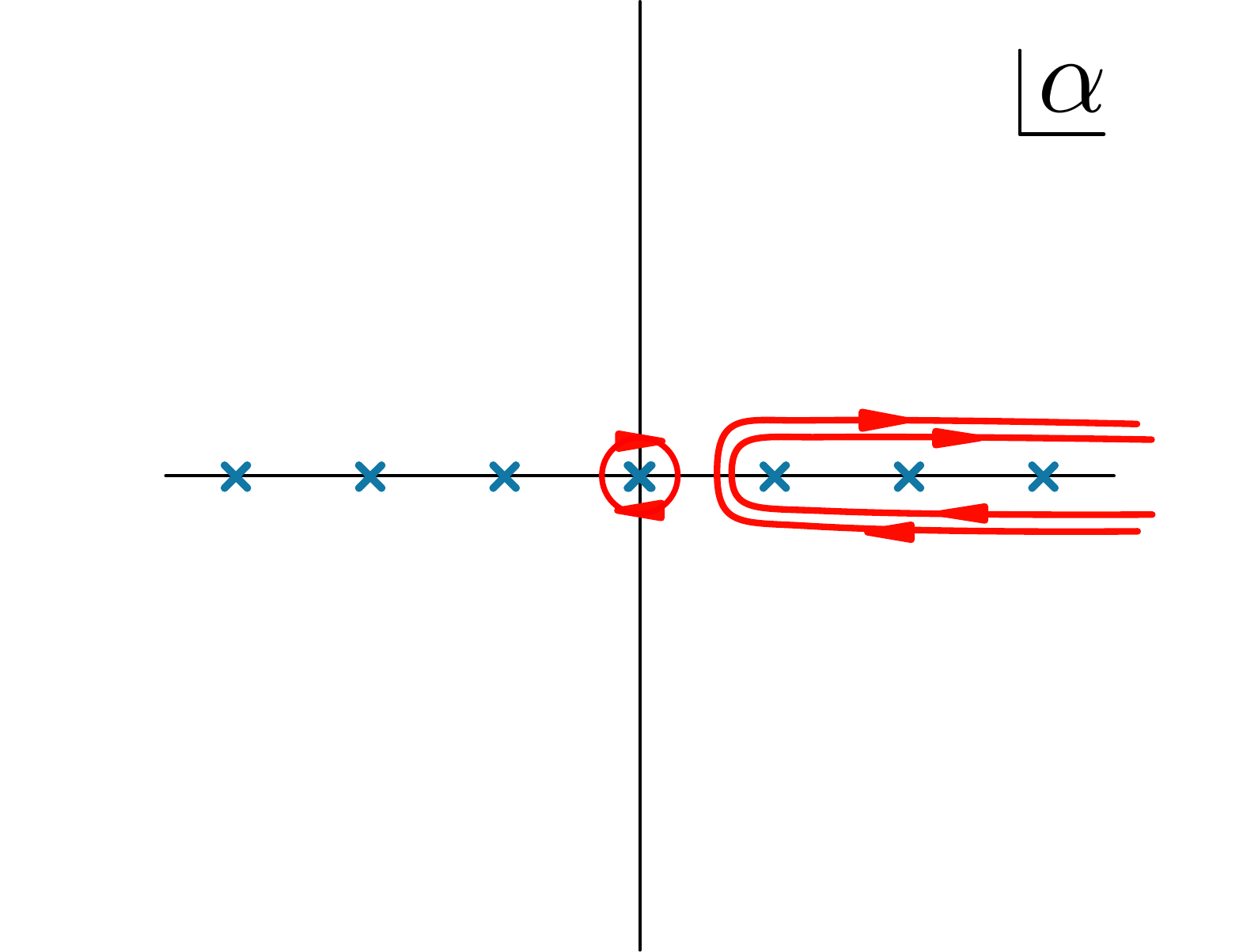}
\caption{\label{fig:alphacont}{{\bf Left:} The integration contour, $\mc C=\mc C_+\cup\mc C_-$, defining $\mathbb W_j$ depicted in red.  Poles in the integrand are depicted as blue ``$\times$"'s. {\bf Middle and right:} Possible deformations for evaluating $\mathbb W_j$.}}
\end{figure}

Let us now state the correspondence: consider fixed connections $A_L$ and $A_R$ which define a non-degenerate metric geometry which is topologically $S^3$.\footnote{That is, in the metric language, still quantized about the $S^3$ saddle.  In the gauge-theory language, not disconnected by a large gauge transformation from the background connections $a_{L/R}$.}  Then the one-loop determinant of a massive scalar field minimally coupled to that background is given by
\beq\label{eq:Zscalar1}
\log Z_{\rm scalar}[A_L,A_R]=\frac{1}{4}\mathbb W_{j}~,
\eeq
where
\beq\label{eq:spooldef}
\mathbb W_j:=i\int_{\mc C}\frac{\dd\alpha}{\alpha}\frac{\cos\alpha/2}{\sin\alpha/2}\mTr_{R_j}\left(\mc P\exp\frac{\alpha}{2\pi}\oint A_L\right)\mTr_{R_j}\left(\mc P\exp\frac{-\alpha}{2\pi}\oint A_R\right)~,
\eeq
is an object we deem the {\it Wilson spool.}  The $\alpha$ contour of integration is $\mc C=\mc C_+\cup\mc C_-$ with $\mc C_-$ and $\mc C_+$ running upwards along the imaginary $\alpha$ axis to the left and to the right, respectively, of the poles at $\alpha=0$, as depicted in Fig.\,\ref{fig:alphacont}. As we will show in Sec.\,\ref{sec:constructionspool}, the integration over $\alpha$ is implementing the sum over worldlines wrapping the compact space. 
The representations appearing in $\mathbb W_j$ are precisely the non-standard $\mf{su}(2)$ representations discussed in Sec.\,\ref{sect:dSrepintro} and  correspond to the mass of a minimally-coupled scalar field via
\beq\label{eq:jtomass}
j=-\frac12\left(1+\sqrt{1-\mathsf{m}^2\ell^2}\right)~.
\eeq

The appeal of \eqref{eq:Zscalar1}-\eqref{eq:spooldef} is that all of its components involve quantities that are precisely defined in Chern-Simons theory. This will allow us to take a step further: we will be able to evaluate
\be
\begin{aligned}\label{eq:logZexpval1}
\gev{\log Z_{\rm scalar}[S^3]}&=\frac{1}{4}\gev{\mathbb W_j}\\
&=\frac{i}{4}e^{ir_LS_{\rm CS}[a_L]+ir_RS_{\rm CS}[a_R]}\int \dd\sigma_L\dd\sigma_R\,e^{i\frac{\pi}{2}r_L\sigma_L^2+i\frac{\pi}{2}r_R\sigma_R^2}\sin^2(\pi\sigma_L)\sin^2(\pi\sigma_R) \\
&\quad ~\times\int_{\mc C}\frac{\dd\alpha}{\alpha}\frac{\cos\alpha/2}{\sin\alpha/2}\chi_{j}\left(\frac{\alpha}{2\pi}(\sigma_L+\ms h_L)\right)\chi_{j}\left(-\frac{\alpha}{2\pi}(\sigma_R+\ms h_R)\right)~,
\end{aligned}
\ee
 which is the gravitational path integral \eqref{eq:Zgravasint} with the Wilson spool inserted. Here $\chi_j(z)$ are the characters of the non-standard representations---either \eqref{char1} or \eqref{eq:Zukchars}---and $\ms h_{L/R}$ are the holonomies of the classical connections $a_{L/R}$:
\beq
\ms h_L=1~,\qquad\ms h_R=-1~.
\eeq
 This is an object that we can systematically compute to any order in $G_N$. It is a prescription for coupling massive fields to dynamical gravity using the Chern-Simons formulation of the dS$_3$ gravity. 

In the following subsections we will scrutinise and derive our proposal by tackling different fronts. We will start in Sec.\,\ref{sec:testofspool} by testing some of the elementary properties of \eqref{eq:spooldef}: we will discuss gauge invariance, and verify that it correctly reproduces the one-loop determinant of a massive scalar field on $S^3$ in the $G_N\rightarrow 0$ limit. In Sec.\,\ref{sec:constructionspool} we will give a first-principles derivation of \eqref{eq:spooldef} and further illustrate that our proposal is not fine-tuned to choices specific to $S^3$, and hence, it is an effective mechanism to couple matter to dynamical gravity in Chern-Simons theory.  Finally, in Sec.\,\ref{sec:quantumspool} we will quantify $\langle \mathbb W_j \rangle_{\rm grav}$, and in particular, its quantum gravity corrections through an expansion in powers of $G_N$.

\subsubsection{Testing the proposal: \texorpdfstring{$\mathbb W_j$}{Wj} on a fixed background.} \label{sec:testofspool}

\paragraph{Gauge Invariance.} Let us briefly remark on the gauge-invariance of \eqref{eq:spooldef}.  A cantankerous reader may object that large gauge transformations require precise quantization of coefficients appearing in Wilson loop operators. However given that we are discussing perturbative quantization about a particular gravitational saddle, we will only require the invariance of \eqref{eq:spooldef} under small gauge transformations.  Indeed, a generic large gauge transformation will shift the saddle, changing the value of either $I_{\rm EH}$ or $I_{\rm GCS}$; gravitationally this is tied to the fact that only gauge transformations connected to the identity are equivalent to diffeomorphisms. Under small gauge transformations $(g_L,g_R)$, (with $g_{L/R}$ single-valued), the traces in $\mathbb W_j$ transform by conjugation, i.e.,
\be
\begin{aligned}
\mTr_{R_j}\left(\mc Pe^{\frac{\alpha}{2\pi}\oint A_L}\right)&~ \rightarrow ~\mTr_{R_j}\left(g_L^{\frac{\alpha}{2\pi}}\circ\mc Pe^{\frac{\alpha}{2\pi}\oint A_L}\circ g^{-\frac{\alpha}{2\pi}}_L\right)~,\\
\mTr_{R_j}\left(\mc Pe^{-\frac{\alpha}{2\pi}\oint A_R}\right)& ~\rightarrow ~\mTr_{R_j}\left(g_R^{-\frac{\alpha}{2\pi}}\circ\mc Pe^{-\frac{\alpha}{2\pi}\oint A_R}\circ g_R^{\frac{\alpha}{2\pi}}\right)~.
\end{aligned}
\ee
These traces are invariant because $g_{L/R}$ are connected to the identity: we can define their exponent, at least formally, as
\beq\label{eq:grouppowerdef}
g_L=\exp\left(i\xi^aL_a\right)\qquad\Rightarrow\qquad g_L^{\frac{\alpha}{2\pi}}:=\exp\left(i\frac{\alpha}{2\pi}\xi^aL_a\right)
\eeq
and similarly for $g_R$.  This definition acts naturally on the $\mf{su}(2)_{L/R}$ representations appearing in the trace. Therefore the Wilson spool is invariant, $\mathbb W_j\rightarrow\mathbb W_j$, under these gauge transformations.

\paragraph{One-loop determinant on $S^3$.} Before diving into the more rigorous derivation  of \eqref{eq:Zscalar1} in the next subsection,  it is instructive to evaluate \eqref{eq:spooldef} and compare it to a one-loop determinant of a massive scalar field on $S^3$. That is, we want to test that
\begin{align}\label{eq:logZ1Lclassical1}
\lim_{G_N\rightarrow0}\frac{\gev{\log Z_{\rm scalar}[S^3]}}{\mc Z_{\rm grav}[S^3]} =\log Z_{\rm scalar}[S^3] = \log \det(-\nabla^2_{S^3}+\mathsf{m}^2\ell^2)^{-1/2}~, 
\end{align}
evaluated using a round $S^3$ metric is actually \eqref{eq:spooldef}, when the connections are given by the classical configuration \eqref{eq:bgconns}-\eqref{eq:holalar}. 

From \eqref{eq:Zscalar1} and \eqref{eq:spooldef} we have\footnote{As reflected by the first equality of \eqref{eq:logZ1Lclassical1}, we can also start from \eqref{eq:logZexpval1} and take the limit $r_L\rightarrow is\rightarrow i\infty$ and $r_R\rightarrow -is\rightarrow-i\infty$. In this limit, the Gaussian integrals localize to $\sigma_{L/R}\sim 0$, and the result is again \eqref{eq:logZ1Lclassical}.}
\begin{align}\label{eq:logZ1Lclassical}
\log Z_{\rm scalar}[S^3]=&\frac{i}{4}\int_{\mc C}\frac{\dd\alpha}{\alpha}\frac{\cos\alpha/2}{\sin\alpha/2}\,\chi_j\left(\frac{\alpha}{2\pi}\ms h_L\right)\chi_j\left(-\frac{\alpha}{2\pi}\ms h_R\right)\nonumber\\
=&-\frac{i}{16}\int_{\mc C}\frac{\dd\alpha}{\alpha}\frac{\cos\alpha/2}{\sin^3\alpha/2}e^{i(2j+1)\alpha}~,
\end{align}
where we used \eqref{eq:holalar} and the explicit expression of the character in \eqref{char1} or \eqref{eq:Zukchars}. 
We can now pull $\mc C_\pm$ towards the real line---depicted as the middle cartoon of Fig.\,\ref{fig:alphacont}---to pick up the third-order poles at $\alpha=\pm 2\pi n$, respectively.  Keeping track of orientations, we find
\be
\begin{aligned}\label{eq:logz123}
\log Z_{\rm scalar}[S^3]&=\sum_{\pm}\mp\frac{\pi}{2}\sum_{n=1}^\infty\frac{d^2}{d\alpha^2}\left.\left(\frac{e^{i(2j+1)\alpha}}{\alpha}\right)\right|_{\alpha=\pm2\pi n}\\
&=\sum_{\pm}\left(-\frac{1}{8\pi^2}\Li3\left(e^{\mp2\pi \mu}\right)\mp\frac{\mu}{4\pi}\Li2\left(e^{\mp2\pi\mu}\right)-\frac{\mu^2}{4}\Li1\left(e^{\mp2\pi\mu}\right)\right)~.
\end{aligned}
\ee
In the second line we used the information for the principal-type representations, where $j=-\frac{1}{2}(1-i\mu)$ and  $\mu^2=\mathsf{m}^2\ell^2-1$; a similar result holds for  the complementary-type representation, where  $j=-\frac{1}{2}(1+\nu)$. We also used the definition of the polylogarithm functions,
\beq
\Li q(x)=\sum_{n=1}^\infty\frac{x^n}{n^q}~.
\eeq

We have written \eqref{eq:logz123} in an inherently symmetric form, however one can also choose to wrap $\mc C_-$ to the right, as in the rightmost cartoon of Fig.~\ref{fig:alphacont}, doubling the contribution of poles on the positive real line, as well as picking up the quadrupole pole at $\alpha=0$, or alternatively using a conspiracy of polylog identities.\footnote{Namely,
\be
\begin{aligned}
\Li1(z)=&\Li1(z^{-1})-\log(z)-i\pi~,\\
\Li2(z)=&-\Li2(z^{-1})-\frac12\log(z)^2-i\pi\log(z)+\frac{\pi^2}{3}~,\\
\Li3(z)=&\Li3(z^{-1})-\frac16\log(z)^3-\frac{i\pi}{2}\log(z)^2+\frac{\pi^2}{3}\log(z)~.
\end{aligned}\ee} This gives
\beq\label{eq:finallogZ}
\log Z_{\rm scalar}[S^3]=\frac{\pi\mu^3}{6}-\frac{1}{4\pi^2}\Li3\left(e^{-2\pi\mu}\right)-\frac{\mu}{2\pi}\Li2\left(e^{-2\pi\mu}\right)-\frac{\mu^2}{2}\Li1\left(e^{-2\pi\mu}\right).
\eeq
 This  makes the comparison to the three-sphere one-loop determinant \eqref{eq:logZfinal}, evaluated independently via heat kernel, very clear. In this regard, a few comments are in order:
 \begin{enumerate}[leftmargin=0.4cm]
 \item Because the above polylogarithms involve sums over $e^{-n2\pi\mu}$, their appearance is the first sign that \eqref{eq:spooldef} packages contributions of Wilson loops wound many times around the $S^3$.  We will comment further on this interpretation below.
     \item In the above computation there is no need to minimally subtract any divergences. As we will see in Sec.\,\ref{sec:constructionspool} there is an $i\epsilon$ prescription in our construction, and it turns out that it is enough to render the entire computation finite.
\item In comparison to the heat kernel method \eqref{eq:logZfinal}, or the methods used in \cite{Anninos:2020hfj,Law:2022zdq}, the finite terms in $\log Z_{\rm scalar}[S^3]$ there agree perfectly with \eqref{eq:finallogZ}. 
\item It is worth comparing in more detail with \cite{Anninos:2020hfj}. In particular we note that the measure and characters in the first line of \eqref{eq:logZ1Lclassical} agree,\footnote{In App.~\ref{app:conventions} we show that certain combinations of our non-standard characters match Harish-Chandra characters of $SO(1,3)$.} albeit the contour prescription is different. One advantage of the procedure here is that \eqref{eq:spooldef} allows us to track the gravitational dependence: we can integrate out the matter field while keeping gravity off-shell.  
 \end{enumerate}
The analysis done here provides a non-trivial test that our proposal is capturing the desired observable. However, one might still be skeptical regarding its physical interpretation (and hence utility in other contexts). We will address this in the next portion by providing a more rigorous derivation of  \eqref{eq:Zscalar1}-\eqref{eq:spooldef}.

\subsubsection{Construction of \texorpdfstring{$\mathbb W_j$}{Wj}}\label{sec:constructionspool}

Let us now construct \eqref{eq:spooldef} under the following guiding principles: we want to find an operator that (i) utilizes the $\mf{su}(2)_L\oplus\mf{su}(2)_R$ structure naturally, (ii) is gauge-invariant, 
(iii) is evaluable inside the gravitational path-integral, and (iv) has a recognizable physical meaning in the classical, $s\rightarrow\infty$, limit.  Focusing on this last point, let us take inspiration\footnote{We also take notable inspiration from the recent application of the DHS formula in \cite{Law:2022zdq}.} from a procedure pioneered by Denef, Hartnoll, and Sachdev (DHS) \cite{Denef:2009kn} for constructing one-loop determinants from quasi-normal mode spectra and generalize this to an expression satisfying the first three points.

The salient point of \cite{Denef:2009kn} is to regard
\beq\label{eq:Zasdet}
Z_{\rm scalar}=\det(-\nabla^2+\mathsf{m}^2\ell^2)^{-1/2}~
\eeq
as a meromorphic function of $\Delta=1+\sqrt{1-\mathsf{m}^2\ell^2}$, where $\mathsf{m}$ is the mass of the scalar field and try to identify its zeroes and poles as $\mathsf{m}^2$ is varied: if these can be identified, then up to an overall analytic function (fixed by asymptotics), $Z_{\rm scalar}$ must be equal to a rational product with the same zeroes and poles.  For a scalar field $Z_{\rm scalar}$ has poles at solutions to $-\nabla^2+\mathsf{m}^2\ell^2=0$. For either a Euclidean black hole or Euclidean de Sitter background, these poles lie precisely on quasi-normal mode frequencies.  An additional constraint comes from the fact that determinants are defined with respect to boundary conditions and demanding regularity of field configurations appearing in \eqref{eq:Zasdet} at the horizon; in Euclidean signature this implies each mode must also be a Matsubara frequency with appropriate horizon temperature, $T_H$.  As a result, $Z_{\rm scalar}$ takes the following form as a function of $\Delta$
\beq\label{eq:originalDHS}
Z_{\rm scalar}=e^{P(z_\ast,\bar z_\ast)}\prod_{z_\ast,\bar z_\ast}\prod_{n\in\mathbb Z}\left(|n|+i\frac{z_\ast(\Delta)}{2\pi T_H}\right)^{-d_{z_\ast}/4}\left(|n|-i\frac{\bar z_\ast(\Delta)}{2\pi T_H}\right)^{-d_{\bar z_\ast}/4}~,
\eeq
where $z_\ast(\Delta)$ and $\bar z_\ast(\Delta)$ are quasi-normal and anti-quasi-normal modes, implicitly dependent on the mass through $\Delta$.  We have allowed for possible degeneracies, $d_{z_\ast}$ and $d_{\bar z_\ast}$, in the spectrum.  Lastly, $e^{P(z_\ast,\bar z_\ast)}$ is an analytic function which can be determined by matching $Z_{\rm scalar}$ to asymptotics in $\mathsf{m}^2$ (e.g by taking $\mathsf{m}^2\rightarrow\infty$ and matching to heat kernel coefficients).

Let us now extend this logic in a manner that makes the action of $\mf{su}(2)_L\oplus\mf{su}(2)_R$ manifest.  We will do this first for the $S^3$ Laplacian and then generalize to curved backgrounds from there.  Two important data points towards this are, firstly, that for the $S^3$ background the $\mf{su}(2)$'s can be represented on scalar functions via left and right acting (or right- and left-invariant, respectively) vector fields, $\{\zeta_a\}_{a=1,2,3}$ and $\{\bar\zeta_a\}_{a=1,2,3}$ such that
\beq
\nabla^2_{S^3}=-2\zeta_a\zeta_a-2\bar\zeta_a\bar\zeta_a=-2c_2^L-2c_2^R~,
\eeq
where $c_2^{L/R}$ is the quadratic Casimir of $\mf{su}(2)_{L/R}$.  The explicit forms of these vector fields can be found in App.\,\ref{app:casimir}.  The second data point is that, as shown in \cite{Castro:2020smu}, scalar fields lie in non-standard representations whose weight-spaces line up with the spectrum of quasi-normal modes on dS$_3$.  This suggests the $\mf{su}(2)_L\oplus\mf{su}(2)_R$ manifest object
\beq\label{eq:Z1Lcasimir}
Z_{\rm scalar}[S^3]=\text{det}\left(2c_2^L+2c_2^R+\mathsf{m}^2\ell^2\right)^{-1/2}~,
\eeq
where we expect the weights of non-standard representations to appear in a similar manner as the quasi-normal modes in the original DHS construction.  
We can now ask, \`a la DHS, treating $Z_{\rm scalar}$ as a meromorphic function of $\Delta$, what function matches the poles of \eqref{eq:Z1Lcasimir}?  It is clear that $Z_{\rm scalar}$ will have a pole for each state of a $\mf{su}(2)_L\oplus\mf{su}(2)_R$ representation, $R_j\otimes R_j$,\footnote{As mentioned in Sec.\,\ref{sec:nonstandardreps}, we must take $j_L=j_R=j$ for a scalar determinant since $j_L-j_R$ is related to the $SO(1,3)$ spin.} satisfying
\beq
j(j+1)=-\frac{\mathsf{m}^2 \ell^2}{4}~.
\eeq
These are precisely the non-standard representations with $j=-\frac{1}{2}(1+\nu)$ for $\mathsf{m}^2\ell^2<1$ and $j=-\frac{1}{2}(1-i\mu)$ for $\mathsf{m}^2\ell^2>1$.  A particular weight, $(\lambda_L,\lambda_R)$ of $R_j\otimes R_j$ can contribute to $Z_{\rm scalar}$ if it leads to a field configuration that is regular at the horizon.  Consider then the parallel transport of a field living in $R_j\otimes R_j$ around a closed curve $\gamma$ enclosing the horizon, $\rho=\frac\pi 2$:
\beq
\Phi_f=R_j\left(\mc P\exp\left(\oint_\gamma a_L\right)\right)\Phi_i\,R_j\left(\mc P\exp\left(-\oint_\gamma a_R\right)\right)~,
\eeq
where the minus sign in front of $a_R$ arises because it is the connection for the right-acting $SU(2)$ which involves inverse group elements.  As mentioned multiple times above, $a_L$ and $a_R$ will possess holonomies ${\ms h}_L$ and ${\ms h}_R$, respectively, around $\gamma$: that is there exists some single-valued group elements $u_{L/R}$ such that
\beq
\mc P\exp\left(\oint_\gamma a_{L/R}\right)=u_{L/R}^{-1}e^{i2\pi{\ms h}_{L/R}L_3}u_{L/R}~.
\eeq
Single-valued-ness of $\Phi$ restricted to the $(\lambda_L,\lambda_R)$ weight sector implies that this weight can contribute a pole to $Z_{\rm scalar}$ if
\beq
\lambda_L{\ms h}_L-\lambda_R{\ms h}_R\in\mathbb Z~.
\eeq
The logic following DHS leads us the following formula for $Z_{\rm scalar}$\footnote{{\bf Erratum:} The expression in \eqref{eq:Zprods} is incorrect due to an extra condition on global regularity of solutions contributing poles to $Z_{\rm scalar}$. A more proper treatment of the scalar one-loop determinant can be found in \cite{Bourne:2024ded}. However the procedures following \eqref{eq:Zprods} apply to the result there and lead to the same Wilson spool, \eqref{eq:S3spool}. The broad conclusions of this section and the rest of the paper remain unchanged.}
\be
\begin{aligned}\label{eq:Zprods}
Z_{\rm scalar}[S^3]=&\prod_{(\lambda_L,\lambda_R)\in R_j\otimes R_j}\prod_{n\in\mathbb Z}\left(|n|-\lambda_L{\ms h}_L+\lambda_R{\ms h}_R\right)^{-1/4}\left(|n|+\lambda_L{\ms h}_L-\lambda_R{\ms h}_R\right)^{-1/4}\\
&\prod_{(\bar\lambda_L,\bar\lambda_R)\in R_{\bar j}\otimes R_{\bar j}}\prod_{n\in\mathbb Z}\left(|n|-\bar\lambda_L{\ms h}_L+\bar\lambda_R{\ms h}_R\right)^{-1/4}\left(|n|+\bar\lambda_L{\ms h}_L-\bar\lambda_R{\ms h}_R\right)^{-1/4}~,
\end{aligned}
\ee
where the second line occurs from noting that since $Z_{\rm scalar}$ is a function only of the Casimir it will receive poles\footnote{Crossterms such as $R_j\otimes R_{\bar j}$ also have the same Casimir however they cannot contribute because the resulting field will have non-trivial spin.} from weights appearing in not only in $R_j\otimes R_j$ but also the ``shadow representations," $R_{\bar j}\otimes R_{\bar j}$ with $\bar j=-1-j$.  This is all up to a potential holomorphic function $e^{P}$ which is trivial for $S^3$ \cite{Denef:2009kn}.

Let us now take the $\log$ of \eqref{eq:Zprods}.  We will write this log in terms of a Schwinger parameter, $\alpha$, as
\beq
\log M=-\int_{\times}^\infty\frac{\dd\alpha}{\alpha}e^{-\alpha M}~,
\eeq
where $\int_\times^\infty$ indicates that we are regulating the UV divergence at $\alpha\sim 0$.  We will discuss this regulator very explicitly in a moment.  Noting that 
\be
\sum_{n\in\mathbb Z}e^{-|n|\alpha}=\frac{\cosh\alpha/2}{\sinh\alpha/2} ~,
\ee
we can perform the integer sum in \eqref{eq:Zprods}; at this step one can infer that  $Z_{\rm scalar}[S^3]$ contains winding (looped) contributions. Finally, the sum over weights can be organized into representation traces, i.e.,
\beq
\sum_{(\lambda_L,\lambda_R)\in R_j\otimes R_j}e^{\alpha\lambda_L{\ms h}_L-\alpha\lambda_R{\ms h}_R}=\mTr_{R_j}\left(e^{\alpha{\ms h}_LL_3}\right)\mTr_{R_j}\left(e^{-\alpha{\ms h}_RL_3}\right)~,
\eeq
which we happily recognize as a Wilson-loop along $a_{L/R}$. Gathering these facts, the log of \eqref{eq:Zprods} becomes
\be
\begin{aligned}\label{eq:spoolhalfway}
\log Z_{\rm scalar}[S^3]=&\frac{1}{4}\int_\times^\infty\frac{\dd\alpha}{\alpha}\frac{\cosh\alpha/2}{\sinh\alpha/2}\Big\{\mTr_{R_j}\left(\mc P e^{-i\frac{\alpha}{2\pi}\oint_\gamma a_L}\right)\mTr_{R_j}\left(\mc P e^{i\frac{\alpha}{2\pi}\oint_\gamma a_R}\right)\\
&\qquad\qquad\qquad\qquad\qquad+\mTr_{R_j}\left(\mc Pe^{i\frac{\alpha}{2\pi}\oint_\gamma a_L}\right)\mTr_{R_j}\left(\mc P e^{-i\frac{\alpha}{2\pi}\oint_\gamma a_R}\right)\\
&\qquad\qquad\qquad\qquad\qquad\qquad\qquad+j\rightarrow \bar j\Big\}~.
\end{aligned}
\ee
Let us now discuss the regulator.  Noting that the first two lines of \eqref{eq:spoolhalfway} can be combined into a single integral over $\int_{\times}^\infty+\int_{-\infty}^{-\times}$, our proposal is to displace the $\alpha$ contour away from the poles at $\alpha=0$ symmetrically via an $i\epsilon$ prescription:
\beq\label{eq:spoolregulator}
\left(\int_{-\infty}^{-\times}+\int_{\times}^{\infty}\right)\frac{\dd\alpha}{\alpha}\,f(\alpha):=\frac{1}{2}\lim_{\epsilon\rightarrow0}\sum_{\pm}\int_{-\infty}^{\infty}\frac{\dd\alpha}{\alpha\pm i\epsilon}f(\alpha\pm i\epsilon) ~.
\eeq
If the integrand, $f$, has no poles at $\alpha=0$ then the Sokhotski-Plemelj formula relates this to the principal part, $\text P\int\frac{d\alpha}{\alpha}f(\alpha):=\lim_{\epsilon\rightarrow0}\left(\int_{-\infty}^{-\epsilon}+\int_{\epsilon}^{\infty}\right)\frac{d\alpha}{\alpha}f(\alpha)$, however for integrands with additional poles, such as in \eqref{eq:spoolhalfway}, these objects differ. As we saw in Sec.\,\ref{sec:testofspool}, our prescription \eqref{eq:spoolregulator} not only gives the correct regulated scalar one-loop determinant in the $G_N\rightarrow0$ limit, but is {\it entirely finite} inside the quantum gravity path-integral.  We take this moment to emphasize that once we make this choice of regulator there are no further ambiguities in any finite $G_N$ calculation.

Lastly as a matter of cosmetic convenience, we redefine\footnote{This is an integration variable redefinition, {\it not} an integration contour deformation.} $\alpha\rightarrow-i\alpha$ to arrive at our final result:
\beq\label{eq:S3spool}
\log Z_{\rm scalar}[S^3]=\frac{i}{4}\int_{\mc C}\frac{\dd\alpha}{\alpha}\frac{\cos\alpha/2}{\sin\alpha/2}\mTr_{R_j}\left(\mc Pe^{\frac{\alpha}{2\pi}\oint a_L}\right)\mTr_{R_j}\left(\mc Pe^{-\frac{\alpha}{2\pi}\oint a_R}\right)~,
\eeq
where again $\mc C=\mc C_-\cup\mc C_+$ is the union of two contours running up the imaginary axis just to the left and right of zero.  Another cosmetic benefit: the two contours also nicely package in the contribution of the shadow representations.

Given that our expression for $\log Z_{\rm scalar}$ on $S^3$ is nicely expressed as gauge-invariant, Wilson loop observables of the background connections, we naturally extend this definition to non-trivial connections, $A_L$ and $A_R$.  Namely, we claim that the one-loop determinant of a scalar field minimally coupled to a background geometry determined by $A_L$ and $A_R$ is given by
\beq\label{eq:logZALAR}
\log Z_{\rm scalar}[A_L,A_R]=\frac{i}{4}\int_{\mc C}\frac{\dd\alpha}{\alpha}\frac{\cos\alpha/2}{\sin\alpha/2}\mTr_{R_j}\left(\mc Pe^{\frac{\alpha}{2\pi}\oint A_L}\right)\mTr_{R_j}\left(\mc Pe^{-\frac{\alpha}{2\pi}\oint A_R}\right)~,
\eeq
which concludes our construction of the Wilson spool.  
One might object that since the construction of \eqref{eq:S3spool} as Wilson loops relied on the $\mf{su}(2)_L\oplus\mf{su}(2)_R$ isometry of $S^3$, there is no reason to expect that we can extend to a  Laplacian on generic curved background (which might have no isometries) as presented in \eqref{eq:logZALAR}.  There is no sleight of hand here, however. In App.\,\ref{app:casimir} we show that by modelling a three-manifold with ``locally tangent three-spheres" we can express its Laplacian as the {\it curved Casimir} of local $\mf{su}(2)_L\oplus\mf{su}(2)_R$ action on its tangent spaces.

\subsubsection{Quantum gravity corrections to \texorpdfstring{$\langle\mathbb W_j\rangle$}{Wj}}\label{sec:quantumspool}

In this last portion we will focus on the quantum version of the Wilson spool, given by \eqref{eq:logZexpval1}, and report on the quantum effects of our proposal. In particular, we will report on the predictions we can make regarding mass renormalization. 

We have 
\be
\begin{aligned}\label{eq:logZexpval2}
\gev{\log Z_{\rm scalar}[S^3]}
&=\frac{i}{4}e^{ir_LS_{\rm CS}[a_L]+ir_RS_{\rm CS}[a_R]}\int \dd\sigma_L\dd\sigma_R\,e^{i\frac{\pi}{2}r_L\sigma_L^2+i\frac{\pi}{2}r_R\sigma_R^2}\sin^2(\pi\sigma_L)\sin^2(\pi\sigma_R) \\
&\quad ~\times\int_{\mc C}\frac{\dd\alpha}{\alpha}\frac{\cos\alpha/2}{\sin\alpha/2}\chi_{j}\left(\frac{\alpha}{2\pi}(\sigma_L+\ms h_L)\right)\chi_{j}\left(-\frac{\alpha}{2\pi}(\sigma_R+\ms h_R)\right)~.
\end{aligned}
\ee
Recall that the gravitational couplings are related to $r_{L/R}$ via 
\beq\label{eq:rLRdef1}
r_L=\hat\delta+is~,\qquad r_R=\hat\delta-is~,\qquad s=\frac{\ell}{4G_N}~.
\eeq
Evaluating \eqref{eq:logZexpval2} exactly as a function of $r_{L/R}$ is cumbersome, hence we will proceed perturbatively in $G_N$, with $\hat\delta$ fixed.  This expansion is systematic and the procedure will include the following steps:
\begin{enumerate}[leftmargin=0.5cm]
\item Beginning with the integral form of \eqref{eq:logZexpval2}, we normalize the Gaussian integrals by $\sigma_{L/R}\rightarrow r_{L/R}^{-1/2}\sigma_{L/R}$ such that the integrand now admits a perturbative expansion in $1/s$ which is ultimately a Taylor expansion in $\sigma_{L/R}$.
\item At any order in $1/s$ perturbation theory, the Gaussian $\sigma_{L/R}$ integrals can be performed.
\item Afterwards the $\alpha$ contour, $\mc C$, can be pulled towards the real line to pick up the poles at $2\pi\mathbb Z_{\neq 0}$.
\item The sum over the poles can then be performed to yield potential polylogarithms.
\end{enumerate}
To implement these steps, we will organize corrections to the scalar one-loop determinant on $S^3$ as a $G_N\ell^{-1}$ expansion:
\beq\label{eq:qGcorrectedlogZ1Lgeneral}
\frac{\gev{\log Z_{\rm scalar}[S^3]}}{\mc Z_{\rm grav}[S^3]}=\log Z_{\rm scalar}[S^3]+\sum_{m=1}^\infty \left(\frac{G_N}{\ell}\right)^{2m}\,(\log Z)_{(2m)}~.
\eeq
The term $(\log Z)_{(2m)}$ encodes the contributions from expanding the numerator and denominator of \eqref{eq:qGcorrectedlogZ1Lgeneral} separately.  Note that only even powers of  $G_N\ell^{-1}$ enter here; this will be more manifest below.  The denominator is easily expanded from \eqref{eq:Zgravfull} and it is given in \eqref{eq:logZgrav-exp-delta}. The expectation value of the spool can be organized as
\beq\label{eq:zsm123}
\gev{\log Z_{\rm scalar}[S^3]}=e^{\frac{\pi\ell}{2G_N}}\int \dd\sigma_L\dd\sigma_Re^{i\frac\pi2(\sigma_L^2+\sigma_R^2)}\int_{\mc C}\frac{\dd\alpha}{\alpha}\frac{\cos\alpha/2}{\sin\alpha/2}\mc I_j[\sigma_L,\sigma_R;\alpha]~,
\eeq
with
\be
\begin{aligned}\label{eq:zsm321}
\mc I_j[\sigma_L,\sigma_R;\alpha]=&\frac{i}{4|\hat\delta+is|}\sin^2\left(\frac{\pi\sigma_L}{\sqrt{\hat\delta+is}}\right)\sin^2\left(\frac{\pi\sigma_R}{\sqrt{\hat\delta-is}}\right)\\
&\qquad\times\chi_j\left(\frac{\alpha}{2\pi}\left(1+\frac{\sigma_L}{\sqrt{\hat\delta+is}}\right)\right)\chi_j\left(\frac{\alpha}{2\pi}\left(1-\frac{\sigma_R}{\sqrt{\hat\delta-is}}\right)\right)~.
\end{aligned}
\ee
 From here it is clear that a small $G_N$ (or large $s$) expansion can be organized as a polynomial expansion of $\mc I_j$ in $\sigma_L$ and $\sigma_R$ which we cast as
\beq
\mc I_j[\sigma_L,\sigma_R;\alpha]=64\pi^4 \left(\frac{G_N}{\ell}\right)^3e^{\frac{\pi\ell}{2G_N}} \,\sigma_L^2\sigma_R^2\, \left(\mc I^{(0,0)}[\alpha]+\sum_{m,n=1}^\infty\mc I^{(m,n)}_j[\alpha]\sigma_L^{2m}\sigma_R^{2n}\right)~.
\eeq
The prefactors are the leading contribution $\mc Z_{\rm grav}[S^3]$, which we normalize by \eqref{eq:qGcorrectedlogZ1Lgeneral}. The first term, $\mc I^{(0,0)}[\alpha]=\frac i4\chi_j\left(\frac{\alpha}{2\pi}\right)^2$, is what we previously evaluated in \eqref{eq:logZ1Lclassical}, giving $\log Z_{\rm scalar}$ on the classical $S^3$.  Each subleading term can be evaluated as moments of the Gaussian $\sigma_{L/R}$ integrals (only even moments can contribute to this expansion).  After doing these integrals we can evaluate the remaining $\alpha$ integration, that is
\beq
\int_{\mc C}\frac{\dd\alpha}{\alpha}\frac{\cos\alpha/2}{\sin\alpha/2}\mc I_j^{(m,n)}[\alpha]~,
\eeq
in the following way: given the form of the characters \eqref{eq:Zukchars} (\eqref{eq:compchar}) appearing in $\mc I_j$, it is easy to see that $\mc I_j^{(m,n)}$ will continue to have poles at $\alpha=\pm 2\pi n$.  We are then free to deform $\mc C$ to wrap these poles along the real axis to pick up their residues, just as we did in evaluating $\mc I^{(0,0)}[\alpha]$.  The result is the sum over $n\in\mathbb Z_{\neq0}$ of the residues of $\mc I_j^{(m,n)}[\alpha]$.  Again, given the form of the characters, \eqref{eq:Zukchars}, the summands inevitably involve Laurent polynomials of $n$ weighted by $e^{i2\pi(2j+1)n}$ and so the end results can be expressed in the form of polylogarithms of various orders.  This entire procedure can be implemented easily on a computer algebra system.  

It is instructive to quantify this expansion at leading order. For concreteness we select the principal-type representations. The first non-trivial correction to \eqref{eq:qGcorrectedlogZ1Lgeneral} is 
\beq\label{eq:OG2correctionsgen}
\begin{aligned}
(\log Z)_{(2)}&=\sum_{\pm}\sum_{i=0}^{3}z^{(2)}_i[\pm\mu]\Li{-i}\left(e^{\mp2\pi\mu}\right)\\
&=z^{(2)}_0[\mu]+2\sum_{i=0}^{3}z^{(2)}_i[\mu]\Li{-i}\left(e^{-2\pi\mu}\right)~,
\end{aligned}
\eeq
with
\be
\begin{aligned}\label{eq:OG2correctionsspec}
z_0^{(2)}[\mu]=&-\left(\frac{8\pi}{3}-\frac4\pi-4i\hat\delta\right)\mu^3-\left(8\pi-\frac{72}{\pi}-12i\hat\delta\right)\mu~,\\
z_1^{(2)}[\mu]=&\left(\frac{8\pi^2}{3}+6-4i\pi\hat\delta\right)\mu^4+\left(8\pi^2-138-12i\pi\hat\delta\right)\mu^2-24~,\\
z_2^{(2)}[\mu]=&-\frac{12\pi}{5}\mu^5+52\pi\mu^3+48\pi\mu~,\\
z_3^{(2)}[\mu]=&-4\pi^2\mu^4-16\pi^2\mu^2~.
\end{aligned}
\ee
It is natural to identify these corrections with a renormalization of the mass of the scalar field\footnote{Because we are only considering the one-loop determinant a possible wavefunction renormalization decouples from this calculation.} (or more naturally, a renormalization of $\mu=\sqrt{\mathsf{m}^2\ell^2-1}$).  To see this let us denote the renormalized mass $\mu_R$ which is related to the bare mass, $\mu$, by
\beq
\mu_R=\mu+\frac{G_N^2}{\ell^{2}} \,\delta^{(2)}_\mu+\ldots~,
\eeq
such that, via direct calculation from \eqref{eq:logZ1Lclassical}, or \eqref{eq:finallogZ}, we obtain at $O(G_N^2)$,
\beq
\left.\log Z_{\rm scalar}[S^3]\right|_{\mu}=\left.\log Z_{\rm scalar}[S^3]\right|_{\mu_R}-\frac{\pi}{2}\frac{G_N^2}{\ell^{2}}\frac{\cosh\left(\pi\mu_R\right)}{\sinh\left(\pi\mu_R\right)}\mu_R^2\delta^{(2)}_\mu+\cdots~.
\eeq
We can then identify quantum gravity corrections, \eqref{eq:qGcorrectedlogZ1Lgeneral}, to this order as a renormalization of the mass
\beq\label{eq:mass-ren}
\delta^{(2)}_\mu=-\frac{2}{\pi\mu_R^2}\frac{\sinh\left(\pi\mu_R\right)}{\cosh\left(\pi\mu_R\right)}(\log Z)_{(2)}~.
\eeq
In the large mass limit, where $\mu^2_R\sim \mathsf{m}_R^2\ell^2\gg1$, we find
\be
\delta^{(2)}_\mu= \frac{48}{5}\mu_R^3\, e^{-2\pi \mu_R}+\cdots~.
\ee
Equations \eqref{eq:OG2correctionsgen} and \eqref{eq:OG2correctionsspec} give {\it predictive} statements about de Sitter quantum gravity, namely how it renormalizes the mass to $O(G_N^2)$.  We emphasize however that the spool allows one, in principle, to calculate this renormalization to all orders in a systematic and finite manner.

Finally, we remark on the peculiar imaginary contributions in  \eqref{eq:OG2correctionsspec}. They are due to the imaginary contribution of the gravitational Chern-Simons term, which is not surprising from Sec.\,\ref{sec:CS-grav-main}:   the action  \eqref{eq:GCS-action} is parity odd. This leads to an imaginary coupling in  Euclidean signature, and its effects were already present  in the loop corrections reported in \eqref{eq:logZgrav-exp-delta}. When coupling to matter, we again encounter imaginary contributions related to $\hat\delta$. We take this as an indication that it is reasonable to simply consider theories with $\hat\delta =0$.

\section{Discussion}\label{sect:disc}

In this paper we revisited the Chern-Simons formulation of three-dimensional de Sitter quantum gravity quantized about its $S^3$ saddle.  Our goal was to address how to couple matter to this theory and compute physically relevant quantities in the gravity path-integral.  Along the way we have established several new results.  

Firstly, we have retooled known exact methods for $SU(2)$ Chern-Simons, namely Abelianisation and $\mc N=2$ supersymmetric localization, to accommodate the complex levels and non-trivial background connections, two ingredients necessary to admit a saddle-point appropriate for a round $S^3$ metric.  These techniques have been applied previously in the literature \cite{Blau:1993tv, Blau:2006gh, Blau:2013oha,Kapustin:2009kz,Marino:2011nm}, however here we have verified their applicability and reproduced known results for the gravity path-integral on $S^3$.  

Secondly, in the interest of capturing single-particle spectra of dS$_3$, we defined an alternative notion of Hermiticity on $\mf{su}(2)$ that allowed us to construct representations that contain necessary features of single-particle representations of de Sitter, namely a continuous parameter to identify with a mass.  These representations were first considered in \cite{Castro:2020smu} for light fields ($\mathsf{m}^2\ell^2<1$) and we have extended the construction include heavy fields ($\mathsf{m}^2\ell^2>1$).  Utilizing these representations we then investigated their role in Wilson loop expectation values.  We verified the validity of the above exact techniques for evaluating these Wilson loops with non-standard representations.  

Finally, having assembled all of the above pieces, we defined a new object, the Wilson spool, which, at the classical ($G_N\to 0$) level reproduces the one-loop determinant of a massive scalar field.  We further showed that this object can be evaluated at any order in $G_N$ perturbation theory and argued that it gives intrinsically finite and predictive quantum gravity corrections to the scalar one-loop determinant.

There is much to this story that remains to be explored.  Let us highlight some of the more pressing open questions below.

\paragraph{Spinning Fields:} So far our discussion has entirely focused on massive scalars and their effective description in Chern-Simons gravity.  However de Sitter space also allows for massive spinning fields that we have so far overlooked.  In terms of the representation theory of $SO(1,3)$ these fields lie in the spinning principal series.  As a question of representation theory, it would be satisfying to enlarge our framework of non-standard $\mf{su}(2)$ representations from Section \ref{sec:nonstandardreps} to encapsulate the spinning principal series.  Doing so will likely require once again altering conditions on Hermiticity and in particular altering the form of the shadow map, $\mc S$, used to build the representations. Incorporating these representations into a Wilson spool that correctly reproduces their sphere partition function poses a further challenge. This is because the sphere partition function of a massive spinning field is more complicated than the simple representation theory of transverse-traceless fields suggests; in addition, one needs to carefully treat zero mode divergences in the path-integral.  This point was emphasized in \cite{Anninos:2020hfj} which also showed that the calculation could be nicely organized into ``bulk" and ``edge" contributions.  With regards to constructing the Wilson spool, it is clear that extra care must be taken in using the DHS formula in order to account for this new ``edge" contribution; we expect \cite{Grewal:2022hlo} to be a helpful guide.  It would be very interesting to understand exactly the role of this edge contribution in the Chern-Simons theory.

\paragraph{Additional topologies:} 
So far all of our calculations have been performed about the $S^3$ saddle of the gravitational path-integral. 
 While this may be seen as the leading saddle in the Euclidean path-integral, it was emphasized in~\cite{Castro:2011xb} that the gravitational path integral includes contributions from a series of additional saddles.  Some of these Euclidean saddles are Lens space geometries, $L(p,q)$; these are quotients of the three-sphere, i.e, $L(p,q) = S^3/\mathbb{Z}_p$. 
 
 At the perturbative level, it would be interesting to extend our Wilson spool construction to these additional $L(p,q)$ topologies. Verifying that $\left\langle\mathbb W_j[M]\right\rangle$, at $G_N\to 0$, reproduces the one-loop determinant of a massive scalar field on a three-manifold $M$ is a non-trivial test of the proposal.\footnote{At the level of the gravitational sector, the derivations in \cite{Castro:2011xb} show that exact Chern-Simons theory results match the graviton determinant at one-loop about each Lens space.} One should also be able to capture systematically the subleading corrections, and verify that they are finite and accessible.  On a technical level, many exact techniques (e.g., Abelianisation) are tailored precisely for considering Wilson loops on Lens spaces (and more generally, Seifert manifolds); still, one must verify that additional gravitational features (i.e., complex level and non-standard representations) do not spoil these methods.
 
There is an interesting non-perturbative aspect to this line of questioning as well. While global dS$_3$ analytically continues to $S^3$, these non-trivial $L(p,q)$ saddles were argued to contribute to the physics of a static patch in \cite{Castro:2011xb}.   
  An important result of their analysis, however, is that the resulting saddle-point sum (the ``Farey tail sum") diverges in a manner that cannot be regulated. It would interesting to  investigate the fate of the Wilson spool under this Farey tail divergence, i.e., loop in matter in the sum. If the perturbative  results hold,  then  the Wilson spool will provide a concrete prescription for extending the Farey tail to gravity+matter via
 \beq
\frac{\sum\limits_M\gev{\log\Zs[M]}}{\sum\limits_M\Zg[M]}= \frac{\sum\limits_M\frac{1}{4}\left\langle\mathbb W_j[M]\right\rangle}{\sum\limits_M\Zg[M]}~.
 \eeq
 An important question is whether the above ratio is finite (despite the saddle-point sum of $\Zg$ itself diverging). If so, it would be an intriguing hint that the inclusion of matter regulates some of the pathologies of dS$_3$ quantum gravity.

\paragraph{Higher-loop matching:}  We have given a prescription for computing quantum corrections to the Wilson spool to all orders, and we provided a gravitational match to a one-loop determinant in the $G_N \rightarrow 0$ limit. Likewise, it would be interesting to gravitationally match the sub-leading corrections. In particular, one would like to verify our mass renormalization formula in \eqref{eq:mass-ren} by evaluating loop corrections due to graviton exchanges in the scalar propagator. 

Another approach is to contrast our corrections against the methods advocated in \cite{Besken:2017fsj,Hikida:2017ehf,Besken:2018zro,DHoker:2019clx}. In their context, Wilson lines in $SL(2,\mathbb{R})$ Chern-Simons theory are used to report on the anomalous dimensions of massive particles on AdS$_3$. It would be interesting to adapt and apply that approach to the $SU(2)$ theory and compare against the results in Sec.\,\ref{sec:quantumspool}.

\paragraph{Edge Modes:} While the philosophy of this paper has concentrated on Chern-Simons gravity as a framework for addressing quantum gravity directly, one obvious advantage to this framework is the possibility of commenting on a possible ``dS/CFT'' dictionary.  This is because Chern-Simons theories exhibit a ``bulk/edge correspondence'' when quantized on manifolds with boundary.  This observation has led to the proposal that the edge-mode spectrum provides a realization of the dS/CFT dictionary \cite{Hikida:2021ese,Hikida:2022ltr} in three dimensions.  In this proposal one prepares the Hartle-Hawking state through path-integration on a Euclidean three-ball and then real-time evolves to future infinity. The resulting wave-function is then dual to the partition function of a Wess-Zumino-Witten (WZW) model living on the future two-sphere. There are significant differences in how the semi-classical limit is realized in \cite{Hikida:2021ese,Hikida:2022ltr} compared to this paper.  Regardless, it is tempting to speculate how the representation theory constructed in this paper fits into this proposed dS/CFT dictionary and what the implications are for the unitarity of the putative dual.  To further illustrate this point, WZW models carry a spectrum generating $\hmf{su}(2)_k$ affine current algebra
\beq\label{eq:affinesu2alg}
[J_n^a,J_m^b]=i\varepsilon^{abc}J^c_{m+n}+k\,m\delta_{m+n}\delta^{ab}~,
\eeq
which are the inheritance of Chern-Simons gauge transformations that have support on the boundary.  States of this CFT can be organized into representations of this algebra. These representations can be built in a standard way starting from representations of the non-affine $\mf{su}(2)$ spanned by $\{J_0^a\}$: given such a representation, $R$, of $\mf{su}(2)$ one demands
\beq
J^a_{n}|v\rangle=0\qquad\forall~~n>0~,\qquad \forall~~|v\rangle\in R~.
\eeq
The affine representation, $\hat R$, is then obtained by the action of arbitrary products of $\{J_{n<0}^a\}$ acting on basis states of $R$.  A subtle point of this construction however is that unitarity of $R$ is not a guarantee of the unitarity of $\hat R$.  This situation is well known to those familiar with the affine $\hmf{sl}(2,\mathbb R)_k$ current algebra: unitary highest-weight representations of $\mf{sl}(2,\mathbb R)$ invariably lead to affine representations containing negative norm states \cite{Dixon:1989cg,Maldacena:2000hw}.  Here a problem potentially arises even for $\hmf{su}(2)_k$ due to the complex level, $k=\delta+is$. That is to say, even if we tailor an inner-product on $R$ such that all states have positive norm (as we did in Sec.\,\ref{sec:nonstandardreps}), $\hat R$ could still contain negative (or even complex) norm states depending on how Hermitian conjugation is promoted to $\hmf{su}(2)_k$.  As a simple example, under the standard definition
\beq\label{eq:naiveJndagger}
\left(J_n^a\right)^\dagger:=\left(J^{a\dagger}\right)_{-n}~,
\eeq
the state $J_{-1}^3|j,0\rangle$ (where $|j,0\rangle$ is the highest-weight state of one of the non-standard representations) has complex norm:
\beq
\left|J_{-1}^3|j,0\rangle\right|^2=\langle j,0|J_1^3J_{-1}^3|j,0\rangle=k\in\mathbb C~.
\eeq
This does not necessarily imply non-unitarity of the boundary CFT: one lesson from Sec.\,\ref{sec:nonstandardreps} is that there is potential freedom in defining inner-products and so one may yet find some suitable replacement of \eqref{eq:naiveJndagger} that ensures positive norm states in the WZW model.  What we can say at this point is that norm-positivity of bulk representations does not guarantee norm-positivity of the WZW model. There are aspects of the dS/CFT proposal in \cite{Hikida:2021ese,Hikida:2022ltr} that suggest non-unitary features of the dual CFT (e.g. imaginary central charge).  Whether or not these features indicate a fundamental non-unitarity or perhaps a weaker form of unitarity (such as norm-positivity under a novel inner-product) requires a careful analysis of the representation theory of \eqref{eq:affinesu2alg} with imaginary level, $k$.  We intend to revisit this question in the near future.

\paragraph{Wilson Lines:} We may also ask if there is a natural gravitational interpretation to cutting open the Wilson loop observables to yield Wilson lines.  Of course Wilson lines are not gauge-invariant in and of themselves: one must append to their endpoints massive degrees of freedom carrying a representation of the gauge group.  It was shown in~\cite{Castro:2020smu} (based on previous applications to AdS$_3$ space \cite{Castro:2018srf}) that de Sitter Wilson lines have a useful gravitational interpretation in their classical limit when their endpoint matter representations satisfy a particular {\it Ishibashi condition}; see \eqref{eq:Ishicond}. As we explain in App.~\ref{app:casimir}, this condition is equivalent to imposing that the endpoint of the Wilson line transforms as a scalar, given the identification of dS$_3$ as a coset space. We would like to know if some of the techniques of this paper can be leveraged in pushing the results of \cite{Castro:2020smu} beyond the classical limit. An obvious obstacle to this program is constructing an appropriate diffeomorphism-invariant path-integral with definite endpoints described by Ishibashi states. Of course much of this conceptual difficulty lies in the absence of a boundary to which to anchor the endpoints of a Wilson line.  As an alternative, we might consider introducing a boundary by hand by cutting open the Euclidean path-integral, say along the hemisphere of the $S^3$: such a procedure is in fact natural for preparing the Hartle-Hawking state.  Evolving this state, Wilson lines anchored to future infinity might present useful applications to cosmological correlators (or to nascent Ryu-Takayanagi-like proposals for entanglement entropy in dS/CFT \cite{Hikida:2022ltr}). An additional hurdle is that it is not clear how to generalize the exact results considered in Sec.~\ref{sec:Wilson-SU2} to open-ended observables. Instead, one might apply a semi-classical expansion and try to compute $G_N$ corrections evaluated explicitly for the non-standard representations. It would be interesting to check whether the quantum corrected Chern-Simons calculation has a natural gravitational interpretation. 

\paragraph{$\Big\langle \text{log}Z_\text{scalar}\Big\rangle$ vs. $\text{log}\Big\langle Z_\text{scalar}\Big\rangle$:}  As constructed, the Wilson spool, $\mathbb W$, is naturally related to $\log Z_\text{scalar}$ of a massive scalar field and as such its expectation value, $\gev{\log\Zs}$, naturally computes quantum gravity corrections to $\log Z_\text{scalar}$.  It is important to contrast this with $\log\gev{ Z_\text{scalar}}$.  

Properly normalizing expectation values with respect to $\mc Z_\text{grav}$, these two objects coincide at first non-trivial order in perturbation theory, $O\left(G_N^2/\ell^{2}\right)$.  However they generically differ at higher orders. In analogy to disorder-averaging, this is similar to the difference between quenched and annealed disorder.\footnote{This is only an analogy: we are not advocating that the Chern-Simons path-integral, at least about a fixed saddle, calculates a disorder-average.} However counter to this analogy with disorder-averaging, inside the Chern-Simons theory $\gev{\log\Zs}$ is somewhat straight-forward to compute, while $\log\gev{\Zs}$ is very difficult to compute: $\Zs$ requires exponentiating the spool, and so it includes arbitrary products of Wilson loop observables.  To avoid contact singularities, one should displace these loops slightly from each other and such a prescription could, in general, allow multiple loops to link together.  Classifying and organizing the links that can appear in the expansion of $\Zs$ is already a highly non-trivial task. Subsequently, evaluating $\Zs$ inside the Chern-Simons path-integral poses another substantial challenge: except for special classes of links (such as torus links) there is a scarcity of efficient techniques for evaluating link invariants, much less techniques that we can trust for complex levels and non-standard representations.  

It is also worth highlighting the corresponding difficulty in the metric formulation, which ultimately stems from a difference in which order we perform the path-integrals.  For $\gev{\log\Zs}$, it is clear that we must perform the scalar path-integral first.  For a scalar field minimally coupled to a metric, $g_{\mu\nu}$, we have a Gaussian action and one can integrate the scalar field out to arrive at $\log\Zs[g_{\mu\nu}]$ as an effective functional of $g_{\mu\nu}$. This can be then evaluated (at least perturbatively) in the remaining gravitational path-integral. On the other hand, for $\gev{\Zs}$, the ordering is ambiguous and one might instead perform the gravity path-integral first.  Indeed we may view $\log\gev{\Zs}\equiv-F_\text{scalar}$ as the scalar free-energy after integrating out gravity.  Even in $G_N$ perturbation theory this is perilous as graviton exchanges can induce irrelevant interactions for the scalar field, leaving the matter path-integral intractable. 

Despite these difficulties in both the Chern-Simons and the metric formulations, we might (wildly) speculate that this quantum-gravity corrected free-energy is related to a sum (with some unspecified measure) of linked spools over all possible $n$-links, i.e.,
\beq
\sum_{n=1}^\infty\frac{1}{4^nn!}\sum_{n\text{ links}}\Big\langle\mathbb W_j[\gamma_1]\,\mathbb W_j[\gamma_2]\,\ldots\mathbb W_j[\gamma_n]\Big\rangle\overset{?}{\sim}e^{-F_\text{scalar}}~.
\eeq
It would interesting to test such a relation in future work. 

\paragraph{The Wilson spool in AdS$_3$:} The focus of this paper has been on the Chern-Simons formulation of three-dimensional quantum gravity with positive cosmological constant. However Chern-Simons gravity has arguably been more fully explored in the context of negative cosmological constant, i.e. AdS$_3$. It is natural to ask if the Wilson spool is a useful object in this context.  Indeed the DHS formula provides a broad construction for one-loop determinants including fields on a black-hole background \cite{Denef:2009kn}. We expect that the Wilson spool can be constructed in a wholly similar way to Section \ref{sect:spool} for the BTZ saddle of AdS$_3$ gravity.  Furthermore, in \cite{Castro:2018srf} it was already pointed out that gravitational Wilson loops wrapping the BTZ horizon arbitrarily many times reproduce $\log\Zs$ of a scalar on a fixed BTZ background in the classical limit
\beq
\left.\sum_{n\neq0}\frac{1}{|n|}\mTr_{R}\mc Pe^{n\oint A_L}\mTr_R\mc P e^{-n\oint A_R}\right|_\text{classical}=\log\Zs[\text{BTZ}]~,
\eeq
for appropriate highest-weight representations of $\mathfrak{sl}(2,\mathbb R)$ (see \cite{Castro:2018srf} for more details).  This echoes \eqref{eq:Zscalar1} and \eqref{eq:spooldef} when the $\alpha$ integrand possesses no additional poles.  Making this suggestive matching more precise is an obvious follow-up.  Compared to $S^3$, Euclidean AdS$_3$ lacks the same library of exact methods for evaluating the spool inside the Chern-Simons path-integral.  Nevertheless, formulating one-loop determinants in the Chern-Simons language may still provide a useful organizing framework for calculating gravitational corrections to one-loop determinants. 

In connection to other directions mentioned above,  AdS$_3$ has an advantage. On AdS$_3$ there is a robust interpretation of the boundary and how to interpret boundary conditions.  In the past, this has meant that it is clear how to anchor the endpoints of Wilson lines \cite{Ammon:2013hba}, which allows a more concrete exploration to our previous questions regarding Wilson lines. If there is a practical definition of the Wilson spool on AdS$_3$ for loops, one can imagine then an appropriately defined open-ended Wilson spool might provide an organizing structure to $1/c$ corrections to CFT correlators that naturally incorporates an intertwining between holomorphic sectors.\footnote{In \cite{Besken:2017fsj,Hikida:2017ehf,Besken:2018zro,DHoker:2019clx} the $1/c$ corrections are treated in a holomorphic manner, where only one copy of the gauge group is discussed. This is a reasonable take since the Wilson line is related to conformal blocks \cite{Besken:2016ooo}. The Wilson spool on dS$_3$ is not obviously holomorphic: it intertwines the left and right gauge groups due to the $\alpha$ countour in its definition. We expect this feature to persist in AdS$_3$, which might resonate with the Ishibashi construction in \cite{Castro:2018srf} and be adequate to describe correlation functions.}

\paragraph{Observational Consequences:} We have presented a calculation for the quantum corrections to $\mbox{log } Z_{\rm scalar}$, which can be interpreted as a mass renormalization. We emphasize that this result is new, and serves as a \emph{prediction} for $3$d de Sitter gravity. The corrections are suppressed in powers of $G_N/\ell$, as might be expected in a theory of quantum gravity. More recently, however,~\cite{Verlinde:2019xfb, Verlinde:2019ade} suggested that certain quantum gravity effects may couple UV and IR scales, and thus be governed by larger scale overall. The basic idea is to consider quantum fluctuations $\delta L$ in lengths related to horizon fluctuations of a causal diamond. In several different models, the expectation value $\left<\delta L \delta L\right>$ was shown to scale like $\ell_P L$ with $L$ an IR scale, rather than $\ell_P^2$ as one might naively expect. The original calculations were coordinate dependent, and it would be interesting to consider a setup where one obtains a coordinate-independent calculation of length fluctuations by considering the natural observable for lengths: a Wilson loop. Then $\left<\delta L \delta L\right>$ should relate to $\left< W W \right>$, where $W$ is a Wilson loop operator. In this paper we have presented an exact calculation for $\left< W \right>$, including the first calculation (to our knowledge) of quantum corrections for a Wilson loop in de Sitter spacetime. It may be interesting to explore an extension to the two-point function to try to connect to this proposal.

\section*{Acknowledgements}
It is a pleasure to thank Dionysios Anninos, Tarek Anous, Frederik Denef,  Dami\'an Galante, Sean Hartnoll, Kurt Hinterbichler,  Austin Joyce, and Marcos Mari\~no for useful discussions. The work of AC and JRF has been partially supported by STFC consolidated grant ST/T000694/1. The work of JRF has been also partially supported by the ERC starting grant GenGeoHoloIC and by Simons Foundation Award number 620869.  IC has been partially supported by the ERC starting grant H2020 ERC StG No.640159. CZ has been partially supported by the ERC Consolidator Grant QUANTIVIOL and a UM Duluth Higholt Professorship, and acknowledges a Heising-Simons Fellowship as part of the “Observational Signatures of Quantum Gravity” collaboration grant 2021-2818. This work is supported by the Delta ITP consortium, a program of the Netherlands Organisation for Scientific Research (NWO) that is funded by the Dutch Ministry of Education, Culture and Science (OCW).

\appendix

\numberwithin{equation}{section}

\section{Conventions}\label{app:conventions}

Let us review several useful bases for the conformal algebra $\mathfrak{so}(1,3)$. One choice is to build it from the antisymmetric generators $L_{AB} = -L_{BA}$, ($A,B=0,1,2,3$) which satisfy the commutation relations
\beq [L_{AB},L_{CD}] = -\eta_{AC} L_{BD} + \eta_{BC} L_{AD} - \eta_{BD} L_{AC} + \eta_{AD} L_{BC}~, \label{eq:Lcommutators}\eeq
where
\begin{comment}(PREV)
\beq \eta_{AB} = \left(\begin{array}{ccc}
-1 & & \\
& 1 & \\
& & \delta_{ij}
\end{array}
\right)~. \eeq
\end{comment}
\beq
\eta_{AB}=\text{diag}(-1,1,1,1)~.
\eeq
This basis acts naturally as (generalized) rotations in embedding space, $\mathbb R^{1,3}$, i.e. as the set of Killing vectors, $\mc L_{AB}=X_A\pa_B-X_B\pa_A$.  This is also a {\it real basis} of the Lie algebra with group elements obtained by exponentiation by a real parameter $\theta^{AB}$:
\beq
g=\exp\left(\theta^{AB}L_{AB}\right)~,
\eeq
and so standard unitary representations of $SO(1,3)$ will come with a inner product compatible with the following Hermiticity condition (we will discuss other choices of Hermiticity in Sec.\,\ref{sec:nonstandardreps}):
\beq
L_{AB}^\dagger=-L_{AB}~.\label{SO13Hermiticity}
\eeq
A useful basis for constructing representations consists of the standard generators $D$, $P_i$, $K_i$, $M_{ij}$ ($i,j=1,2$) of the Euclidean conformal algebra, which are related to $L_{AB}$ above as
\begin{align}
    L_{AB} = \left(\begin{array}{ccc}
    0 & \frac{1}{2}(P_i-K_i) & D\\
    -\frac{1}{2}(P_i-K_i) & M_{ij} & -\frac{1}{2}(P_i+K_i)\\
    -D & \frac{1}{2}(P_i+K_i) & 0
    \end{array}\right)~.\label{eq:conformaldecomp}
\end{align}
Starting from \eqref{eq:Lcommutators} it is easy to check that the Euclidean conformal generators satisfy the commutation relations
\begin{equation}
\begin{aligned}
 [D,P_i] &= P_i~, \\
[D,K_i] &= -K_i~,\\
[K_i,P_j] &= 2(\delta_{ij} D - M_{ij})~,\\
[P_i, M_{jk}] &= \delta_{ij} P_j - \delta_{ik} P_j~,\\ 
[K_i, M_{jk}] &= \delta_{ij} K_j - \delta_{ik} K_j~,\\
[M_{ij},M_{kl}] &= -\delta_{ik} M_{jl} + \delta_{jk} M_{il} - \delta_{jl} M_{ik} + \delta_{il} M_{jk}~.
\end{aligned}
\end{equation}
Another useful basis making the $\mf{so}(1,2)$ subalgebra manifest is given by $\{\mc J_a,\mc P_b\}_{a,b=1,23}$ related to $L_{AB}$ as
\be
\begin{aligned}
\mc J_1=&iL_{13}~,\qquad \mc J_2=-iL_{23}~,\qquad \mc J_3=L_{12}~,\\
\mc P_1=&iL_{01}~,\qquad\mc P_2=iL_{02}~,\quad\mc P_3=-L_{03}~,
\end{aligned}
\ee
and obeying
\be\label{eq:alg1234}
\begin{aligned}
[\mc J_a,\mc J_b]=&\epsilon_{abc}\eta^{cd}\mc J_d~,\\
[\mc J_a,\mc P_b]=&\epsilon_{abc}\eta^{cd}\mc P_d~,\\
[\mc P_a,\mc P_b]=&-\epsilon_{abc}\eta^{cd}\mc J_d~,
\end{aligned}
\ee
with $\eta=\text{diag}(1,1,-1)$. 
\paragraph{Map to $\mf{su}(2)_L\oplus\mf{su}(2)_R$.}
Now we want to map this algebra to $\mf{su}(2)\oplus\mf{su}(2)$.  To do so it is first useful to Wick-rotate the $\{\mathcal J_a,\mathcal P_b\}$ basis, to $\mf{so}(4)$ via
\be
\begin{aligned}
    \mc J_1&=i\mathbb{P}_1~,\quad
    \mc J_2=i \mathbb{J}_2~,\quad
    \mc J_3=-\mathbb{P}_3~,\\
    \mc P_1&=\mathbb{J}_1~,\quad
    \mc P_2=\mathbb{P}_2~,\quad
    \mc P_3=i\mathbb{J}_3~,
\end{aligned}
\ee
with $\{\mathbb{J}_a,\mathbb{P}_b\}$ obeying
\be
\begin{aligned}
[\mathbb{J}_a,\mathbb{J}_b]=&-\epsilon_{abc}\delta^{cd}\mathbb{J}_d~,\\
[\mathbb{J}_a,\mathbb{P}_b]=&-\epsilon_{abc}\delta^{cd}\mathbb{P}_d~,\\
[\mathbb{P}_a,\mathbb{P}_b]=&-\epsilon_{abc}\delta^{cd}\mathbb{J}_d~.
\end{aligned}
\ee
Lastly the splitting to $\mf{su}(2)\oplus\mf{su}(2)$ is given by
\beq
\mathbb{P}_a=i\left(L_a-\bar L_a\right)~,\qquad \mathbb{J}_a=i\left(L_a+\bar L_a\right)~,
\eeq
with $\{L_a\}$ and $\{\bar L_a\}$ generating the two $\mf{su}(2)$'s:
\beq
[L_a,L_b]=i{\epsilon_{ab}}^c\,L_c~,\qquad [\bar L_a,\bar L_a]=i{\epsilon_{ab}}^c\,\bar L_c~,\qquad [L_a,\bar L_b]=0~.
\eeq
For the sake of ease of comparison, let us rewrite this map in terms of the conformal basis: 
\be
\begin{aligned}
   D&=-(L_3+\bar L_3)~,\\
   M&=i(L_3-\bar L_3)~,\\
   P_1&=L_-+\bar L_-~,\\
   P_2&=i\left(L_--\bar L_-\right)~,\\
   K_1&=-\left(L_++\bar L_+\right)~,\\
   K_2&=i\left(L_+-\bar L_+\right)~,
\end{aligned}
\ee
where $M=\frac{1}{2}\varepsilon^{ij}M_{ij}$ and $L_\pm=L_1\pm i L_2$ as usual.  We briefly point a couple of features of this mapping.  Firstly, representations that are labelled by $\{L_3,\bar L_3\}$ eigenvalues, $(j_L,j_R)$, can equivalently be labelled by their conformal weight, $\Delta$, and spin, $s$, the eigenvalues of $ D$ and $-i M$, respectively, via
\beq\label{eq:appDStojs}
\Delta=-j_L-j_R~,\qquad s=j_L-j_R~,
\eeq
which facilitates how we will utilize $\mf{su}(2)$ representations to describe particle content on dS$_3$ described in Sec.\,\ref{sect:dSrepintro}.  Secondly, the $\mf{so}(1,3)$ quadratic Casimir maps to twice the sum of the $\mf{su}(2)$ Casimirs:
\be
\begin{aligned}
c_2^{\mf{so}(1,3)}&= D(2- D)+\sum_{i=1,2} P_i K_i+ M^2\\
&=-2\left((L_3)^2+L_3+L_-L_++(\bar L_3)^2+\bar L_3+\bar L_-\bar L_+\right)\\
&=-2c_2^{\mf{su}(2)_L}-2c_2^{\mf{su}(2)_R}~.
\end{aligned}
\ee
Lastly we point out that the Hermiticity of a real $\mf{so}(1,3)$, \eqref{SO13Hermiticity}, is unnatural from the point of view of the $\mf{su}(2)\oplus\mf{su}(2)$:
\beq
(L_3)^\dagger=-\bar L_3~,\qquad (L_\pm)^\dagger=-\bar L_\pm~.
\eeq
This stems from the fact that $\mf{so}(1,3)$ and $\mf{su}(2)\oplus\mf{su}(2)$ are not isomorphic {\it as real algebras}.  Instead we have identified the generators in their common complexification, $\mf{sl}(2,\mathbb C)$.

While we do not utilize the following result, it is amusing that one can form the $\mf{so}(1,3)$ representation characters from the $\mf{su}(2)\oplus\mf{su}(2)$ characters defined in Sec.\,\ref{sec:nonstandardreps}. More explicitly, from  \eqref{eq:charp},  we see that the following combination gives
\be \chi_{j_L}(z_L) \chi_{j_R}(z_R) + \chi_{\bar j_L}(z_L) \chi_{\bar j_R}(z_R) = \frac{e^{2\pi i j_L z_L +2\pi i j_R z_R}+e^{2\pi i\bar j_Lz_L + 2\pi i\bar j_Rz_R}}{(1-e^{-2\pi i z_L})(1-e^{-2\pi i z_R})}~,\ee
where $\bar j_{L/R}=-1-j_{L/R}$.  
Writing $q=e^{-i\pi(z_L+z_R)}$ and $w=e^{i\pi(z_L-z_R)}$
\be \chi_{j_L}(z_L) \chi_{j_R}(z_R) + \chi_{\bar j_L}(z_L) \chi_{\bar j_R}(z_R) = \frac{w^s q^\Delta + w^{-s} q^{\bar \Delta}}{(1-w^{-1}q)(1-wq)}~,\ee
where $\Delta$ and $s$ are identified as \eqref{eq:appDStojs} and $\bar \Delta = 2-\Delta$. This is in fact the Harish-Chandra character for the $\mf{so}(1,3)$ spinning principal series~\cite{Basile:2016aen}.  For $s=0$ this matching follows from the observation that a $q$-expansion of the Harish-Chandra character lies up with the frequencies and degeneracies of quasi-normal modes \cite{Anninos:2020hfj} and this quasi-normal expansion is precisely realized as the weight spaces of our non-standard representations.  However, to be clear, for $s\neq 0$, the above matching is simply a suggestive observation: as of yet, we have not constructed non-standard $\mf{su}(2)_L\oplus\mf{su}(2)_R$ representations corresponding to the spinning principal series.

\section{Non-Abelian localization}\label{sec:locApp}

Non-Abelian localization is a method developed in \cite{Beasley:2005vf,Beasley:2009mb} for computing partition functions and expectation values of Wilson loops of $SU(2)_k$ Chern-Simons theory on a Seifert manifold.  In this case, $M$ is a circle bundle over a Riemann surface and so admits a locally free $U(1)$ action that rotates the $S^1$ fibers. \cite{Beasley:2009mb} showed that for Wilson loops whose curve, $\gamma$, wraps a circle fiber (a Seifert loop), the stationary phase approximation to the path-integral is exact leading to the following expectation value for finite dimension dimensional representations of $SU(2)$:
\beq \label{eq:Wloop3}
\big\langle W_{j}[\gamma]\big\rangle = \frac{1}{2\pi i} \mathrm{e}^{-\frac{i \pi(1+(2j+1)^2)}{2(k+2)}} \int_{-\infty}^{+\infty} dx ~  \mathrm{ch}_{j} \left( \mathrm{e}^{\frac{i3\pi}{4}} \frac{x}{2\pi} \right) \sinh^2 \left( \mathrm{e}^{\frac{i\pi}{4}} \frac{x}{2} \right) \mathrm{e}^{-\frac{k+2}{8\pi} x^2 } ~,
\eeq
where $\mathrm{ch}_{j}(x)=\sin(\pi(2j+1)x)/\sin(\pi x)$ is the $SU(2)$ character associated to the finite-dimensional representation with highest-weight $j$. (We use this notation to distinguish it from the characters defined in Sec.\,\ref{sec:nonstandardreps}).

The main idea leading to \eqref{eq:Wloop3} is to recast the 
expectation value as a symplectic integral of the canonical form 
\beq \label{SymplPathIntW}
Z(\epsilon)=\int_X \exp \left(\Omega - S/\epsilon \right)~,
\eeq
which has a cohomological interpretation and which localises onto the critical points of a classical action $S$. In the new form \eqref{SymplPathIntW} of the path-integral, $X$ is a symplectic manifold with symplectic form $\Omega$, and a Lie group $H$ acts on $X$ in a Hamiltonian fashion with moment map $\mu:X\to \mathfrak{h}^*$, where $\mathfrak{h}^*$ is dual to the Lie algebra $\mathfrak{h}$ of $H$ and where $S=(\mu,\mu)/2$.   
The two key ideas to obtain the symplectic description \eqref{SymplPathIntW} of the Seifert loop path-integral are:
\begin{enumerate}[leftmargin=0.4cm]
    \item The Chern-Simons path-integral gets recast as a symplectic integral by %effectively 
    employing a shift symmetry that acts on the %Chern-Simons 
    gauge field $A$.  This decouples one of its components from the path-integral, thereby reducing this to the integral of a two-dimensional theory. 
    \item When the path-integral includes a Wilson loop, it is necessary to rewrite the classical description \eqref{Wilsondef} as a path-integral over an auxiliary bosonic field $U$ attached to the curve $\gamma$ and coupled to the connection $A$ as a background field.
\end{enumerate}
This second point follows from the Borel-Weil-Bott theorem, realizing an irreducible representation $R$ (with highest-weight $\lambda$) as the quantization of the co-adjoint orbit $\mc O_\lambda\subset\mf g^\ast$ passing through $\lambda.$  $R$ is thus identified with the space of holomorphic sections of a unitary pre-quantum line bundle over $\mc O_\lambda$ carrying a canonical flat connection $\Theta_\lambda$.  This leads to the description of the Wilson loop as a path-integral over an auxiliary field $U:\gamma\longrightarrow\mc O_\lambda$, which reads
\beq \label{eq:Wloop4a} 
W_R[\gamma] = \int \mathcal{D} U ~ \exp \left[ i cs_R (U;A|_\gamma) \right] ~,\qquad 
cs_R(U;A|_\gamma) = \oint_\gamma U^*(\Theta_\lambda (A)) ~.
\eeq
It is this rewriting of \eqref{Wilsondef} as a symplectic path-integral, \eqref{eq:Wloop4a}, that imposes an obstruction for using this method for Chern-Simons gravity.  Namely, the Borel-Weil-Bott theorem requires $R$ to be a finite dimensional irreducible representation of $G$ and the status of the non-standard representations from Sec.\,\ref{sec:nonstandardreps} under this theorem is unknown.  
Thus, despite the obvious similarities of the final integral formula for $\langle W_R\rangle$, \eqref{eq:Wloop3}, to the formula arrived through Abelianisation and supersymmetric localization, \eqref{eq:WexpAb}, we do not rely on this method in this paper.

\section{The \texorpdfstring{$S^3$}{S3} heat kernel}\label{app:HK}

In this appendix we show how the one-loop determinant of a massive scalar field can be written as a worldline path-integral.  Importantly, we show explicitly that this determinant utilizes an infinite sum of worldlines wrapping the sphere multiple times.  To this end we will use the formalism of the heat kernel.

To briefly remind the reader, the heat-kernel is defined formally as
\beq
K_{\mathsf{m}^2}(x,y;\beta):=\langle x|e^{-\beta(-\nabla^2+\mathsf{m}^2\ell^2)}|y\rangle~,
\eeq
which can be used to assign meaningful expressions to, e.g., functional determinants in Euclidean signature
\be\label{eq:ZK1}
\begin{aligned}
\log\det(-\nabla^2+\mathsf{m}^2\ell^2)&=\mtr\log(-\nabla^2+\mathsf{m}^2\ell^2)\\
&=-\int_0^\infty \frac{\dd\beta}{\beta}\int d^dx\sqrt{g(x)}\,K_{\mathsf{m}^2}(x,x;\beta)~,
\end{aligned}
\ee
or a Green's function
\be
G_{\mathsf{m}^2}(x,y)=\frac{1}{-\nabla^2+\mathsf{m}^2\ell^2}=\int_0^\infty \dd\beta\,K_{\mathsf{m}^2}(x,y;\beta)~, 
\ee
up to regularization.  More precisely we define $K_{\mathsf{m}^2}$ as the solution to the heat equation with a delta-function localized initial source, i.e.,
\be\label{eq:Km2heateq}
(\nabla^2_{(x)}-\mathsf{m}^2\ell^2)K_{\mathsf{m}^2}(x,y;\beta)=\frac{d}{d\beta}K_{\mathsf{m}^2}(x,y;\beta)~,
\ee
and
\be
\lim_{\beta\rightarrow0}K_{\mathsf{m}^2}(x,y;\beta)=\frac{1}{\sqrt{\det g(x)}}\delta^d(x-y)~.
\eeq
On $S^3$, because this defining equation and initial condition are spherically symmetric, $K_{\mathsf{m}^2}$ can only depend on the geodesic arclength, $\theta$, between $x$ and $y$.  We can express this length conveniently in embedding space coordinates, where $X,Y \in \mathbb{R}^4$, as $\theta=\arccos(X\cdot Y)$.  

There are multiple routes for solving for $K_{\mathsf{m}^2}(\theta;\beta)$ on $S^3$ exactly; for example, through solving \eqref{eq:Km2heateq} as an ordinary differential equation in $\theta$ or through analytic continuation from $\mathbb H^3$. Here, we will take a circuitous route, with our aim being to relate $K_{\mathsf{m}^2}$ to a world-line path integral. For this approach, we  first express $K_{\mathsf{m}^2}$ as a formal sum over eigenfunctions of $\nabla^2_{S^3}$, that is
\beq
K_{\mathsf{m}^2}(x,y;\beta)=e^{-\beta \mathsf{m}^2\ell^2}\sum_{l=0}^\infty\sum_{\vec{\mathsf{m}}}\mc Y_{l,\vec{\mathsf{m}}}^\ast(x)e^{-\beta l(l+2)}\mc Y_{l,\vec{\mathsf{m}}}(y)~,
\eeq
where $\mc Y_{l,\vec{m}}$ are a complete set of hyper-spherical harmonics.  We make this expression look line  a  path-integral we need to replace our discrete summands with continuous variables.  To this end, we introduce two auxilliary variables, $\alpha$ and $p$, such that we can replace the discrete quadratic Casimir with a Gaussian:
\beq\label{eq:Km2introducealpha}
K_{\mathsf{m}^2}(x,y;\beta)=e^{-\beta(\mathsf{m}^2\ell^2-1)}\sum_{l=0}^\infty\sum_{\vec{\mathsf{m}}}\mc Y_{l,\vec{\mathsf{m}}}^\ast(x)\int\frac{\dd\alpha \dd p}{2\pi}e^{i\alpha(p-l-1)}e^{-\beta p^2}\mc Y_{l,\vec{\mathsf{m}}}(y)~.
\eeq
 Next, we make use of the identity
\beq
\sum_{\vec{\mathsf{m}}}\mc Y_{l,\vec{\mathsf{m}}}^\ast(x)\mc Y_{l,\vec{\mathsf{m}}}(y)=\frac{d_l}{V_{S^3}}\mc P_l(\cos\theta)~,
\eeq
where $d_l=(l+1)^2$ is the degeneracy of $l$ eigenvalues, $V_{S^3}=2\pi^2$ is the volume of $S^3$, and  $\mc P_l(x)$ is the hyper-spherical Legendre polynomial.  The generating function of these  polynomials is
\beq
\frac{1-t^2}{(1-2tx+t^2)^2}=\sum_{l=0}^\infty\,d_l\,\mc P_l(x)\,t^l~,
\eeq
which leads us to
\begin{align}\label{eq:Kmalpha-p}
K_{\mathsf{m}^2}(\theta;\beta)=&-\frac{e^{-\beta(\mathsf{m}^2\ell^2-1)}}{8\pi^2}\int \dd p\int\frac{\dd\alpha}{2\pi i}\frac{\cos(\alpha/2)\sin(\alpha/2)}{\sin^2\left(\frac{\alpha+\theta}{2}\right)\sin^2\left(\frac{\alpha-\theta}{2}\right)}e^{i\alpha p}e^{-\beta p^2}~.
\end{align}
At this point we are ready to perform a trick: we will cast the integral over $p$ as a path integral of a point particle. To do so, we introduce an auxiliary worldline, parameterized by $\tau\in[0,1]$, and introduce two dynamical variables: the position $X(\tau)$ and momentum $P(\tau)$.  The degree of freedom $X$ will act as a Lagrange multiplier forcing $\dot P=0$, and hence $P(\tau)=p$. More concretly, we can write \eqref{eq:Kmalpha-p} as 
\begin{align}
K_{\mathsf{m}^2}(\theta;\beta)=&-\mc N\,e^{-\beta(\mathsf{m}^2\ell^2-1)}\int\left.\mc DX\right|_{X_i=0}^{X_f=0}\left.\mc DP\right|_{P_i=P_f}\nonumber\\
&\qquad\qquad\times\int\frac{\dd\alpha}{2\pi i}\frac{\cos(\alpha/2)\sin(\alpha/2)}{\sin^2\left(\frac{\alpha+\theta}{2}\right)\sin^2\left(\frac{\alpha-\theta}{2}\right)}e^{i\int\limits_0^1\dd\tau\,\alpha\, P+i\int\limits_0^1\dd\tau\,X\,\dot P-\beta\int\limits_0^1\dd\tau\,P^2}~.
\end{align}
The factor $\mc N$ parameterizes numerical ambiguities in the definition of the measures $\mc DX\mc DP$, which we will be imprecise about.  Note that in the measure we impose Dirichlet boundary conditions  on $X$, i.e., $X(0)=X(1)=0$.  This allows us to integrate the action by parts, and subsequently integrate out $P$, which gives
\beq\label{eq:KasWPI2}
\begin{aligned}
K_{\mathsf{m}^2}(\theta;\beta)=&-\mc Ne^{-\beta(\mathsf{m}^2\ell^2-1)}\int \left.\mc DX\right|_{X_i=0}^{X_f=0}\\
&\qquad\qquad\times
\int\frac{\dd\alpha}{2\pi i}\frac{\cos(\alpha/2)\sin(\alpha/2)}{\sin^2\left(\frac{\alpha+\theta}{2}\right)\sin^2\left(\frac{\alpha-\theta}{2}\right)}e^{-\frac{1}{4\beta}\int\limits_0^1\dd\tau\left(\dot X+\alpha\right)^2}~.
\end{aligned}
\eeq
We are now in a position to give a new interpretation to $\alpha$.  The integration over $X$ began life with Dirichlet boundary conditions, however integrating over $\alpha$ endows $X$ with winding degrees of freedom.  To be explicit about this let us note that the $\alpha$ integral has second order poles\footnote{We have been cavalier about the $\alpha$ contour in \eqref{eq:Km2introducealpha} however we can more properly utilize $\alpha$ as a Lagrange multiplier via
\beq
\delta(p-l-1)=\int_{-\infty}^\infty \frac{d\alpha}{2\pi} \cos(\alpha(p-l-1))
\eeq
and prescribe $+i\epsilon$ to the positive exponential and $-i\epsilon$ to the negative exponential.  This is equivalent to writing $\int_\mc C \frac{d\alpha}{2\pi}e^{i\alpha(p-l-1)}$ where $\mc C$ consists of a contours running $\epsilon$ above and below the real axis with opposite orientation.} at $\alpha=\pm\theta+2\pi n$.  The residue about these poles is given by
\beq
K_{\mathsf{m}^2}(\theta;s)=\mc N\frac{e^{-\beta(\mathsf{m}^2\ell^2-1)}}{4\pi \beta}\int \left.\mc DX\right|_{X_i=0}^{X_f=0}\sum_{n\in\mathbb Z}\frac{\theta+2\pi n}{\sin\theta}e^{-\frac{1}{4\beta}\int\limits_0^1\dd\tau\left(\dot X+\theta+2\pi n\right)^2}~.
\eeq
Next, we write $(4\pi \beta)^{-1}$ as the free path-integral over two more scalar degrees of freedom, $X^2$ and $X^3$, with Dirichlet boundary conditions. This gives 
\beq\label{eq:K-pi}
K_{\mathsf{m}^2}(\theta;\beta)=\mc N\,e^{-\beta(\mathsf{m}^2\ell^2-1)}\int\sum_{n\in\mathbb Z} \mc D\vec X_{(n)}\,\frac{\theta+2\pi n}{\sin\theta}e^{-\frac{1}{4\beta}\int_0^1d\tau\,\dot {\vec{X}}_{(n)}^2}~,
\eeq
where $\vec X_{(n)}=(X^1_{(n)}, X^2,X^3)$ and $X^1_{(n)}\equiv X+(\theta+2\pi n)\tau$. The boundary conditions on 
$\vec X_{(n)}$ are
\beq
\vec X_{(n)}(0)=\vec 0~,\qquad \vec X_{(n)}(1)=(\theta+2\pi n,0,0)~.
\eeq
The expression in \eqref{eq:K-pi}  is precisely the exact worldline path-integral of a massive scalar field on $S^3$ expressed in a set of Riemann normal coordinates, see e.g. \cite{Bastianelli:2017xhi}. However, one important feature is that this expression includes a sum over a saddles that wind around the geodesic arclength, $\theta$ (this contribution was missed in \cite{Bastianelli:2017xhi}).  This is the price to pay for replacing our discrete basis of eigenfunctions by a continuous path-integral and ultimately has its root in the simple fact that $S^3$ is compact.

Now let's look at the one-loop determinant. As stated in \eqref{eq:ZK1} it is related to the heat kernel via
\beq\label{eq:lZasKcoin}
\begin{aligned}
\log Z_{\rm scalar}&\sim\frac{1}{2}\int_0^\infty\frac{\dd\beta}{\beta}\int d^3x\sqrt{g(x)}K_{\mathsf{m}^2}(x,x;s)\\&\sim \frac{1}{2}V_{S^3}\int_0^\infty \frac{\dd\beta}{\beta}\lim_{\theta\rightarrow 0}K_{\mathsf{m}^2}(\theta;\beta)~,
\end{aligned}
\eeq
where $\sim$ indicates that we should implicitly include the regulator, $R_\epsilon(\beta)=e^{-\epsilon^2/4\beta}$ which tames the $\beta\to0$ (UV) behavior.  Taking the coincident limit of \eqref{eq:KasWPI2}, one finds
\beq\label{eq:Kcoinc}
K_{\mathsf{m}^2}(0;\beta)=\mc Ne^{-\beta(\mathsf{m}^2\ell^2-1)}\int\left.\mc DX\right|_{X_i=0}^{X_f=0}\int \frac{d\alpha}{2\pi i}\frac{\cos(\alpha/2)}{\sin^3(\alpha/2)}e^{-\frac{1}{4\beta}\int\limits_0^1\dd\tau\left(\dot X+\alpha\right)^2}~.
\eeq
The integral over $\alpha$ is straightforward to evaluate: it reduces to residues at the poles $\alpha=2\pi n$. With this we have
\beq
K_{\mathsf{m}^2}(0;\beta)=\mc Ne^{-\beta(\mathsf{m}^2\ell^2-1)}\int\sum_{n\in\mathbb Z}\left.\mc DX\right|_{X(0)=0}^{X(1)=2\pi n}\left(-\frac{1}{2\beta}+\frac{\dot X^2}{4\beta^2}\right)e^{-\frac{1}{4\beta}\int\limits_0^1d\tau\dot X^2}~.
\eeq
And finally, performing the path integral over $X$  gives
\beq\label{eq:Kfinal}
K_{\mathsf{m}^2}(0;\beta)=-\frac{1}{2\pi}\frac{e^{-\beta(\mathsf{m}^2\ell^2-1)}}{\sqrt{4\pi \beta}}\sum_{n\in\mathbb Z}\left(-\frac{1}{2\beta}+\frac{\pi^2n^2}{\beta^2}\right)e^{-\frac{\pi^2n^2}{\beta}}~.
\eeq
Here ${\cal N}$ has been adjusted such that we recover \eqref{eq:Kmalpha-p}. 

The next step is to insert \eqref{eq:Kfinal} into \eqref{eq:lZasKcoin}, and perform the integral over $\beta$. There are two aspects to keep in mind. First, the integral over $\beta$ converges for $\mathsf{m}^2\ell^2>1$, and so we will write $\mu^2=\mathsf{m}^2\ell^2-1$ which we assume to positive. Second, we will introduce our UV regulator: we will replace the  $n=0$ term by $R_\epsilon(\beta)=e^{-\epsilon^2/4\beta}$ (all other terms in the sum over $n$ are finite). Incorporating this we find
\be
\begin{aligned}
\log Z_{\rm scalar}=&\frac{1}{2}V_{S^3}\int_0^\infty\frac{\dd\beta}{\beta}\frac{e^{-\mu^2 \beta}}{(4\pi \beta)^{3/2}}\left(e^{-\epsilon^2/4\beta}+2\sum_{n=1}^\infty\left(1-\frac{2\pi^2n^2}{\beta}\right)e^{-\frac{\pi^2n^2}{\beta}}\right)\\
=&\frac{1}{2}V_{S^3}\left(\frac{e^{-\epsilon|\mu|}}{2\pi \epsilon^3}(|\mu|\epsilon+1)-\frac{1}{4\pi^4}\sum_{n=1}^\infty e^{-2\pi n|\mu|}\left(\frac{1}{n^3}+\frac{2\pi |\mu|}{n^2}+\frac{2\pi^2\mu^2}{n}\right)\right).
\end{aligned}
\ee
Taking the limit $\epsilon\rightarrow 0$, we find the following one-loop determinant:
\begin{multline}
    \label{eq:logZfinal}
\log Z_{\rm scalar}[S^3]=\frac{\pi}{2\epsilon^3}-\frac{\pi \mu^2}{4\epsilon}+\frac{\pi\mu^3}{6}\\-\frac{1}{4\pi^2}\left(\text{Li}_3(e^{-2\pi\mu})+2\pi\mu \text{Li}_2(e^{-2\pi\mu})+2\pi^2\mu^2\text{Li}_1(e^{-2\pi\mu})\right)~,
\end{multline}
where we used the definition of the polylogarithm 
\beq
\text{Li}_a(x)=\sum_{n=1}^\infty \frac{x^n}{n^a}~,
\eeq
and $V_{S^3}=2\pi^2$. This answer was independently arrived at by \cite{Anninos:2020hfj} which our answer matches. 
It is also worth noticing that under the replacement $\mu\rightarrow i\nu$, with $\nu^2=1-\mathsf{m}^2\ell^2$ and $\mathsf{m}^2\ell^2<1$, \eqref{eq:logZfinal} is real: this gives the one-loop determinant of a light scalar field.

\section{The curved Casimir}\label{app:casimir}

In this appendix we add details to the claim from Sec.\,\ref{sect:spool} that the one-loop determinant on a round $S^3$ is equivalent to a determinant over quadratic Casimirs
\beq
\det\left(-\nabla^2_{S^3}+\mathsf{m}^2\ell^2\right)=\det\left(2c_2^{(L)}+2c_2^{(R)}+\mathsf{m}^2\ell^2\right)~.
\eeq
We will then show that this result can be generalized, expressing the Laplacian on a curved three-dimensional manifold (away from the $S^3$ saddle) as the Casimir of local $\mf{su}(2)_L\oplus\mf{su}(2)_R$ action.

We begin by noting that $S^3$ is globally diffeomorphic to $SU(2)$ with the map (in the fundamental representation) given by
\beq
(\rho,\tau,\varphi)\rightarrow\,g_{S^3}=\left(\begin{array}{cc}\cos\rho\,e^{i\tau}&\sin\rho\,e^{-i\varphi}\\-\sin\rho\,e^{i\varphi}&\cos\rho\,e^{-i\tau}\end{array}\right)~.
\eeq
The isometry group $SU(2)_L\times SU(2)_R/\mathbb Z_2$ acts naturally on this geometry via left and right action,\footnote{With $(-1,-1)$, generating the $\mathbb Z_2$, acting trivially.} given by
\beq
(g_L,g_R):g_{S^3}\rightarrow g_Lg_{S^3}g_R^{-1}~.
\eeq
This isometry group is generated by left (right) acting Killing vectors, $\{\zeta_a\}$ and $\{\bar\zeta_a\}$
\beq\label{eq:S3KVdef}
\zeta_a^{\mu}\pa_\mu g_{S^3}=-L_ag_{S^3}~,\qquad \bar\zeta_a^{\mu}\pa_\mu g_{S^3}=g_{S^3}L_a~,
\eeq
given explicitly by 
\be
\begin{aligned}
\zeta_1=&-i\frac{\cos\phi_-}{\sin 2\rho}\pa_++\frac i2\sin\phi_-\pa_\rho+i\cot2\rho\cos\phi_-\,\pa_-~,\\
\zeta_2=&i\frac{\sin\phi_-}{\sin2\rho}\pa_++\frac i2\cos\phi_-\pa_\rho-i\cot2\rho\sin\phi_-\,\pa_-~,\\
\zeta_3=&i\pa_-~,\\
\bar\zeta_1=&-i\frac{\cos\phi_+}{\sin 2\rho}\pa_-+\frac i2\sin\phi_+\pa_\rho+i\cot2 \rho\cos\phi_+\pa_+~,\\
\bar\zeta_2=&-i\frac{\sin\phi_+}{\sin2\rho}\pa_--\frac i2\cos\phi_+\pa_\rho+i\cot2\rho\sin\phi_+\pa_+~,\\
\bar\zeta_3=&-i\pa_+~,
\end{aligned}
\ee
where $\phi_\pm=\tau\pm\varphi$.  The left (right)-Maurer Cartan forms,\footnote{These should be not be confused with the $\sigma$'s appearing as integration variables in Abelianisation or localization, e.g., \eqref{eq:WexpAb}.}
\beq
\sigma^aL_a=-\dd g_{S^3}g_{S^3}^{-1}~,\qquad \bar\sigma^aL_a=g^{-1}_{S^3}\dd g_{S^3}~,
\eeq
are dual to these Killing vectors: $\sigma^a_\mu\zeta_b^\mu=\bar\sigma^a_\mu\bar\zeta_b^\mu=\delta^a_b$.  Note that $\sigma^a$ and $\bar\sigma^a$ are related by conjugation.  As a result either serves as a valid metric frame, where
\beq
g_{\mu\nu}=-\frac{1}{2}\mTr_f\left(\sigma_\mu\sigma_\nu\right)=-\frac{1}{2}\mTr_f\left(\bar\sigma_\mu\bar\sigma_\nu\right)~,
\eeq
and the associated metric is the round three-sphere metric
\beq\label{eq:S3metric}
g_{\mu\nu}\dd x^\mu \dd x^\nu=\dd\rho^2+\cos^2\rho\,\dd\tau^2+\sin^2\rho\,\dd\varphi^2~.
\eeq
It is also easy to verify that the forms satisfy the Maurer-Cartan structure equations
\beq
\dd\sigma+\sigma\wedge\sigma=0~,\qquad \dd\bar\sigma+\bar\sigma\wedge\bar\sigma=0~.
\eeq
By direct computation we discover that the $\mf{su}(2)_L\oplus\mf{su}(2)_R$ quadratic Casimir, expressed as vector fields, is precisely the scalar Laplacian on the background \eqref{eq:S3metric}, determined by $\sigma$ (or $\bar\sigma$). That is,
\beq\label{eq:S3castolap}
-2\left(c_2^{(L)}+c_2^{(R)}\right)=-2\left(\delta^{ab}\zeta_a\zeta_b+\delta^{ab}\bar\zeta_a\bar\zeta_b\right)=\nabla^2_{S^3}~.
\eeq
In fact each Casimir separately satisfies $c_2^{(L)}=c_2^{(R)}=-\frac{1}{4}\nabla^2_{S^3}$. 

Now we want to generalize this to curved three-manifolds, expressing $\nabla^2_M$ as a Casimir acting on representations of a {\it local} $\mf{su}(2)_L\oplus\mf{su}(2)_R$ action.  To this end, it is useful to model a three-manifold, $M$, by ``locally tangent Euclidean de Sitter spaces" (as opposed to tangent $\mathbb R^3$ vectors spaces, as typical in Riemannian geometry) on which the $\mf{su}(2)$'s act naturally.\footnote{More formally, we are modeling $M$ as a {\it Cartan geometry} based on the homogeneous space $G/H=SO(1,3)/SO(1,2)$ for which $A_L\oplus A_R$ is the Wick rotation of its Cartan connection, $\mathscr A$.  We will not need any heavy machinery from this formalism, however it does give us a nice framework for organizing the thoughts of this appendix.  See \cite{Wise:2006sm} for a friendly review.} 
Actually, it is somewhat useful to first think about modelling $M$ in {\it Lorentzian} signature, where dS$_3$ admits a natural quotient structure: dS$_3=SO(1,3)/SO(1,2)$.  A basis of $\mf{so}(1,3)$ making this manifest is $\{\mc J_a,\mc P_a\}_{a=1,2,3}$  in \eqref{eq:alg1234}. 
The subgroup $\mf{so}(1,2)=\text{span}\{\mc J_a\}$ consists of boosts/rotation while $\mf{so}(1,3)\ominus\mf{so}(1,2)=\text{span}\{\mc P_a\}$ are translations.  The statement, ``dS$_3=SO(1,3)/SO(1,2)$'' is simply equivalent to the statement that ``points'' are objects that are stabilized by boosts/rotations but not by translations.  A function on this quotient is a scalar when it is boost/rotation invariant:
\beq
\Phi\text{ is a scalar on dS$_3$}\qquad\Leftrightarrow\qquad \mc J_a\Phi=0~.
\eeq
Using the map from $\mf{so}(1,3)$ to $\mf{su}(2)\oplus\mf{su}(2)$ in App.\,\ref{app:conventions}, this Wick rotates to Euclidean signature as the $\mf{su}(2)_L\oplus\mf{su}(2)_R$ condition
\beq\label{eq:Ishicond}
(L_3-\bar L_3)\Phi=(L_+-\bar L_-)\Phi=(L_--\bar L_+)\Phi=0~,%\qquad L_\pm=L_1\pm iL_2~,
\eeq
which is precisely the {\it Ishibashi condition} from \cite{Castro:2018srf,Castro:2020smu}.  This is a concrete reason why the $\mf{su}(2)_L\oplus\mf{su}(2)_R$ Ishibashi states are naturally connected with de Sitter geometry \cite{Castro:2020smu}.

Returning to our manifold, $M$, since we are describing the gravity path-integral about the $S^3$ saddle, we will let $M$ be diffeomorphic to $S^3$.  We can then model $M$ as ``locally de Sitter'' through a map $\mc G:S^3\rightarrow SU(2)$ locally satisfying the Ishibashi condition, \eqref{eq:Ishicond}.  The curvature of $M$ is expressed through the coupling to $A_L$ and $A_R$, in the following way.  We fix a fiducial point, $x_0\in S^3$ and a fiducial group element $g_0\in SU(2)$.  We require that $\mc G$ is determined at another nearby point, $x$, via parallel transport along a curve $\gamma:x_0\rightarrow x$ while maintaining the Ishibashi condition
\beq\label{eq:Gparalleltrans}
\mc G(x)=\mc P\exp\left(\int_{x_0}^xA_L\right)g_0\,\mc P\exp\left(-\int_{x_0}^x\tilde A_R\right)~,\qquad\tilde A_R=\Sigma_{\text{Ish}}\,A_R\,\Sigma_{\text{Ish}}^{-1}~,
\eeq
where $\Sigma_{\text{Ish}}\left(\cdot\right)\Sigma_{\text{Ish}}^{-1}$ is the intertwiner from $\mf{su}(2)_R\rightarrow \mf{su}(2)_L$ determined by \eqref{eq:Ishicond}.  Details of this map can be found in \cite{Castro:2020smu}.  We can now follow the recipe at the beginning of this appendix to build vector fields,  $\{\mc L_a=\mc L_a^\mu\pa_\mu\}_{a=1,2,3}$ and $\{\barmc L_a=\barmc L_a^\mu\pa_\mu\}_{a=1,2,3}$, corresponding to local $\mf{su}(2)$ action:
\beq\label{eq:localsu2action}
\mc L_a^\mu\pa_\mu\mc G(x)=-L_a\mc G(x)~,\qquad\barmc{L}_a^{\mu}\pa_\mu\mc G(x)=\mc G(x)L_a~.
\eeq
Differentiating equation \eqref{eq:Gparalleltrans}, we find
\be\label{eq:curvedLadef}
\begin{aligned}
&\mc L_a^\mu\mTr_f\left((A_{L\mu}-\mc G \tilde A_{R\mu}\mc G^{-1})L_b\right)=-\frac{1}{2}\delta_{ab}~,\\
&\barmc L_a^{\mu}\mTr_f\left((\mc G^{-1}A_{L\mu}\mc G-\tilde A_{R\mu})L_b\right)=\frac{1}{2}\delta_{ab}~.
\end{aligned}
\ee
When $A^a_{L\mu}-(\mc G\tilde A_{R\mu}\mc G^{-1})^a$ is invertible this can be solved for $\mc L_a$ and $\barmc L_a$.  Invertibility also implies that we can regard
\beq\label{eq:MCformdef}
\sigma_\mu=-\left(A_{L\mu}-\mc G\tilde A_{R\mu}\mc G^{-1}\right)~,\qquad \bar\sigma_\mu=\left(\mc G^{-1}A_{L\mu}\mc G-\tilde A_{R\mu}\right)~,
\eeq
as co-frames defining a non-degenerate metric on $M$
\beq\label{eq:Gmetric}
g^{M}_{\mu\nu}=-\frac{1}{2}\mTr_f\left(\sigma_\mu\sigma_\mu\right)=-\frac{1}{2}\mTr_f\left(\bar\sigma_\mu\bar\sigma_\mu\right)~,
\eeq
with $\mc L$ and $\barmc L$ their frames.  Because $\sigma$ and $\bar\sigma$ are dual one-forms they still satisfy Maurer-Cartan structure equations
\beq\label{eq:MCstructure}
\dd\sigma^a+i\varepsilon^{abc}\sigma^b\wedge\sigma^c=\dd\bar\sigma^a+i\varepsilon^{abc}\bar\sigma^b\wedge \bar\sigma^c=0~,
\eeq
which can easily be verified by noticing that differentiating \eqref{eq:Gparalleltrans} and using \eqref{eq:MCformdef} implies $\dd\mc G=-\sigma^aL_a\mc G=\mc G\bar\sigma^aL_a$.  It is important to note that $\sigma$ and $\bar\sigma$ {\it identically satisfy} \eqref{eq:MCstructure} without imposing any additional flatness conditions on the connections, $A_L$ and $A_R$.  Instead we should regard them as a special choice of coframe.

We now use $\mc L_a$ and $\barmc L_a$ to build a {\it curved Casimir}, $\mc C_2$, as
\beq\label{eq:CC}
\mc C_2=\delta^{ab}\mc L_a\mc L_a+\delta^{ab}\barmc L_a\barmc L_b~.
\eeq
In what follows we will show that $\mc C_2$ is equivalent to the Laplacian associated to the metric, \eqref{eq:Gmetric}.  The argument is fairly simple: given a set of frames, $E_a^\mu$, and associated coframes, $e^a_\mu$, for a metric, $g_{\mu\nu}$, the scalar Laplacian can be written as
\beq
\nabla^2_{g}=\delta^{ab}E_aE_b+\delta^{ab}\text{div}(E_a)E_b~,
\eeq
where the divergence of a vector field, $\text{div}(X)$, is defined implicitly through the volume form, $\Omega_e=e^1\wedge e^2\wedge\ldots \wedge e^d$, as $\mathscr L_X\Omega_e=\text{div}(X)\Omega_e$ (here $\mathscr L_X$ is the Lie-derivative). Given that both $\sigma^a$ and $\bar\sigma^a$ are valid coframes (up to normalization) for $g_{\mu\nu}^{M}$ it follows that
\beq
\nabla^2_{M}=-2\delta^{ab}\left(\mc L_a\mc L_b+\text{div}(\mc L_a)\mc L_b+\barmc L_a\barmc L_b+\text{div}(\barmc L_a)\barmc L_b\right)~.
\eeq
Lastly we note that for co-frames satsifying Maurer-Cartan structure equations, \eqref{eq:MCstructure}, their associated frames are divergenceless:
\beq
\mathscr L_{\mc L_a}\Omega_\sigma=\dd\,i_{\mc L_a}\Omega_\sigma=\frac{1}{2}\varepsilon_{abc} \dd\left(\sigma^b\wedge\sigma^c\right)=i\varepsilon_{abc}\varepsilon^{bdf}\sigma^d\wedge\sigma^f\wedge\sigma^c=i\varepsilon_{abc}\varepsilon^{bdf}\varepsilon^{dfc}\,\Omega_\sigma=0~,
\eeq
and so the Laplacian is exactly the curved Casimir, \eqref{eq:CC},
\beq
\mc C_2=-\frac{1}{2}\nabla^2_{M}~.
\eeq
This gives us confidence that after deforming $A_L$ and $A_R$ away from their $S^3$ saddle-point value, the Wilson spool, \eqref{eq:spooldef}, still computes a determinant of a Laplacian on the geometry created by $A_L$ and $A_R$.  Again, we emphasize that in the course of this construction, no flatness conditions have been imposed on $A_L$ and $A_R$: we have instead used local $\mf{su}(2)\oplus\mf{su}(2)$ action to find a special basis of frames for which the Laplacian takes a simple form.\footnote{In fact, a simple counting of constraints implies that, under mild assumptions, a given metric generically has a basis of divergenceless frames in three dimensions.  Namely the divergenceless conditions, $\text{div}(E_a)=0$, is a set of $d$ first order partial differential equations for a set of $d^2-\frac{d(d+1)}{2}=\frac{d(d-1)}{2}$ unknowns (the ``matrix elements'' of $E^\mu_a$ subtracted by their $O(d)$ redundancy).  In $d=3$ these equations are precisely determined and so divergenceless frames generically exist and will be unique up to a specification of boundary conditions (such as regularity at the fiducial point, $x_0$).}

%\bibliographystyle{JHEP}
%\bibliography{dSWilsonRefs.bib}

\providecommand{\href}[2]{#2}\begingroup\raggedright\endgroup

\end{document}